\documentclass[namedreferences,optionalrh,hyperref]{spr-sola}

\usepackage[utf8]{inputenc}
\usepackage{graphicx}
\graphicspath{{figures/}{./}{../figures/}}
\usepackage{tabularx}

\usepackage{newtxtext}
\usepackage{mathastext}

\usepackage[T1]{fontenc}

\usepackage{color}
\usepackage[shortlabels]{enumitem}
\usepackage{subfig} %
\usepackage{datetime}

\usepackage{pifont}%
\newcommand{\cmark}{\ding{51}}%
\newcommand{\xmark}{\ding{55}}%

\usepackage{siunitx}
\newcommand{\lnum}[1]{\num[group-digits = integer,group-separator = {,}]{#1}}

\newcolumntype{C}[1]{>{\centering\arraybackslash}p{#1}}
\newcolumntype{L}[1]{>{\raggedright\arraybackslash}p{#1}}
\newcolumntype{R}[1]{>{\raggedleft\let\newline\\\arraybackslash\hspace{0pt}}m{#1}}
\newcolumntype{^}{>{\currentrowstyle}}

\newcommand{\cmd}{\begingroup
  \catcode`_=13 
  \cmdint}
\newcommand{\cmdint}[1]{%
  {\scantokens{#1\noexpand}}%
  \endgroup
}

\usepackage[printonlyused,nohyperlinks]{acronym}

\definecolor{mpg}{rgb}{0.,0.48627451,0.4745098}
\definecolor{mpgdark}{rgb}{0.,0.36470588, 0.35588235}
\definecolor{linka}{rgb}{0.0,0.0,0.7}
\definecolor{linkb}{rgb}{0.0,0.0,0.5}
\definecolor{linkc}{rgb}{0.0,0.0,0.3}
\definecolor{cmt}{rgb}{0.5,0.0,0.0}
\definecolor{todo}{rgb}{0.6,0.4,0.8}
\definecolor{shade}{rgb}{0.7,0.7,0.7}
\definecolor{akl}{rgb}{0.6,0.2,0.0}
\definecolor{jcdti}{rgb}{0.6,0.2,0.6}
\definecolor{sn}{rgb}{0.8,0.2,0.6}
\definecolor{sks}{rgb}{0.8,0.1,0.0}
\definecolor{pnb}{rgb}{1.0,0.49,0.0}

\newcommand{\colfig}[3][1.0]{
  \begin{figure}
    \centering
    \includegraphics[width=#1\linewidth,clip=TRUE]{#2}
    \caption{#3}
    \label{#2}
  \end{figure}
}
\newcommand{\colfigwide}[3][1.0]{
  \begin{figure*}
    \centering
    \includegraphics[width=#1\linewidth,clip=TRUE]{#2}
    \caption{#3}
    \label{#2}
  \end{figure*}
}
\newcommand{\colfigtwo}[4][1.0]{
  \begin{figure*}
    \centering
    \includegraphics[width=#1\linewidth,clip=TRUE]{#2}
    \hfill
    \includegraphics[width=#1\linewidth,clip=TRUE]{#3}
    \caption{#4}
    \label{#2}
  \end{figure*}
}

\newcommand{\docdb}[1]{\ac{docdb} document \href{https://www.mps.mpg.de/solar-physics/sunrise/documents}{\cmd{#1}} \cite[]{#1}}
\newcommand{\aplsrd}{\href{https://www.csbf.nasa.gov/documents/gondola/820-PG-8700.0.1A\%20Gondola\%20Structural\%20Design\%20Req.\%2011.4.pdf}{820-PG-8700.0.1a}}

\newcommand{\ten}[1]{10\textsuperscript{#1}}
\newcommand{\fig}[1]{Figure~\ref{#1}} %
\newcommand{\tab}[1]{Table~\ref{#1}} %
\newcommand{\sref}[1]{Section~\ref{#1}} %

\newcommand{\sunrise}{\textsc{Sunrise}}
\newcommand{\sunrisei}{\textsc{Sunrise~i}}
\newcommand{\sunriseiandii}{\textsc{Sunrise~i} and \textsc{ii}}
\newcommand{\sunriseii}{\textsc{Sunrise~ii}}
\newcommand{\sunriseiii}{\textsc{Sunrise~iii}}
\newcommand{\hinode}{\textit{Hinode}}
\newcommand{\halpha}{H$\alpha$}
\newcommand{\arcsec}{''}
\newcommand{\arcmin}{'}
\newcommand{\spc}{\hbox{S$^3$PC}}

\newcommand{\caii}{Ca\,\textsc{ii}}

\newcommand{\caiihk}{Ca\,\textsc{ii}\,H\&K}
\newcommand{\ki}{K\,\textsc{i}}

\newcommand{\fei}{Fe\,\textsc{i}}

\newcommand{\mgi}{Mg\,\textsc{i}}

\newcommand{\dg}{$^\circ$}

\newcommand{\scip}{\ac{scip}}

\begin{document}

\begin{frontmatter}

\title{\sunriseiii{}: Overview of Observatory and Instruments}

\author[addressref={mps,aalto},corref,email={lagg@mps.mpg.de}]{\inits{A.}\fnm{Andreas}~\lnm{Korpi-Lagg}\orcid{0000-0003-1459-7074}}
\author[addressref={mps},email={gandorfer@mps.mpg.de}]{\inits{A.}\fnm{Achim}~\lnm{Gandorfer}\orcid{0000-0002-9972-9840}}
\author[addressref={mps},email={solanki@mps.mpg.de}]{\inits{S.~K.}\fnm{Sami~K.}~\lnm{Solanki}\orcid{0000-0002-3418-8449}}
\author[addressref={iaa,S3PC},email={jti@iaa.es}]{\inits{J.~C.}\fnm{Jose~Carlos}~\lnm{del~Toro~Iniesta}\orcid{0000-0002-3387-026X}}
\author[addressref={naoj,unitokyo,grad},email={yukio.katsukawa@nao.ac.jp}]{\inits{Y.}\fnm{Yukio}~\lnm{Katsukawa}\orcid{0000-0002-5054-8782}}
\author[addressref={apl},email={pietro.bernasconi@jhuapl.edu}]{\inits{P.}\fnm{Pietro}~\lnm{Bernasconi}\orcid{0000-0002-0787-8954}}
\author[addressref={kis},email={thomas.berkefeld@leibniz-kis.de}]{\inits{T.}\fnm{Thomas}~\lnm{Berkefeld}}
\author[addressref={mps},email={feller@mps.mpg.de}]{\inits{A.}\fnm{Alex}~\lnm{Feller}\orcid{0009-0009-4425-599X}}
\author[addressref={mps},email={riethmueller@mps.mpg.de}]{\inits{T.~L.}\fnm{Tino~L.}~\lnm{Riethmüller}\orcid{0000-0001-6317-4380}}
\author[addressref={inta,S3PC}]{\inits{A.}\fnm{Alberto}~\lnm{Álvarez-Herrero}\orcid{0000-0001-9228-3412}}
% ...

\address[id={mps}]{Max-Planck-Institut für Sonnensystemforschung, Justus-von-Liebig-Weg 3, 37077 Göttingen, Germany}
\address[id={iaa}]{Instituto de Astrofísica de Andalucía, CSIC, Glorieta de la Astronomía s/n, 18008 Granada, Spain}
\address[id={naoj}]{National Astronomical Observatory of Japan, 2-21-1 Osawa, Mitaka, Tokyo 181-8588, Japan}
\address[id={apl}]{Johns Hopkins University Applied Physics Laboratory, 11100 Johns Hopkins Road, Laurel, Maryland, USA}
\address[id={kis}]{Institut für Sonnenphysik (KIS), Georges-Köhler-Allee 401a, 79110 Freiburg, Germany}
% ...
\address[id={S3PC}]{Spanish Space Solar Physics Consortium (\href{https://s3pc.es}{\spc})}
\address[id={inta}]{Instituto Nacional de T\'ecnica Aeroespacial (INTA), Ctra. de Ajalvir, km. 4, E-28850 Torrejón de Ardoz, Spain}
% ...
\address[id={unitokyo}]{Department of Earth and Planetary Science, The University of Tokyo, 7-3-1, Hongo, Bunkyo-ku, Tokyo 113-0033, Japan}
% ...
\address[id={aalto}]{Aalto University, Department of Computer Science, Konemiehentie 2, 02150 Espoo, Finland}
% ...
\address[id={grad}]{Department of Astronomical Science, The Graduate University for Advanced Studies (SOKENDAI), 2-21-1 Osawa, Mitaka, Tokyo 1818588, Japan}
% ...

\date{Received \today; \currenttime}

\begin{abstract}
    In July 2024, \sunrise{} completed its third successful science flight. The \sunriseiii{} observatory had been upgraded significantly after the two previous successful flights in 2009 and 2013, to tackle the most recent science challenges concerning the solar atmosphere. Three completely new instruments focus on small-scale physical processes and their complex interaction from the deepest observable layers in the photosphere up to chromospheric heights. Previously poorly explored spectral regions and lines are exploited to paint a  three-dimensional picture of the solar atmosphere with unprecedented completeness and level of detail.
    
    The full polarimetric information is captured by all three instruments to reveal the interaction between the magnetic fields and the hydrodynamic processes. Two slit-based spectropolarimeters, the \acl{susi} (\acs{susi}) and the \acl{scip} (\acs{scip}), focus on the \acl{nuv} (309--417\,nm) and the \acl{nir} (765--855\,nm) regions respectively, and the imaging spectropolarimeter \acl{tumag} (\acs{tumag})  simultaneously obtains maps of the full \acl{fov} of $46\times46$\,Mm$^2$ in the photosphere and the chromosphere in the visible (525 and 517\,nm). The instruments are operated in an orchestrated mode, benefiting from a new \acl{islid} (\acs{islid}), with the \acl{cws} (\acs{cws}) providing the autofocus control and an image stability with a \acl{rms} value smaller than 0.005\arcsec{}. A new gondola was constructed to significantly improve the telescope pointing stability, required to achieve uninterrupted observations over many hours.
    
    \sunriseiii{} was launched successfully on 10 July 2024, from the \aclu{esrange} of the \acl{ssc} near Kiruna (Sweden). It reached the landing site between the Mackenzie River and the Great Bear Lake  in Canada after a flight duration of 6.5 days. In this paper, we give an overview of the \sunriseiii{} observatory and its instruments.
\end{abstract}



\end{frontmatter}

\section{Introduction}

\sunrise{} is the most powerful solar telescope ever to have left the ground. It escapes the disturbing influence of air turbulences by floating in the stratosphere above more than 99\% of the Earth's atmosphere. This location opens up the important \ac{nuv} spectral range for study, which is hard to access from ground. During two successful flights in 2009 and 2013, \sunriseiandii{}  demonstrated the potential of the observatory. The  magnetic field and velocity data and the \ac{uv} images of highest spatial and temporal resolution unaffected by atmospheric turbulence provided new insights into small-scale magnetic fields in the photosphere of the Sun. 

In 2016, the work on \sunriseiii{} began. The observatory has been upgraded considerably, with three new science instruments, improved light distribution and image stabilization systems, and a new gondola. This expands the scientific scope of \sunrise{} significantly, especially since it is now possible to seamlessly probe a greater range of heights in the solar atmosphere, ranging from the solar surface to the middle chromosphere. To achieve this goal, the new science instruments now cover the solar spectrum from the \ac{nuv} to the \ac{nir}. The \acf{susi}, a novel \ac{nuv} spectropolarimeter built by the \acl{mps} (\acsu{mps}, Germany), explores and exploits the very rich, but, due to poor accessibility from the ground, so far little studied spectral range between 309 and 417\,nm. The \acf{scip}, a \ac{nir} spectropolarimeter, built jointly by the \acl{naoj} (\acsu{naoj}, Japan), the \acl{s3pc}\footnote{The \ac{s3pc} is a consortium formed by the Spanish research institutes \ac{iaa}, \ac{inta}, \ac{unival}, \ac{upm}, and \ac{iac} under the leadership of \acs{iaa} (\href{https://s3pc.es}{https://s3pc.es}).} (\acsu{s3pc}, Spain) and \acsu{mps}, specializes on the chromospheric \caii{} infrared lines and on the upper photospheric \ki{} lines around 770\,nm, poorly accessible from the ground. The \acf{tumag}, a very fast tunable imaging spectropolarimeter able to access multiple spectral lines and built by \ac{s3pc}, delivers the high time and spatial resolution full \ac{fov} information in both, the chromospheric and the photospheric layers, quasi-simultaneously.

A completely new gondola, designed and built by the \acl{apl} (\acsu{apl}, USA),  provides the pointing stability allowing the science instruments to benefit from the absence of a day-night cycle to obtain ultra-long time series. In combination with the \ac{cws}, contributed by the \acl{kis} (\acsu{kis}, Germany), an unprecedented pointing accuracy better than 0.005\arcsec{} \ac{rms} is achieved over multiple hours, corresponding to a distance on the Sun of only 3.5\,km. 
In addition, the \ac{cws} controls the optimum focus of the telescope in closed-loop and provides the information for the correction of a residual coma. 
The new instrument suite required a complete redesign of the \acf{islid}, containing not only the mirrors and beam splitters to feed the science instruments, but also the fast tip/tilt mirror controlled by the \ac{cws} for
compensating residual
image motion not corrected by the gondola. A powerful \acf{ics} houses a multi-node workstation and the \ac{dss}, capable of controlling all the instruments autonomously, storing science and \ac{hk} data, and responsible for the communication and commanding between ground station and observatory during the 6.5 days long flight.

\sunriseiii{} was launched on 10 July 2024, at the very end of the 40-day long launch window, defined by two boundary conditions: the sunset-free time around  northern hemisphere summer solstice, and the presence of circumpolar stratospheric winds transporting \sunriseiii{} westwards from \aclu{esrange} near Kiruna, Sweden, to north-western Canada at a float altitude of 33--37\,km (see \sref{sunriseiii2024}). During the flight, \sunriseiii{} completed its scientific program on a zero-pressure \acf{ldb} of \acsu{nasa}'s scientific balloon program (\acl{bpo}, \acsu{bpo}), operated by the \acf{csbf}.

With this paper, we describe the scientific goals of the mission, and present an overview of the \sunriseiii{} hardware and its technical specifications. 
The paper is structured as follows: We start with the history of  the \sunrise{} project (\sref{history}), followed by a description of the scientific topics to be addressed with \sunriseiii{} (\sref{science}). The observatory with all its subsystems is described in \sref{instruments}, including the thermal concept. \sref{testing} summarizes the most important tests and the path towards flight readiness. We conclude the paper with a brief overview of the  scientific performance and the data concept and policy in \sref{performance}, and present a summary in \sref{outlook}.

\section{\sunrise{} History}\label{history}

The \sunrise{} project is one of the long-term scientific pillars of \ac{mps}. It started already in the beginning of this millennium at the \ac{mpae} in Katlenburg-Lindau (Germany), which was renamed to \ac{mps} in 2004.
In this section, we briefly recapitulate the history of the \sunrise{} observatory and its science, including the successful flight of \sunriseiii{} in 2024.

\subsection{\sunriseiandii{}}

The \sunrise{} observatory is composed of a telescope with a 1\,m aperture. For the first two flights, it was equipped with two scientific instruments, the \ac{imax} and the \ac{sufi}. Additional important subsystems were the \ac{cws} providing the signal needed to stabilize images to the necessary precision, the \ac{islid}, the on-board computer and data storage system, and a protective gondola. The instrumentation flown during the first flight of \sunrise{} in June 2009 has been described by \cite{barthol11,martinezpillet11,gandorfer11}; and \cite{berkefeld11}. 

Two successful long-duration science flights on a zero-pressure stratospheric balloon took place in June 2009 (under solar activity minimum conditions; referred to as \sunrisei{}) and in June 2013 (at a relatively high activity level; \sunriseii{}). The missions were operated and launched from \acs{esrange} \acf{ssc} in Kiruna, Sweden. After launch, the balloon slowly drifted westward, reaching northern Canada after 5 days, where the payload landed safely without major damage. The entire payload was successfully recovered after both flights. Overviews of the two flights (\sunriseiandii{}),  the instrumentation, the data gathered, and of some of the early science are given by \cite{solanki10,solanki17}, 
respectively. \sunriseiandii{} provided data that are diffraction-limited at wavelengths as short as 300\,nm, making these the highest resolution solar images until then. They also provided the highest resolution solar data at 214\,nm and at 279.5\,nm and excellent polarimetric precision in the visible wavelengths. 

The first scientific results from the two flights were published in special issues of  \textit{The Astrophysical Journal Letters} \cite[see][]{solanki10}  and \textit{Astrophys. J. Suppl. Ser.} \cite[see][]{solanki17,solanki17a}, respectively.

\subsection{\sunriseiii{}}\label{sunriseiii-22}

The data from \sunriseiandii{} were mainly sensitive to the solar photosphere and the lowest layers of the chromosphere. They could not provide detailed insight into the influence of the small-scale dynamical and magnetic photospheric phenomena on the overlying layers of the atmosphere, in particular the chromosphere. 
\sunriseiii{} carries dedicated instruments that  spectropolarimetrically sample both the photospheric sources and the chromospheric response in a multitude of spectral lines over a broad wavelength range from about 309 to 855\,nm. 

The \sunriseiandii{} experience as well as general advances in solar physics have shown the need for seeing-free observations at high resolution that make simultaneous use of narrow-band polarimeters, broad-band imagers, and scanning slit-spectropolarimeters. These instruments complement each other by, on the one hand, allowing to follow rapidly evolving features,
while, on the other hand, providing detailed diagnostics and the 3-dimensional stratification of the relevant physical parameters in the solar atmosphere, including the magnetic field. Such a carefully balanced set of instruments was chosen to form the core of the new \sunrise{} payload.

The \acl{susi} \cite[\acsu{susi}, see \sref{susi},][]{Feller2020,susi25} on \sunriseiii{} images and spectropolarimetrically explores the wavelength range 309--417\,nm at high spatial resolution. 
The possibilities of this wavelength range are exciting, as it provides a large number of lines that can sample the chromosphere, and also resolve weak magnetic fields using the Hanle effect. Furthermore, the high density of spectral lines allows analysing their polarization signals jointly to increase the \ac{s2n} ratio and also enables sampling many heights in the solar atmosphere simultaneously. 

The \acl{tumag} \cite[\acsu{tumag}, see \sref{tumag},][]{alvarez-herrero2022,tumag25} performs narrow-band imaging polarimetry in the visible, employing the \fei~525.02/525.06\,nm photospheric line pair, with the first line being highly magnetically sensitive (effective Landé factor $g=3$),  plus the \mgi~517.27\,nm line,  sampling the boundary between the chromosphere and photosphere. \ac{tumag} allows dynamic events to be followed and connections across the full \ac{fov} of $46\times46$\,Mm$^2$ to be made. 

Spectropolarimetry in two of the \caii{} infrared triplet lines, carried out by the \acl{scip} \cite[\acsu{scip}, see \sref{scip},][]{katsukawa2020,scip25}, provides diagnostics with a proven capability of deducing the 3D structure of the chromospheric magnetic field through \ac{nlte} inversions. In addition, the line core is formed particularly high in the solar atmosphere, and with the high Zeeman sensitivity and photon flux, a good \ac{s2n} ratio can be obtained already at rather short exposure times. Together, these attributes allow high precision polarimetry at high spatial and temporal resolution. The temperature minimum heights above the photosphere are probed with the \ki{} 769.9/766.5 \,nm lines, which have extremely high velocity sensitivities, and are poorly accessible from the ground due to telluric blends.

However, the full power of \sunriseiii{} lies in simultaneously probing (through both spectropolarimetry and imaging) the \ac{nuv}, while simultaneously carrying out narrow-band polarimetric imaging in the visible and spectropolarimetry in the infrared. Together, this makes \sunriseiii{} a unique observatory, with capabilities going far beyond those that were available during the first two flights. 
One shortcoming of previous flights was the relatively short continuous time series, due partly to the limited time for which the gondola could maintain stability. This has been countered by improvements in the internal image stabilization system, the \ac{cws} \cite[]{cws25}, and by employing a new gondola \cite[see \sref{gondola},][]{gondola25}.

The new instrumentation and the improvements in the image stabilization required a complete new design for the \ac{islid} and the \ac{pfi} platform (see \sref{pfi}). 
A redesign of the \ac{hrw}, which limited the length of time series during the latter part of the \sunriseii{} mission, ensures that it poses no such limitations during the flight of \sunriseiii{}. A new \ac{ics} (see \sref{ics}) was developed to cope with the increased complexity of the observatory, and the significantly higher data rates from the scientific instruments. Consequently, also the harness concept and the onboard network infrastructure had to be renewed.

\sunriseiii{} had an unsuccessful flight on 10 July 2022, caused mainly by non-ideal launch conditions. The mission had to be aborted only 6~hours after the launch, because the launch-lock mechanism, designed to hold the telescope in place in the gondola during the launch and the ascent/descent phases, did not withstand the unduly strong acceleration during the release of \sunriseiii{} from the launch vehicle. Strong winds at an altitude of 100 to 300\,m above the launch pad resulted in an approximately ten times higher release shock than  typical for previous launches of balloon payloads. The bolts of the launch-lock mechanism were not designed for this shock and sheared off, allowing the telescope to move freely in elevation. Unfortunately, the telescope got stuck at an elevation of $\approx\,$60\dg{}, and all attempts to move the telescope during the short excursion into the stratosphere failed. The landing of the observatory, still in Sweden close to the Norwegian border, was rather smooth and the observatory could be recovered after 10 days with marginal damage to the scientific payload.

As part of the preparations for the re-flight in 2024, the observatory was cleaned of any dirt accumulated during the 10 days spent on a Swedish hillside, refurbished and thoroughly tested. This included re-coating the main and the secondary mirrors (M1 and M2) as well as the folding mirrors M3 and M4, a repair of the tilt mechanism for the etalon within \ac{tumag}, the high-precision re-alignment of the telescope and the scientific instruments, and the detailed testing of all scientific instruments and mechanisms. In addition, the launch-lock mechanism was strengthened to withstand even rougher launch conditions than those seen during the 2022 launch.

\subsubsection{\sunriseiii{} flight in 2024}
\label{sunriseiii2024}

\begin{figure}
  \centering
  \includegraphics[width=1.00\textwidth]{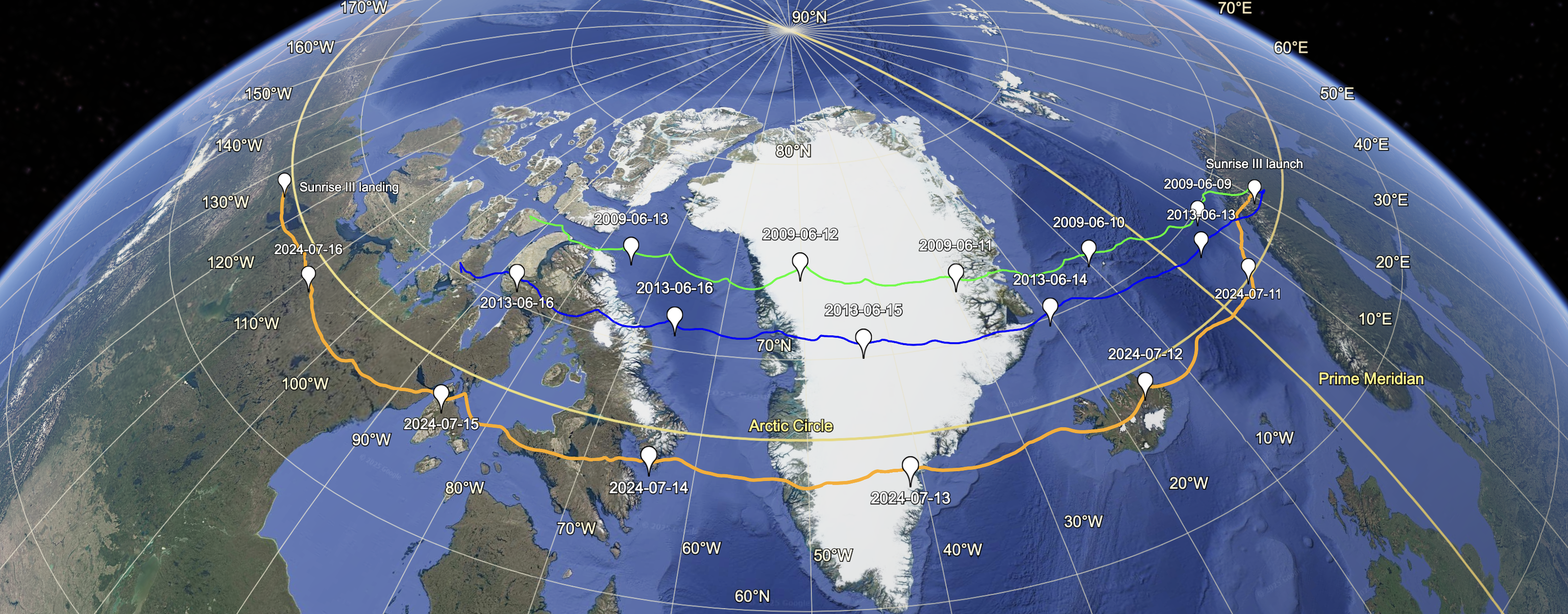}\\
  \includegraphics[width=1.00\textwidth]{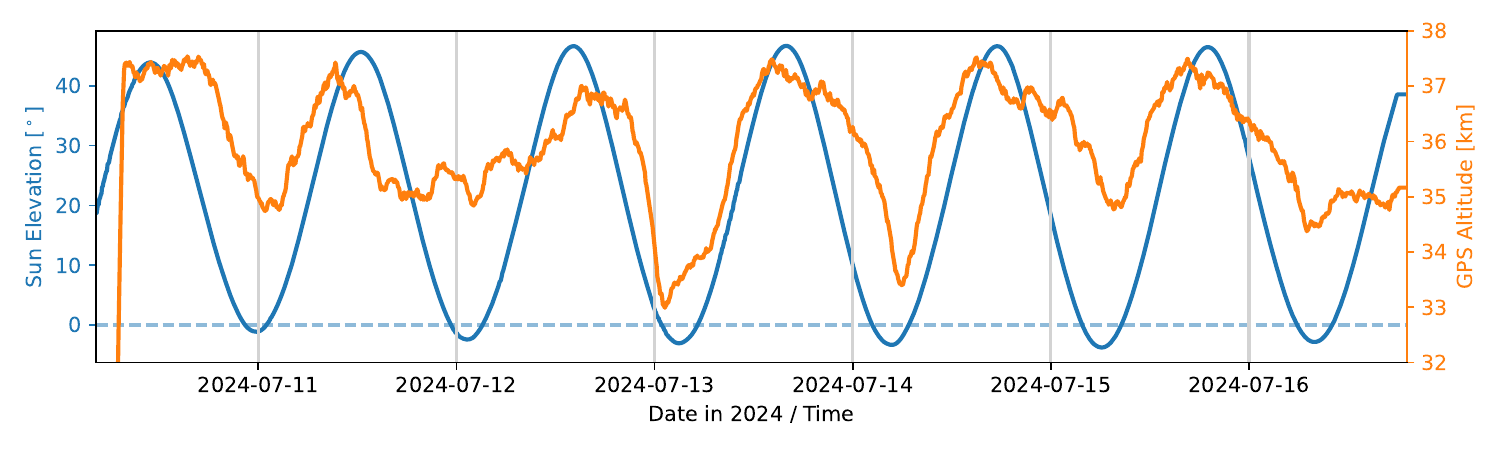}
  \caption{\textsl{Top}: Trajectories of the three successful \sunrise{} flights: \sunrisei{} (\textsl{green}), \sunriseii{} (\textsl{blue}), and \sunriseiii{} (\textsl{orange}). \textsl{Bottom}: Sun elevation angle (\textsl{blue}) and flight altitude obtained from the gondola \acs{gps} receiver (\textsl{orange}) for \sunriseiii{}. Background map: Google Earth (data attribution: Google Landsat / Copernicus Data SIO, NOAA, U.S. Navy, GEBCO IBCAO U.S. Geological Survey INEGI).}  
\label{Sunrise_trajectories}
\end{figure}

Exactly two years after the 2022 mishap, \sunriseiii{} started on its second science flight on 10 July 2024, at 04:22:40\,UTC from \ac{esrange} under \ac{csbf} flight number 740N (see \fig{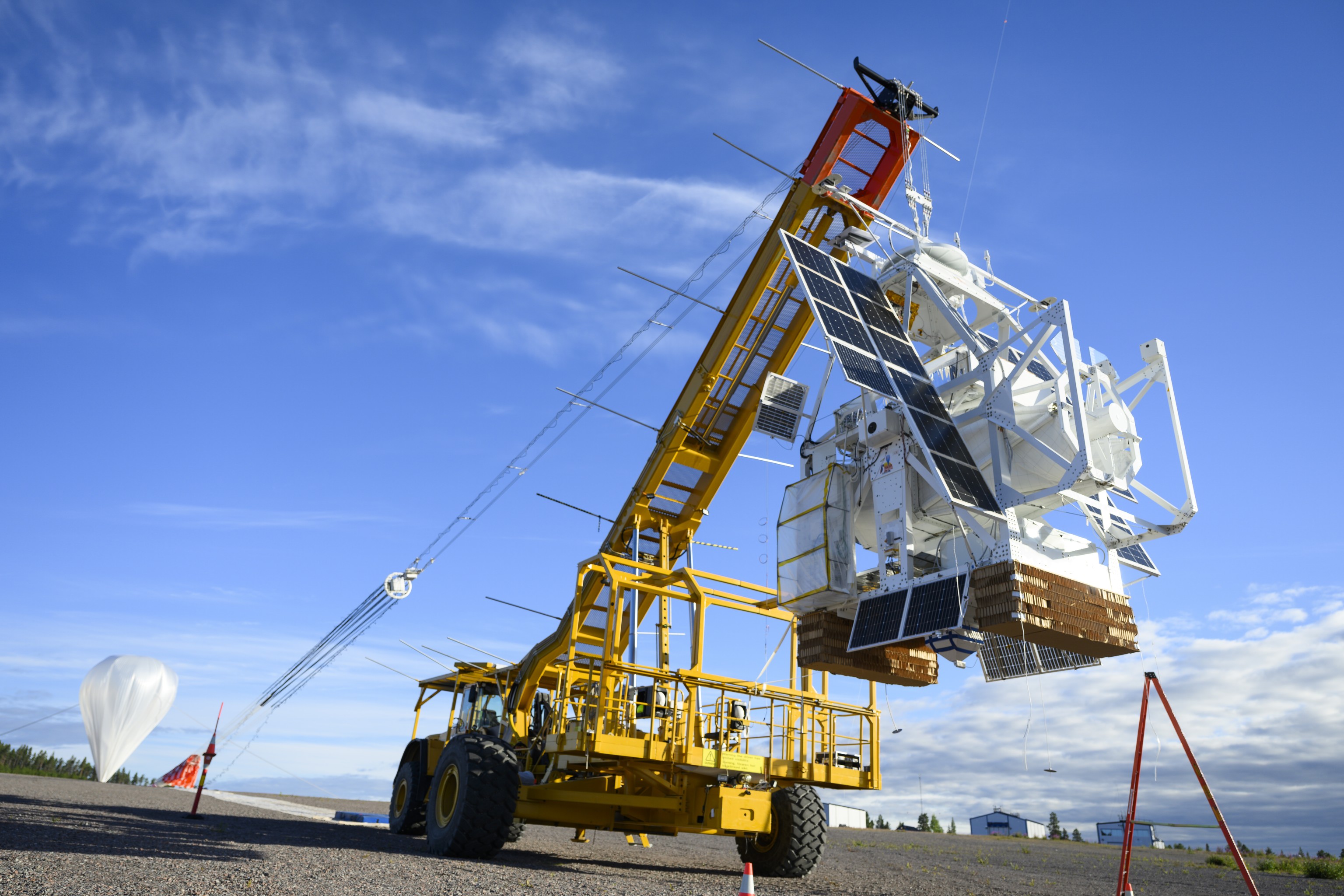}). The release from the launch vehicle was again quite rough, but the reinforcements in the launch-lock mechanism were effective and kept the telescope in place. At 07:34\,UTC the float altitude of 37.5\,km was reached, and the stratospheric float winds transported \sunriseiii{} to the west on a trajectory at a latitude between 64.5\dg{} and 67\dg{}, significantly more southward than the previous two successful flights. The \sunriseiii{} flight trajectories are displayed in \fig{Sunrise_trajectories} (\textsl{top} panel), along with the information about the solar elevation and the flight altitude (\textsl{bottom} panel).

\colfig{SunriseIII_2024-launch.jpg}{\sunriseiii{} a few minutes before the launch on 10 July 2024 on the balloon pad of \ac{esrange} space center. The balloon is already fully inflated, the red and white parachute is attached to the flight train, that stretches over the launch vehicle to the observatory. Image courtesy: \ac{ssc} / Mattias Forsberg.}

Directly after reaching float altitude, the commissioning phase was started at the \ac{goc}, from where all the science operations were executed (see \sref{goc}). All subsystems were operational and in good condition, so that on 10 July at 17:05\,UTC the first scientific program could be started. All pre-defined programs could be executed according to the priorities defined in the daily meetings of the \ac{sswg}. Despite the southward trajectory, \sunriseiii{} did not see any sunset, but the telescope elevation was negative for about 2 to 4 hours around local midnight (the horizon at float altitude is at $-4.9$\dg{}, see \textit{blue  solid} line in \fig{Sunrise_trajectories}, \textsl{bottom}). During this time the observations were affected by the absorption and seeing caused by observing the Sun through the Earth's atmosphere. 

Science operations ended on 16 July, 2024 at 15:00\,UTC. Flight termination occurred at 18:20:54\,UTC at the position 66\dg{}07'47''\,N 127\dg{}12'06''\,W.
The observatory impacted on 16 July, 2024, at 19:09:00\,UTC west of Great Bear Lake in Canada (66\dg{}28'34.8''\,N 127\dg{}00'00''\,W), marking a total flight duration of 6 days, 14 hours and 46 minutes (total float time: 6 days, 10 hours, 47 minutes). The landing site was located in a forest, making the recovery operations more difficult than expected. The recovery team first secured the data vaults, and completed the dismantling of the observatory on 29 July 2024 with the final helicopter transport from the landing site to the airport of Norman Wells. 
A winged airplane continued the transport to Yellowknife Airport 
on 05 August 2024, where the observatory was loaded into two shipping containers, marking the end of the recovery operation. The data vaults were transported to \ac{mps} by the recovery team.

The observatory worked extremely stably and reliably. Two technical issues interrupted scientific observations for a short time. The power supply by the photovoltaic system stopped working twice, forcing \sunriseiii{} to run on batteries. Both times, the battery charging could be reestablished and normal operation resumed. The gondola \ac{pcs} had only one major pointing loss on the last day of the mission, exceeding the scientists' high expectations. The full potential and quality of the \sunriseiii{} dataset can only be assessed after the data have been reduced.

\subsubsection{Scientific performance during the 2024 flight}

During the 6.5-day-long flight, \sunriseiii{} recorded more than 200\,TByte of data. Using the preflight-defined sequences of science observations and calibrations described in \sref{timeline}, and exploiting all the various instruments observing modes, we targeted a wide variety of solar features. The observatory executed successfully all high-priority predefined observing programs. Additionally, through daily science planning meetings, we performed many additional observations based on thorough analysis of the real-time solar features.

The total campaign duration from launch to landing was 155.77\,h. 
First light in flight occurred 3.5\,h after launch with the opening of the aperture door, which stayed open for 97.7\% of the remaining flight duration  until termination. For 102.7\,h (68\%) the \ac{cws} was in state 1, indicating that stable locking and auto-focus was established, either on a solar feature, or on an F2-filter wheel target for calibration purposes (see \sref{islid}) For the remaining 24.6\% of the time the observatory was either in flatfielding mode or in other calibration modes (e.g., taking dark images). The total time of science observation was 68\,h. Other non-observing times covered periods of negative solar elevation (a result of the southern trajectory) and setup procedures (e.g., finding the right target for the observations).

The solar targets covered the quiet Sun, plage and network field regions, a coronal hole, sunspots, and flaring regions. Center to limb variations were recorded, including observations at the limb. All instrument modes were executed, like the full spectral scan of the \ac{susi} instrument over the spectral region from 309--417\,nm, the interlaced observation between the photospheric \fei{} 525\,nm line and the chromospheric \mgi{}b 517.3\,nm line by \ac{tumag}, or the rapid scan mode by \ac{scip}. While \ac{tumag} always observed with the full \ac{fov} of 63\arcsec{}$\times$63\arcsec{}, \ac{susi} and \scip{} varied their scanning region from sit\,\&\,stare observations over narrow, repetitive scans, to 60\arcsec{} large scans, always using the 60\arcsec{} full slit length, and within the \ac{fov} of \ac{tumag}.

\section{\sunriseiii{} Science}\label{science}

\subsection{Science Objectives}

\sunriseiii{} is a high-spatial resolution solar observatory.
The large-aperture solar telescope combined with the state-of-the-art image stabilization allows \sunriseiii{} to resolve solar features down to a size of 60\,km at the shortest wavelength, which is within a factor of two of the theoretical resolution limit of the National Science Foundation's \acl{dkist} \cite[\acs{dkist},][]{DKIST:2020,DKIST:2021} at 600\,nm. The stratospheric vantage point above 99\% of the Earth's atmosphere provides an image stability over multiple hours not achievable from the ground, similar to space-borne instruments, like the \ac{sotsp} and the \ac{sotnfi} of the \ac{sot} aboard \hinode{} \cite[]{kosugi2007,tsuneta08}, the \ac{hmi} on \acl{sdo} \cite[\acs{sdo}:][]{pesnell:12,schou12,scherrer12} or the \ac{phi} on \acl{solo} \cite[]{solanki20,mueller20}. 
Compared to space missions, balloon-borne observatories have less restrictions on payload mass and data rate (data can be stored on-board and recovered), and can rely on standard, off-the shelf components. This allows \sunriseiii{} to be equipped with significantly more complex and versatile instrumentation than possible for a space mission at relatively low costs \cite[see][for a comparison of state-of-the-art instrumentation for spectropolarimetry]{Iglesias2019}. The space-like observing conditions allow to simultaneously cover wavelengths from the \ac{nuv} to the \ac{nir} for nearly the whole 6.5-day long flight. This makes \sunriseiii{} well suited to address important topics in the physics of solar magnetism and its manifestations in the lower solar atmosphere.

\subsubsection{The crucial role of the magnetic field} \label{CrucialMag}

Magnetic fields in the solar atmosphere are the root cause of the many phenomena that together constitute solar activity \cite[e.g.,][]{solanki06a}. Hereby, the interaction of the magnetic field with the solar plasma is of fundamental importance, since it is the key to the conversion of energy between its mechanical, magnetic, radiative, and thermal forms. 
The interaction between the turbulent solar convection \cite[]{nordlund09} and the magnetic field leads to a rich variety of magnetic structures in the photosphere at many scales.

Equally fascinating, but less understood is the solar chromosphere \cite[e.g.,][]{lagg:17,carlsson19}, directly overlying the photosphere. There the magnetic energy density, the thermal energy density and the kinetic energy density in waves and flows are all of roughly equal magnitude, so that they can be easily transformed into each other. This leads to a very dynamic environment, with many complex interactions. These give rise to structures as diverse and dynamic as canopies of magnetic field, jet-like spicules, and prominences of dense material dangling on magnetic field lines high up in the thin  atmosphere. 

Mechanical and magnetic energy is injected into the solar atmosphere in the turbulent and dense photospheric layers. Most of this energy is deposited in the chromosphere, where it heats the gas and drives its dynamics, and much of the rest is converted into another form there. Therefore, it has become increasingly clear in the past few years that to understand the flow of energy in the solar atmosphere, both photosphere and chromosphere must be studied simultaneously. In particular, the magnetic field in these two layers must be investigated in detail at the highest achievable resolution (because many of the key interactions take place at small scales), but also with the best possible sensitivity (to catch the weak magnetic flux that dominates a large part of the quiet Sun).

\subsubsection{The complex magnetic field in the photosphere}

The magnetic field in the solar photosphere shows a very complex and diverse structure \cite[e.g.][]{solanki93}. Magnetic flux concentrations of kG strength appear in a broad range of sizes: from sunspots, with diameters reaching up to \lnum{50000}\,km, to small flux concentrations on a scale of 30--100\,km \cite[]{stein12}. In addition to these strong-field features, turbulent  magnetic fields pervade the photosphere. 

The flux concentrations on the small-scale end of the size distribution have field strengths of 1--2\,kG \cite[]{stenflo73}. The first flight of \sunrise{} (\sunrisei{}) provided the first spatially resolved observations of such small-scale flux concentrations with kG fields in the quiet Sun internetwork \cite[]{lagg10,martinezgonzalez12}. State-of-the-art \ac{mhd} simulations and \sunrise{} data have also shown that the magnetic elements are extremely dynamic. Thus, the convective intensification of such magnetic features in the quiet Sun often only lasts a minute or two and they constantly break up or merge with other features \cite[e.g.,][]{requerey14,requerey17a,anusha17}. To follow these complex dynamics, a near-perfect  and time-independent image quality is required, which is difficult to be achieved from the ground. 

The large magnetic features such as dark pores and sunspots also display a host of dynamic small-scale structures \cite[]{solanki03a}. Numerical simulations indicate that magnetic filamentation is the key to understanding energy transport in sunspot umbrae and the surrounding penumbrae \cite[]{schuessler06,rempel09,rempel11,rempel12}. \sunriseiii{} allows testing the predictions from the simulations (and hence the validity of the simplifications underlying them) by spectropolarimetrically investigating the structure and dynamics of sunspots on their intrinsic spatio-temporal scales, thus revealing the mode(s) of energy transport acting in them. 

Besides the magnetic flux concentrations there are mixed-polarity magnetic features all over the solar surface, constituting a  turbulent magnetic field \cite[]{danilovic10a,borrero17,bellot19}. The field strengths in these features are often quite low. This, together with the mixing of magnetic polarities at very small scales, implies that such fields are very hard to catch using the Zeeman effect generally employed to measure solar magnetic fields. The Hanle effect, a quantum interference effect, which is most potent in the \ac{uv}, is the method of choice for examining the properties of turbulent fields, as it does not suffer from the problems besetting the Zeeman effect \cite[]{stenflo94}. The Hanle effect combined with the stable conditions offered by a balloon-platform is ideal for finally determining the true distribution of the so far hidden magnetic energy proposed to reside in the quiet Sun \cite[e.g.,][]{trujillobueno04}.

To probe both the strong, more organized, and the weak, more turbulent fields, a combination of Hanle and Zeeman diagnostics in multiple spectral lines is needed. Meeting this goal of \sunriseiii{} requires instruments probing additional wavelengths to those covered by \sunriseiandii{}.

\subsubsection{Turbulent magnetoconvection: origin and fate of magnetic flux}\label{magnetoconvection}

The emergence of magnetic flux in the photosphere is frequent and widespread \cite[e.g.,][]{centeno07,martinezgonzalez07,thornton11}. Bipolar structures appear over a large range of sizes, from sunspot-forming, coherent magnetic flux bundles in the tens of Mm range \cite[e.g.,][]{stenflo12,cheung14,chen17a} down to scales barely detectable with current solar telescopes. These small-scale emergences happen continuously in every single granule, and various aspects of their turbulent nature were already uncovered by the \sunriseiandii{} flights \cite[]{danilovic10,guglielmino12,anusha17,centeno17,smitha17}. \sunriseiii{} observations add the height information about the physical conditions during these emergences and allow to determine their role in shaping the higher atmospheric layers of the Sun. The  high activity level during the 2024 flight and the long time series not only provide detailed statistics of these tiny emergences, but also increase the likelihood of capturing the rise of a larger bipolar magnetic bundle and its complex, small-scale interaction with the turbulent convective cells \cite[]{weber23}.

The processes by which the magnetic flux is removed from the solar surface are even less well known, mainly because much of it is thought to happen at very small spatial scales. Flux removal has to take place at the same, very high average rate as the emergence of new flux, to keep the total flux in balance. It is expected that all the magnetic flux in the quiet Sun is replaced within 1--3 days \cite[]{schrijver97,lamb13}. Although there are a number of processes which can remove flux, only two of these are accessible to observations: submergence of $\Omega$-shaped magnetic loops to below the surface and ejection of U-shaped loops from the Sun \cite[]{zwaan87,cameron11}. Magnetic reconnection is likely to play an important role in both flux removal processes \cite[]{vanBallegooijen89} and consequently provides a considerable amount of energy for local heating and acceleration of plasma jets \cite[]{thaler23}. Observational studies of flux removal have not delivered any clear results favouring one or the other mechanism \cite[]{bellot19}. Flux removal and associated processes generally take place on or below the spatial scale of the individual flux elements and over a large range of heights, making observations at high spatial resolution and height coverage mandatory. \sunriseiandii{} already revealed a large number of cancellation events \cite[]{centeno17,anusha17}. However, only a spectropolarimeter sampling a number of spectral lines simultaneously, together with a vector  magnetograph, allows the full three-dimensional evolution of the magnetic field in a cancelling event to be followed and thus the various possible scenarios of flux removal to be distinguished.  

\subsubsection{The upper solar atmosphere}\label{upperatmosphere}

The atmospheric layers above the solar photosphere are of fascinating beauty, but still full of mystery when it comes to the understanding of the underlying physical processes.
The chromosphere with its poorly studied magnetic field has increasingly moved into the focus of solar physics in recent years \cite[]{lagg:17,delacruz:17,molnar19,dasilva22,rouppevandervoort23,kuridze24}. Its structure and dynamics are characterized by the interaction of plasma flows, waves, radiation and magnetic field. Because the plasma-$\beta$, i.e., the ratio between thermal and magnetic energy density, may drop below unity in parts of the chromosphere, and the kinetic energy density in shock waves can reach the thermal energy density, the chromosphere provides the appropriate conditions for the easy conversion of energy from one of these forms to another.

How the chromospheres and coronae of the Sun and other stars are heated is one of the outstanding open problems of solar and stellar physics.  Although on average only 8000\,K hot, the chromosphere, by dint of its higher density, requires a factor of 2--10 larger energy input than the corona to maintain its temperature \cite[]{withbroe77,anderson89}. The key lies in the  magnetic field with its decisive role in heating large parts of the chromosphere and the corona \cite[]{judge24}. 
At the solar surface, the magnetic field interacts strongly with turbulent convection and is moved around by the flows, imparting the energy from the flow to the field. Slow motions of the plasma braid magnetic field lines, inducing currents in the upper solar atmosphere \cite[]{parker83,klimchuk06,pontin20}.
Rapidly varying surface motions excite a multitude of \ac{mhd} wave modes that carry energy as they propagate upwards \cite[]{jess09,vanBallegooijen11}. Interaction of an existing magnetic structure with nearby opposite polarity fields can lead to magnetic reconnection in the lower atmosphere. This heats the local plasma, which is then injected into the upper regions in the form of small \cite[]{chitta17,chitta20}, or larger jets \cite[][see also \fig{SCIP_Ca8498}]{shibata07,matsumoto23}. 
Self-interactions of undulated flux tubes can also induce magnetic reconnection in the emerging flux region, manifesting themselves as Ellerman bombs \cite[]{georgoulis02,vissers15,rouppe16,kawabata24}.
Finally, magnetic vortex flows with twisted magnetic fields, so-called solar tornadoes, have been identified as an additional source of energy transfer into the upper atmosphere \cite[]{wedemeyer12}.

\colfig{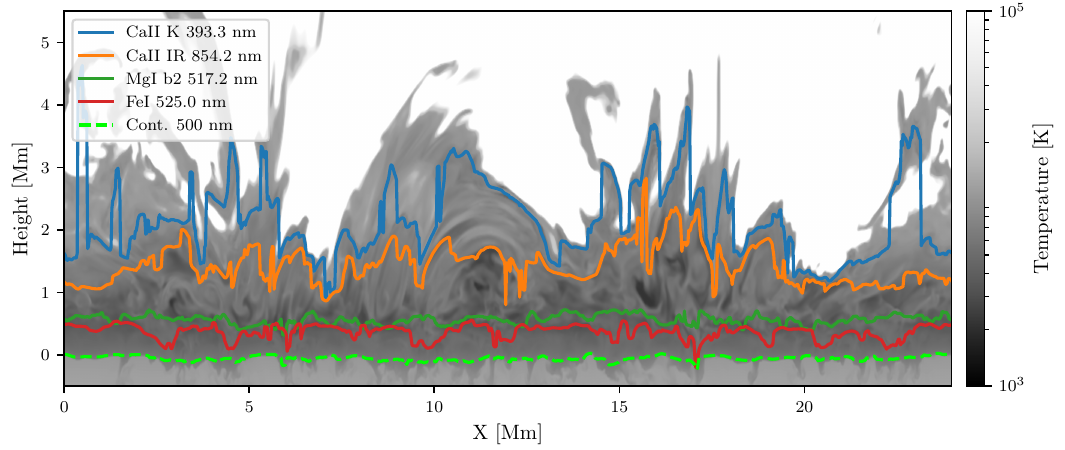}{Vertical cut showing the temperature stratification in a \acs{muramce} simulation. The heights for optical depth unity for the continuum at 500\,nm and for selected spectral lines observed by the \sunriseiii{} instruments are indicated by the \textsl{colored solid} lines.}

Signatures of some of these processes have been observed. For example, various modes of \ac{mhd} waves were found in the solar atmosphere \cite[]{jess09,mcintosh11}, although energy estimates may be uncertain \cite[e.g.,][]{yadav21}. For braiding, mainly indirect observational evidence has been reported \cite[]{peter04}, except for the rare cases of direct observations \cite[]{cirtain13,chitta22}, which still leave important questions unanswered. 
To determine the contribution of these processes to chromospheric and coronal heating, knowledge of the physical quantities (temperature, density, velocity, and magnetic field) in the photosphere and in the chromosphere is required. Recent \ac{mhd} simulations with the chromospheric extension of the \acl{muram} code \cite[\acsu{muramce}:][]{przybylski22} demonstrate the need for the high spatial, height, and temporal resolution of such measurements to characterize the fine-structured and strongly dynamic chromospheric plasma \cite[]{rouppevandervoort23,delaCruzRodriguez24}. A snapshot of the temperature stratification from a \ac{muramce} simulation is shown in \fig{MURaMCE-Sunrise}. It illustrates the small-scale nature of the chromospheric features, and how strongly the optical-depth-unity surface varies in height, indicated for some of the spectral lines observed by the \sunriseiii{} instruments.

Most chromospheric spectral lines have a rather limited sensitivity to the magnetic field, which is generally weak in the chromosphere compared with its photospheric counterpart \cite[]{delacruz:17}.
The \ac{nuv}, observed by the \ac{susi} spectropolarimeter, contains a large number of magnetically sensitive spectral lines, increasing the sensitivity of the chromospheric magnetic field measurements when many lines are observed simultaneously \cite[]{riethmueller19}. Small scans over narrow \ac{fov}  provide the necessary high cadence to capture the dynamics and, with long and constant-quality time series of such small scans, the signature of the various \ac{mhd} wave modes. A unique feature of observations from the stratosphere is the absence of atmospheric refraction, which allows obtaining co-spatial and co-temporal information also in the \ac{nir} \caii{} line observed by \ac{scip}. 
Time-distance diagrams like the one shown in \fig{SCIP_Ca8498} \cite[adopted from Fig. 5 of][]{matsumoto23} for this spectral line can be created from these small scans for all relevant atmospheric parameters and at multiple height layers. 
By embedding these scans in the high-cadence maps of the photosphere and the chromosphere provided by \ac{tumag}, \sunriseiii{} provides the magnetic field information at high spatial and height resolution, high cadence and high \ac{s2n}, and thus hopefully helping to solve many mysteries of the chromospheric magnetic field.

\colfig{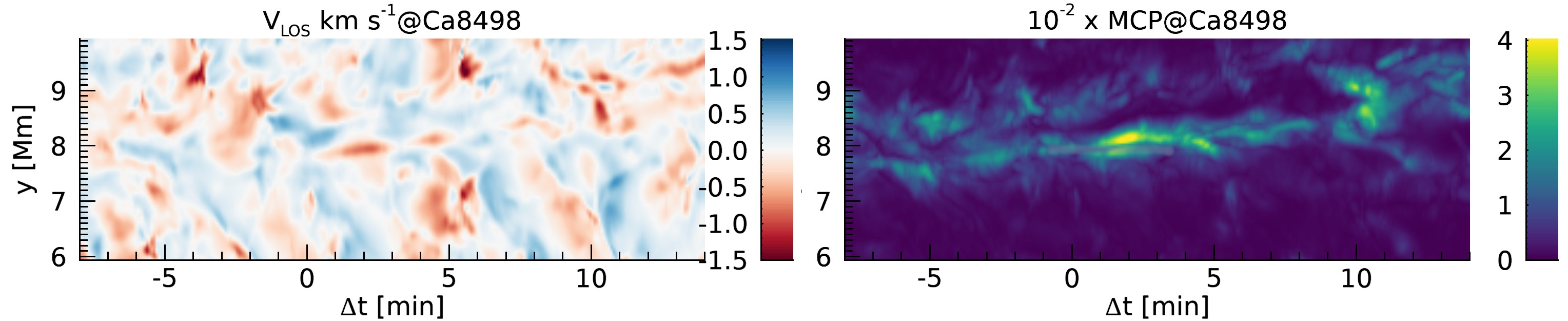}{Simulated \ac{scip} observation in the \caii{} 849.8\,nm line showing the temporal evolution of a chromospheric jet \cite[adopted from][Figure 5]{matsumoto23}. The time-distance diagram of the \acl{los} velocity ($v_{\rm LOS}$, \textsl{left}) and the maximum amplitude of circular polarization (MCP, \textsl{right}) demonstrate the highly dynamic nature of the chromosphere.}

\subsubsection{The small-scale dynamo}\label{dynamo}

The continuous dissipation of magnetic flux and its ejection from the Sun requires its constant replenishment by magnetic-field generating or amplifying processes. In the current understanding, a large-scale dynamo is operational in the solar convection zone, and this mechanism, powered by non-uniform rotation, convection and buoyancy, results into the generation of large bipolar regions \cite[for an overview see][]{charbonneau:23}. 
For the magnetism in the quiet Sun, 3-dimensional \ac{mhd} simulations make the exciting prediction that a small-scale turbulent dynamo operates in the Sun \cite[]{petrovay93,cattaneo99,voegler07,rempel14,hotta15,khomenko17,rempel:23,warnecke:23}.
The relevance of a small-scale dynamo for the fields measured in the solar atmosphere is still controversial \cite[]{stenflo12a,martinezsykora19,ebert24}, and an observational proof of its existence is extremely difficult \cite[]{danilovic10b,shchukina11,buehler13,lites14,danilovic16}.  Tests carried out so far suffer from non-ideal, sometimes inhomogeneous datasets. Time series at a constant very high spatial resolution, reliably covering all photospheric heights are needed to address this question.

Another promising observational approach relies on the technique of local helioseismology \cite[]{gizon05,gizon10}, allowing to probe the convection in the layers immediately below the surface \cite[]{singh16,korpilagg22,waidele23}. To achieve this aim, long (three hours or more) time series of velocity and magnetic field at highest spatial resolution are needed. Providing such time series at constant  quality is a unique feature of \sunriseiii{} and cannot be obtained from the ground, due to the ever-present and fluctuating image degradations caused by turbulence in the Earth's atmosphere.

\subsubsection{Exploration of the polarized solar spectrum in the \acs{nuv}}

The polarization properties of the solar spectrum are poorly known over most of the wavelength interval between 300 and 400\,nm. At these short wavelengths, it becomes increasingly challenging to obtain high-resolution data from the ground, and basically impossible to do so with a good \ac{s2n} ratio, owing to the strong absorption and Rayleigh scattering by the air molecules in the Earth's atmosphere. The strong absorption requires prohibitively long integration times when approaching the short-wavelength end of this spectral range, while the Rayleigh scattering introduces significant stray light contamination that strongly dilutes the intrinsic polarization signals. Both limitations can only be addressed by observing from above most of the Earth's atmosphere, as is the case for \sunriseiii{}. No space mission has yet covered these wavelengths at high spatial resolution, so that \sunriseiii{}  provides the first high resolution spectropolarimetric data in this wavelength band, opening up new discovery space, and enhancing our understanding of the so-called `second solar spectrum' \cite[]{stenflo97,gandorfer05}. 

In spite of being poorly studied, the \ac{nuv} is of particular interest for a number of reasons:
Firstly, the \ac{nuv} contributes dominantly to the irradiance variability of the Sun (i.e. to the variation of the Sun's total radiative output), and is a strong driver of the chemistry of the Earth's stratosphere \cite[]{krivova06,haigh10,solanki13}. The Sun's \ac{uv} variability is prototypical of that displayed by other stars, which can have an important influence on the habitability of exoplanets.%

Secondly, the \ac{nuv} part of the solar spectrum harbours a very high density of spectral lines, allowing for the use of many-line Zeeman and Hanle diagnostics \cite[]{solanki84,riethmueller19}. The line density between 300 and 400\,nm is more than four times larger than that between 500 and 600\,nm, so that the information content of a far larger number and variety of spectral lines can be made use of than in the visible or infrared. \cite{riethmueller19} demonstrated also that the polarization signals of many of the spectral lines are very strong, with peak values between 30 and 40\% even for the small flux concentrations in magnetically quiet areas of the Sun.

In addition, the \ac{nuv} harbours many spectral lines whose diagnostic potential is largely unexplored. To assess this potential, we computed the formation heights (the heights where the opacity at the given wavelength reaches unity) of $\approx\,$\lnum{60000} spectral lines in the \ac{susi} wavelength range from the Kurucz line-list database \cite[]{kurucz95}, and present the result in \fig{formheights}. The computations were performed using the radiative transfer code RH \cite[]{uitenbroek01,pereira15} with a FALC model atmosphere \cite[]{fontenla93}, and all spectral lines were treated in \ac{lte}, except the \caiihk{} lines which were treated in \ac{nlte}. \fig{formheights} demonstrates that, in sharp contrast to the handful of chromospheric lines that can be found in the visible and near-infrared, more than 150 lines form above 600\,km, i.e., are clearly chromospheric.  
Because of the large number of chromospheric lines in the \ac{uv}, and their widely varying thermodynamic sensitivities, this spectral range opens a new window to studying the chromospheric structure, dynamics and magnetic field and offers seamless height coverage from the deepest photospheric layers up to heights of \lnum{2000}\,km.

\colfig{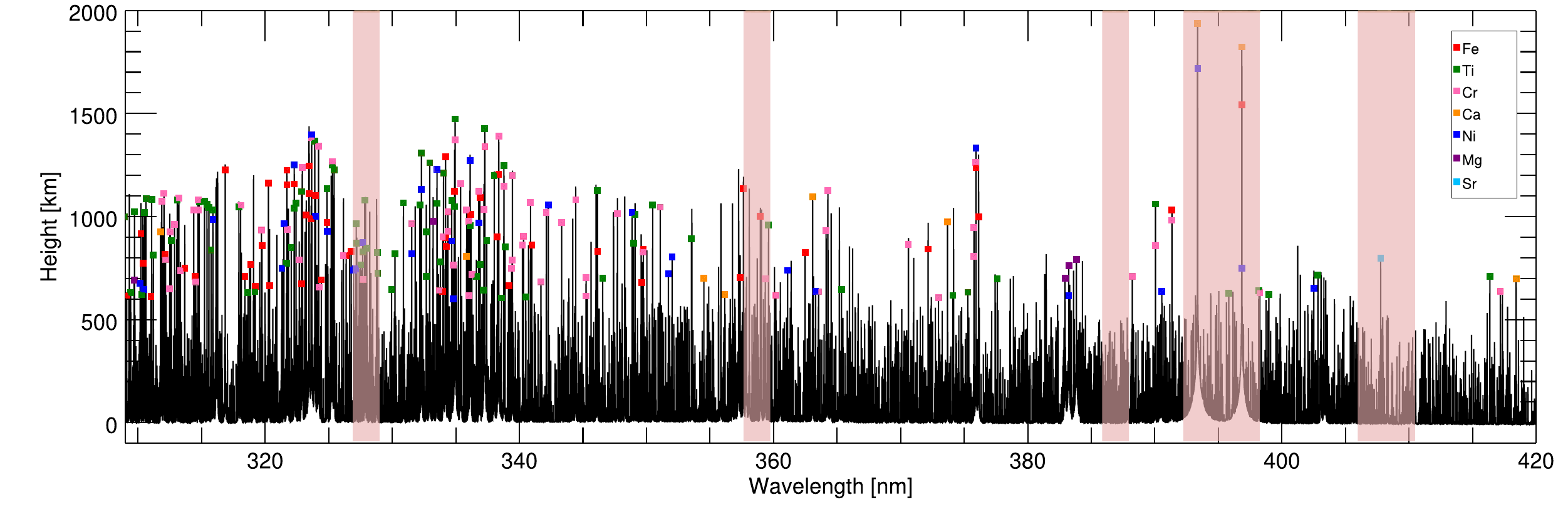}{Formation height of $\approx\,$\lnum{60000} spectral lines in the \ac{susi} spectral range from 309 to 417\,nm. The highest forming lines (above 600\,km) of a few selected species are indicated by \textsl{colored} symbols. The shaded regions indicate the six most requested \ac{susi} spectral windows, with two windows overlapping at the \caiihk{} lines around 390\,nm.}

\subsection{Preparing for \sunriseiii{} Science and Operation Planning}\label{sec:sswg}

Parallel to the hardware development phase, the \sunrise{} partner institutes established the \aclu{sswg} (\acsu{sswg}). The main task of this working group was to develop an observation strategy for the flight, which exploits fully the unique capabilities of the observatory, and satisfies the demands by the international community of solar researchers. Prior to the 2022 flight, more than 60
science proposals requesting access to data or specific observations were collected,
and nine proposals were accepted within the \href{https://www.solarnet-project.eu/acess}{SOLARNET Transnational Access} program.
Since much of the science of \sunriseiii{} is unique, the science proposals could be used unaltered as the basis for the observations during the 2024 flight.

\subsubsection{Timeline concept}\label{timeline}

A timeline concept was developed to operate all science instruments and the \acf{pcs} in an orchestrated mode controlled by the \ac{ics}. This concept ensures not only a nearly autonomous operation of \sunriseiii{} (important if real-time communication with the observatory is disturbed or slow), but also the efficient use of the limited observing opportunities. The aim was to gather data with a given instrumental setup and solar target that fulfil the requirements of as many as possible of the observing proposals submitted. Data from a single observing program  enable a variety of analyses with different scientific goals.

The timeline consists of individual observing blocks, concatenating the scientific observing modes of every instrument, defined by specific settings (e.g., \ac{fov}, wavelength, scan-speed), with the 
necessary calibration measurements required to ensure that the recorded data can be reduced on ground with the best possible accuracy. 

\begin{figure}%
    \centering
    \subfloat{{\includegraphics[width=\linewidth]{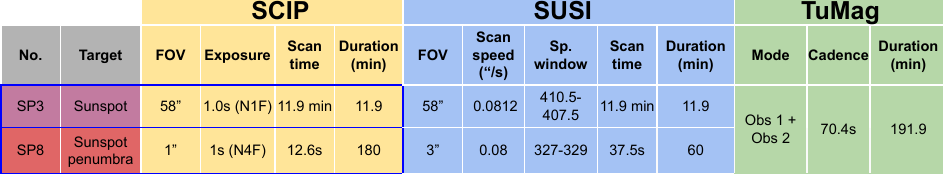}}}\\
    \subfloat{{\includegraphics[width=\linewidth]{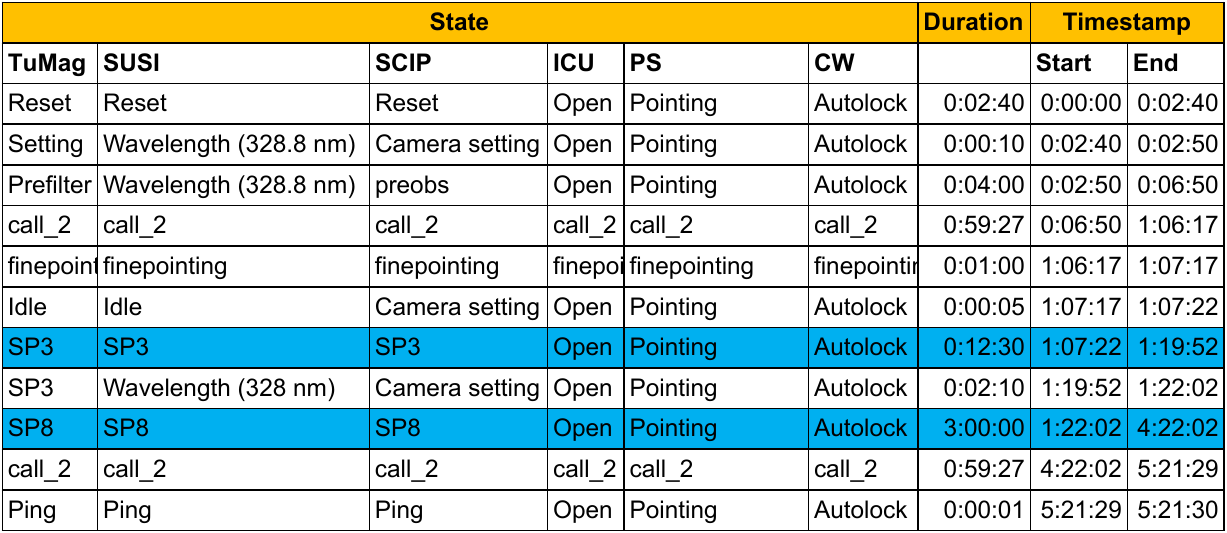}}}
    \caption{Timeline `SP\_2'  defining the operation of the three science instruments, the \ac{ics}, the \ac{pcs}, and the \ac{cws} to study the fine structure of a penumbral filament. }
    \label{obsblock}
\end{figure}

\fig{obsblock} shows, as an example, the observing block `SP\_2', defining an observation dedicated to study the penumbral fine structure. The top panel of \fig{obsblock} is an excerpt of the science planning sheets: A 12-minute full-\ac{fov} context scan by \ac{scip} and \ac{susi} (mode `SP3') is followed by a 3-hour long observation (`SP8') with the spectrograph slits placed along a penumbral filament: 
\ac{scip} provides narrow 1\arcsec{} scans with a time resolution of 12.5\,s, while the \ac{susi} slit is moved over a slightly wider region of 3\arcsec{} in 37.5\,s. During the whole time, \ac{tumag}  observes the full \ac{fov}, quasi-simultaneously in the photosphere and the chromosphere. The bottom panel of \fig{obsblock} presents the details of the observing block `Spot \#2' to be executed by the \ac{ics}. It contains the commands for the instrument settings, and two long calibration measurements (`call\_2', containing, e.g., dark current measurements, flat fields, stray light target measurements) surrounding the two  science observing modes `SP3' and `SP8'.

Similar observing blocks of lengths between 2 and 8 hours were compiled for observations of the quiet Sun close to solar disc centre, sunspots, plage regions, emerging flux regions, the poles, active regions, center-to-limb variations, flares, and coronal holes. A six hours long full spectral scan in the so far unexplored \ac{nuv} region is a first of its kind. The observing blocks have been prioritized by the \ac{sswg}, but the final decision on the order of their execution and the detailed target selection was performed shortly before and during the flight in daily science planning meetings. This  allowed to react to the prevailing solar conditions, observatory parameters like solar elevation (important for observations in the \ac{uv}) or image rotation speed (see \sref{obsbc}), and the instrument performance. The so-defined sequence of observing blocks constitutes the timeline, which was executed autonomously by the \ac{ics} without requiring commanding from the ground station. However, to optimize telescope and instrument performance, as well as to fine-adjust the pointing to an interesting solar feature, a real-time communication with \ac{hk} value and thumbnail image transfer to the observatory was available throughout the whole flight (see \sref{comm}).

All observing blocks were tested on ground with the observatory in flight configuration. The timeline concept also allowed to handle unplanned interruptions to the timeline (e.g. because of a possible instrumental problem or loss of pointing), and a resumption after the problem is solved. At any time, all the instruments could also be operated `manually', i.e., by direct commands from the \ac{goc} through the \ac{ics} (see \sref{comm}). The \docdb{SR3-MPS-HO-SI000-001} summarizes all observing blocks, timelines and scientific ideas.

\subsubsection{Observational boundary conditions}\label{obsbc}

The optimal placement of the timelines and the observing blocks required to take into account not only the current state of the Sun, but also two important observational boundary conditions in the daily science planning meetings: the elevation of the telescope, and the image rotation inherent for alt-azimuth mounted telescopes.

Telescope elevation: Although \sunriseiii{} floated at an altitude between 33 and 37\,km, there was still some residual absorption by the Earth's atmosphere. \ac{sufi} observations from the first two flights of \sunrise{} showed that for low telescope elevations ($\le10$\dg, i.e. around local midnight), the ozone layer affects the sky transparency at the short end of the \ac{susi} wavelength range.
The measured variation of the photon flux from lowest  (3\dg{}) to the highest (42\dg{}) elevation by \ac{sufi} during the \sunrisei{} flight was a factor of 2 at a wavelength of 313\,nm. At 388\,nm, the photon flux did not show any elevation dependence. The start and end times of  \ac{susi} observations at the shortest wavelengths therefore need to be scheduled accordingly.

Image rotation: The image rotation speed for \sunriseiii{} was below $\pm$10\dg{}/h for the 2024 flight trajectory from Kiruna to Canada (see \fig{siii_sspps}). The center of rotation is defined by the lock-point of the \ac{cws}, lying within $\pm15$\arcsec{} to the center of the \ac{tumag} \ac{fov}. The scan mechanisms of \ac{scip} and \ac{susi} allowed to put the slit of both instruments independently so that the center of the image rotation is on the spectrograph slit for both instruments. This was of special relevance when performing observations in sit and stare mode (i.e., without moving the slit), or when scanning a very narrow \ac{fov} in the sub-arcsecond to arcsecond range. Such scans were placed at times when the image rotation is small, in order to maximize the spatial coherence over a large portion of the slit for the whole duration of the time series (typically in the range of 1 hour, see green shaded areas in \fig{siii_sspps}). A small image rotation during such a scan is acceptable, as it results in a `circular' scan, with spatially coherent information at the slit center, and an increasing \ac{fov} towards the slit ends, accompanied with a seamless decrease of the \ac{s2n} ratio along the slit.

\subsubsection{Co-observations with \sunriseiii{}}\label{coobs}

The \ac{sswg} invited  all major ground-based and space-borne high- or medium-resolution solar observatories to co-observe with \sunriseiii{}.
All contacted leaders of the observatories agreed to prepare their facilities for co-observations and run specific observing programs. 
Of special relevance was the support by the world's largest solar telescope, the National Science Foundation's  \acl{dkist} \cite[\acsu{dkist}:][]{DKIST:2020}, and other large-aperture solar telescopes distributed over the globe: the \acl{gst} \cite[\acsu{gst}, U.S.:][]{goode:10,cao:10} operating at the \ac{bbso}, \aclu{gregor} and \acl{vtt} on Tenerife \cite[\acsu{vtt}, Spain:][]{schmidt:12,soltau91}, the \acl{sst} with its new \ac{ao} system on La Palma \cite[\acsu{sst}, Spain:][]{scharmer:24}, the \acl{nvst} \cite[\acs{nvst}, China:][]{NVST:2014}, \acl{mast} \cite[\acsu{mast}, India:][]{MAST:2017}, the \acl{dst} \cite[\acsu{dst}, U.S.:][]{DST1998}, and the \acl{smart} \cite[\acsu{smart}:][]{Hida2004} at the Hida observatory (Japan). 
Highly complementary data sets were obtained by the space-based observatories \hinode{} \cite[]{kosugi2007} and \acl{iris} \cite[\acsu{iris}:][]{dePontieu:14}, for which a special \sunriseiii{}/\ac{ihop} was executed during the flight, and by the new full-disk \halpha{} observatory \acl{chase} \cite[\acsu{chase}:][]{chase:22}, and the space-based instruments \acl{suit} \cite[\acsu{suit}:][]{SUIT2023} on board Aditya-L1 \cite[]{AdityaL1}, \acl{suvi} \cite[\acsu{suvi}:][]{SUVI2022} on board GOES-R, and \acl{sutri} \cite[\acsu{sutri}:][]{SUTRI2023}.

Unfortunately, the \acl{alma} \cite[\acsu{alma}:][]{wootten:09,wedemeyer:16}, which was ready to support the 2022 flight, was not available for the 2024 flight because of the observatory setup not allowing for solar observations. Similarly, no co-observations with \acl{solo} \cite[\acsu{solo}:][]{mueller20} were possible due to the orbital configuration (Solar Orbiter was on the far side of the Sun at the time of the \sunriseiii{} flight).

The coordination of the co-observation during the flight was based on a website with live-updates on the current and the planned observations of \sunriseiii{}, including the accurate pointing and timing information  and recent images from the \aclu{sdo} \cite[\acsu{sdo}:][]{pesnell:12}. Co-pointing of \sunriseiii{} and the co-observing facilities was based on these images, with overlaid \acp{fov} of the \sunriseiii{} instruments (see \fig{planningtool}). 

\subsection{The \acf{sppt}}\label{sppt}

\colfig{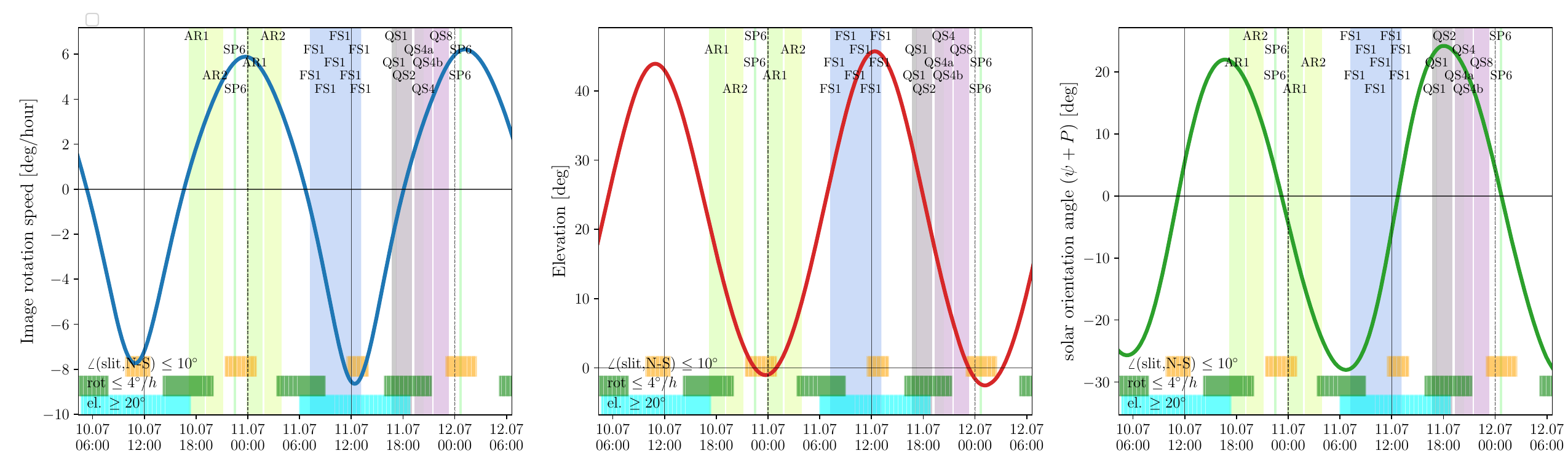}{Image rotation speed (\textsl{left panel}), telescope elevation angle (\textsl{middle}) and slit orientation angle with respect to the solar North-South direction (\textsl{right}) for the first 48 hours of the 2024 flight as produced by \ac{sppt} (see \sref{sppt}).  The \textsl{horizontal bars} at the bottom indicate the periods with high telescope elevation (\textsl{cyan bars}), low image rotation speed (\textsl{green bars}) and a slit orientation parallel to the solar rotation axis (\textsl{orange bars}). The \textsl{vertical bars} indicate the observing modes defined in the science planning meeting prior to the launch.}

\colfig{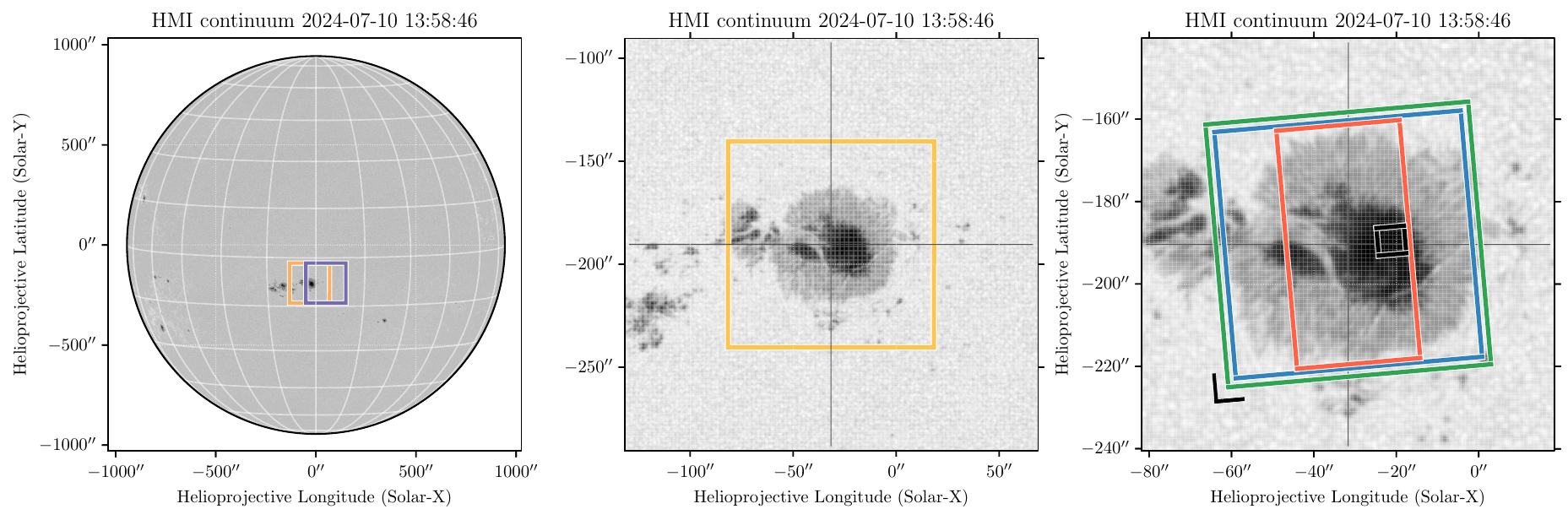}{Snapshot from the \ac{sppt} observation planning page. The most recent solar images (here \ac{sdo}/\ac{hmi}) are used to define the pointing of Sunrise. The tool presents the \acp{fov} of the individual instruments (\textsl{green} \ac{tumag}, \textsl{red} \ac{scip}, \textsl{blue} \ac{susi}, \textsl{black} \ac{cws}, \textsl{right panel}), taking into account the image rotation at the planned start time of the observation. The \textsl{orange} and the \textsl{blue rectangle} in the \textsl{left} panel indicate the location on the Sun at the time of the observation planning and the planned execution time, respectively.}

The observational boundary conditions described in \sref{obsbc} 
require a thorough placement of the pre-defined observing blocks to construct the final timeline. The solar conditions needed to be taken into account in real-time, using the data from multiple solar observatories, most importantly from \ac{sdo}. The versatility of the \sunriseiii{} instrumentation and the large number of solar features to be addressed during the relatively short flight (duration: 6.5 days) relied on a sophisticated planning tool, the Python software package \aclu{sppt} (\acsu{sppt}) developed explicitly for the \sunriseiii{} flight. All information required for the planning is displayed in the context of the \sunriseiii{} instrumentation (e.g. by showing their \acp{fov} on the real-time images of the Sun, see \fig{planningtool}). Observing blocks could be loaded and shuffled around to minimize conflicts of their placement with respect to the observational boundary conditions described in \sref{obsbc}. 
\fig{siii_sspps} shows an example of the planning for the first  observing blocks produced by \ac{sppt} after the pre-flight science planning  meeting. Conflicts of the individual observing modes with respect to the observational boundary conditions are clearly marked by \ac{sppt}, allowing to optimize the placement of the observing blocks in the final timeline.

Additionally, \ac{sppt} is equipped with a thumbnail display and analysis tool, showing the real-time images from all \sunriseiii{} instruments during the flight with the important capability of determining the offset between the different instruments. This enabled to monitor the position of the \ac{scip} and the \ac{susi} slits during the 2024 flight to achieve near-perfect alignment, a unique feature of \sunriseiii{} allowing to combine full-Stokes spectropolarimetric data in the \ac{nir} and the \ac{nuv}. After the planning phase, the tool presented the final timeline of the observations, and fed the current and planned pointing of \sunriseiii{} to the co-observer website.

\section{The \sunriseiii{} Observatory}\label{instruments}

The \sunriseiii{} observatory is the payload of a \ac{nasa}/\ac{csbf} operated \acf{ldb} flight from \ac{esrange} (Sweden) to Canada. It is suspended from a 34.43 million cubic feet (\lnum{975000}\,m$^3$) Helium balloon and flight train, the latter consisting of a parachute and the so-called ladder, a set of steel cables connected on one end to the parachute and on the other end via a spreader plate to the gondola suspension point.
The design of the  \sunriseiii{} observatory is shown in \fig{siii_design}, and some key parameters are listed in \tab{tab:observatory}. It is composed of four major building blocks\textbf{}, as illustrated in \fig{fig:siii_overview}:

\begin{figure*}
\includegraphics[width=\linewidth]{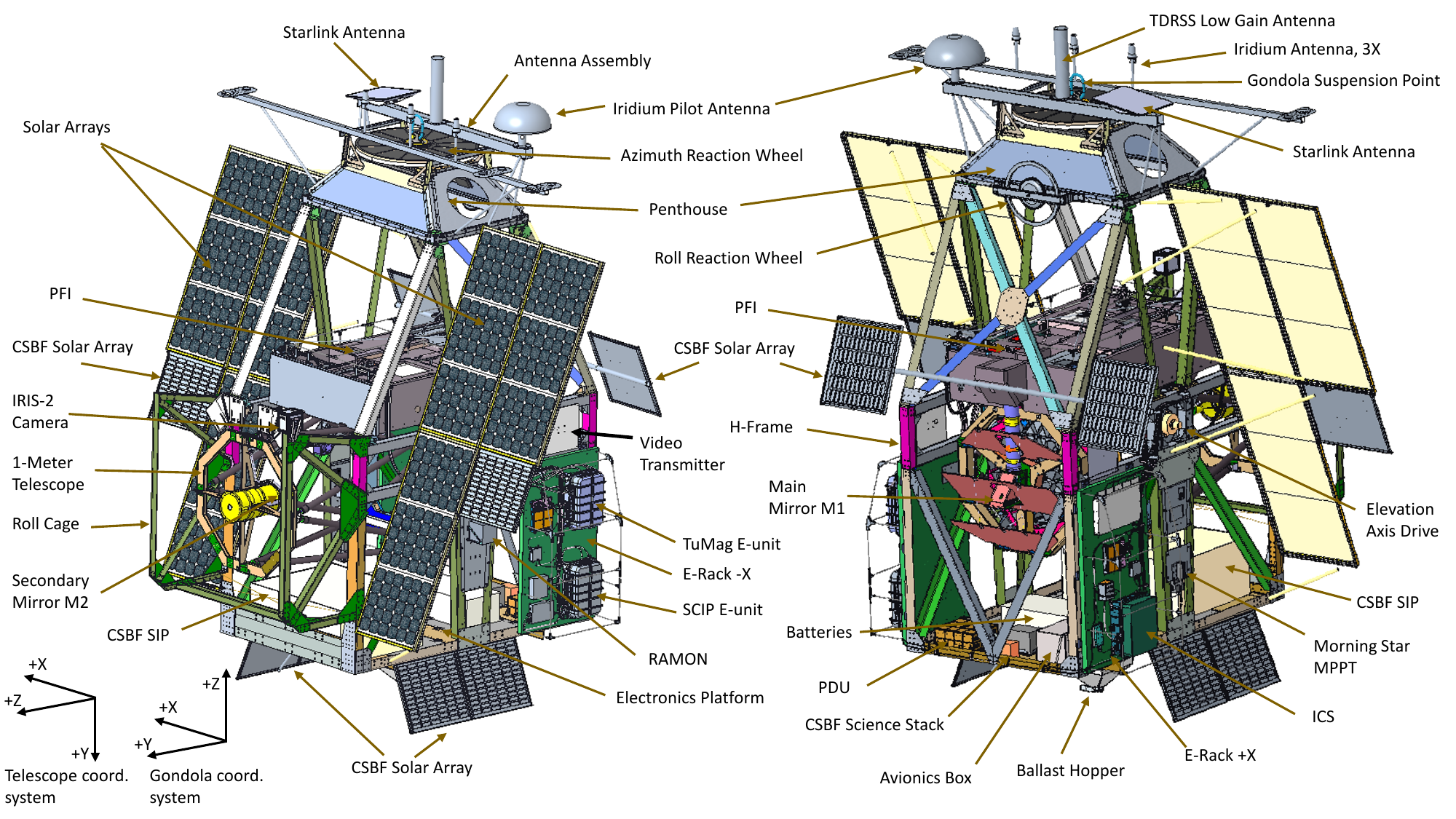}
\caption{Design drawing of the \sunriseiii{} observatory. The telescope is located in the middle of the gondola and is oriented horizontally in the plotted configuration. The \textsl{gray-purple box} on top of the telescope is the \acf{pfi} platform housing the science instruments. Instrument control electronics are housed on two panels on the lower back part of the gondola. The gondola control and power electronics units, as well as the \ac{csbf} \acf{sip} are housed on the bottom electronics platform. The \textsl{lower left corner} of the figure shows the two separate coordinate systems used by the telescope and instruments (\textsl{on the left}) and the gondola (\textsl{on the right}). The telescope coordinate system rotates about the $x$-axis relative to the gondola one depending on its elevation angle.
}
\label{siii_design}
\end{figure*}

\begin{table*}
\caption{\sunriseiii{} key parameters.}
\label{tab:observatory}
\begin{tabular}
{L{0.15\textwidth}L{0.78\textwidth}}\hline
Parameter & Description \\ \hline
Flight details & Launch: 10 July 2024, \acs{esrange} (Sweden); 
6.5 day uninterrupted solar observations; stratospheric 33--37\,km float altitude; landing: west of Great Bear Lake, Canada.
\\
& \acs{nasa}/\acs{csbf} \acs{ldb} 34.43 million cubic feet Helium balloon ($\approx\,$\lnum{975000}\,m$^3$).
\\ \hline
Mass & 
Science payload: 828\,kg; gondola: \lnum{1269}\,kg; \ac{csbf} equipment: 275\,kg;
equipment
above / below pin: 442\,kg / 103\,kg; ballast: 272\,kg; balloon film: \lnum{2330}\,kg; required Helium mass (10\% lift) $\approx\,$\lnum{1065}\,kg.
\\ \hline
Dimensions & 
Gondola: height (without ballast, solar panels): 523\,cm; width without solar panels: 196\,cm, with solar panels: 540\,cm; length (including roll cage): 419\,cm.
\\
& Flight train: suspension ladder: 22.5\,m; parachute length: 73\,m (diameter: 48.5\,m); balloon size at float altitude: `pumpkin' shape, width 134\,m, height 103\,m.
\\ \hline
Power & 
Solar panels with 1.8\,kW peak power, buffered by 4 lead batteries (\lnum{1728}\,Wh). \\
& Max. power consumption during ascent: \lnum{1031}\,W, at float altitude (science): $\approx\,$750\,W.
\\ \hline
Key features & 
Gregory-type telescope with 1-meter main mirror, 24.2\,m focal length. \\
& Pointing stability: $<$3\arcsec{} \ac{rms} (gondola); $\le$0.005\arcsec \ac{rms} (internal image stabilization). \\ \hline
Science & 3 scientific instruments; wavelength 309--855\,nm; full Stokes spectropolarimetry; diffraction-limited spatial resolution: 57\,km (@309\,nm) to 154\,km (@855\,nm).
\\ \hline
Data  & 
On-board data storage: approx. 240\,TByte net storage on 48 \acsp{ssd}. \\
Communication & In-flight communication via satellite (max. 8\,Mbit/s); \acs{los} downlink-only during ascent and first hours of flight (12.5\,Mbit/s).
\\ \hline
\end{tabular}
\end{table*}

\begin{figure}
    \centering
    \includegraphics[width=\columnwidth]{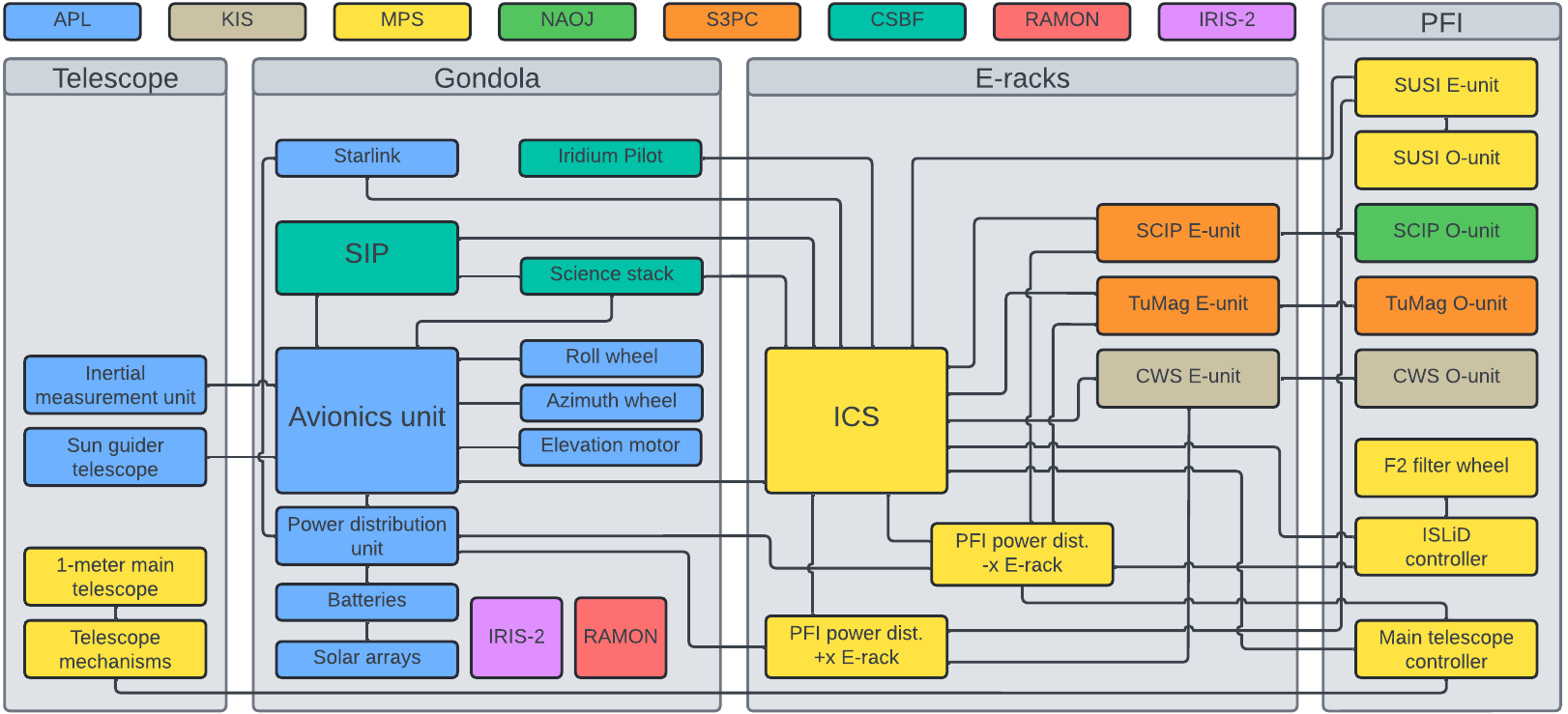}
    \caption{Schematic illustrating the four major building blocks of the \sunriseiii{} instruments and infrastructures. The \textsl{coloring} indicates the responsible institute or team.}
    \label{fig:siii_overview}
\end{figure}

\begin{enumerate}[(i)]
  \item The heart of the observatory is the telescope with its light-weighted Zerodur\textsuperscript{\textregistered} mirror with a diameter of 1\,m in a Gregory configuration and an altitude-azimuth mounting (see \sref{telescope}). The altitude (or elevation) is controlled by a motor acting on the horizontal axis of the telescope perpendicular to its optical axis, while the azimuth pointing is achieved by rotating the whole gondola (see \sref{gondola}). Attached to the telescope are the \acf{sgt} (see \sref{sunguider}) and the \acf{imu} (see \sref{imu}). Feedback from these devices is used to determine the drive signals for the elevation, azimuth, and roll motors.
  \item The \aclu{pfi} (\acsu{pfi}, see \sref{pfi}) is the box housing the \acf{islid} with the \ac{cws} (\sref{islid} and \sref{cws}), and the three science instruments \ac{susi} (\sref{susi}), \ac{tumag} (\sref{tumag}) and \ac{scip} (\sref{scip}). It also contains the \ac{islid} controller and mechanism (the F2 filter wheel) and the \ac{mtc}.
  \item The gondola provides the housing for all components of the observatory and connects the observatory to the flight train with the parachute and the balloon (see \sref{gondola}). It acts as a protection from the shocks during launch, flight termination, parachute inflation, and finally the landing. During the flight, the gondola provides power to the entire observatory and the stable pointing that allows the scientific instruments to obtain several hours of uninterrupted observations of specific targets on the Sun. Finally, it also contains the hardware responsible for the communication with the ground station.
  \item The \acfp{erack} are  mounted on the $+x$ and $-x$ side of the gondola (see gondola coordinate system in \fig{siii_design}). They provide the platform for the electronic units (see \sref{eunits}), which control all subsystems of the observatory. The central unit is the \acf{ics}, responsible for executing the commands from the ground station, running a coordinated observation timeline for all instruments by sending commands to their individual electronic units, and for reporting \ac{hk} values and thumbnail data to the ground station.
\end{enumerate}

\subsection{Gondola}\label{gondola}

The \sunriseiii{} gondola structure provides the stable platform enabling the precision telescope pointing towards the Sun, and keeps the scientific instrumentation safe during launch and landing. 
Photovoltaic arrays generate the required power for the electronics and keep the system's batteries fully charged during science operations. %
The two previous flights of \sunrise{} have proven that maintaining a constant pointing and tracking of the Sun for extended periods of more than a few hours is a very challenging task due to pendulum motion of the gondola and occasional wind shear which introduced significant disturbances, even in the stratosphere. The \acf{apl} provided a new gondola for \sunriseiii{} featuring a state-of-the-art attitude control system \cite[see \fig{siii_design} and][]{brianalvarez:2023}, that during the 2024 6.5 days of flight provided precise and nearly uninterrupted stable and constant pointing at the Sun.

\colfig{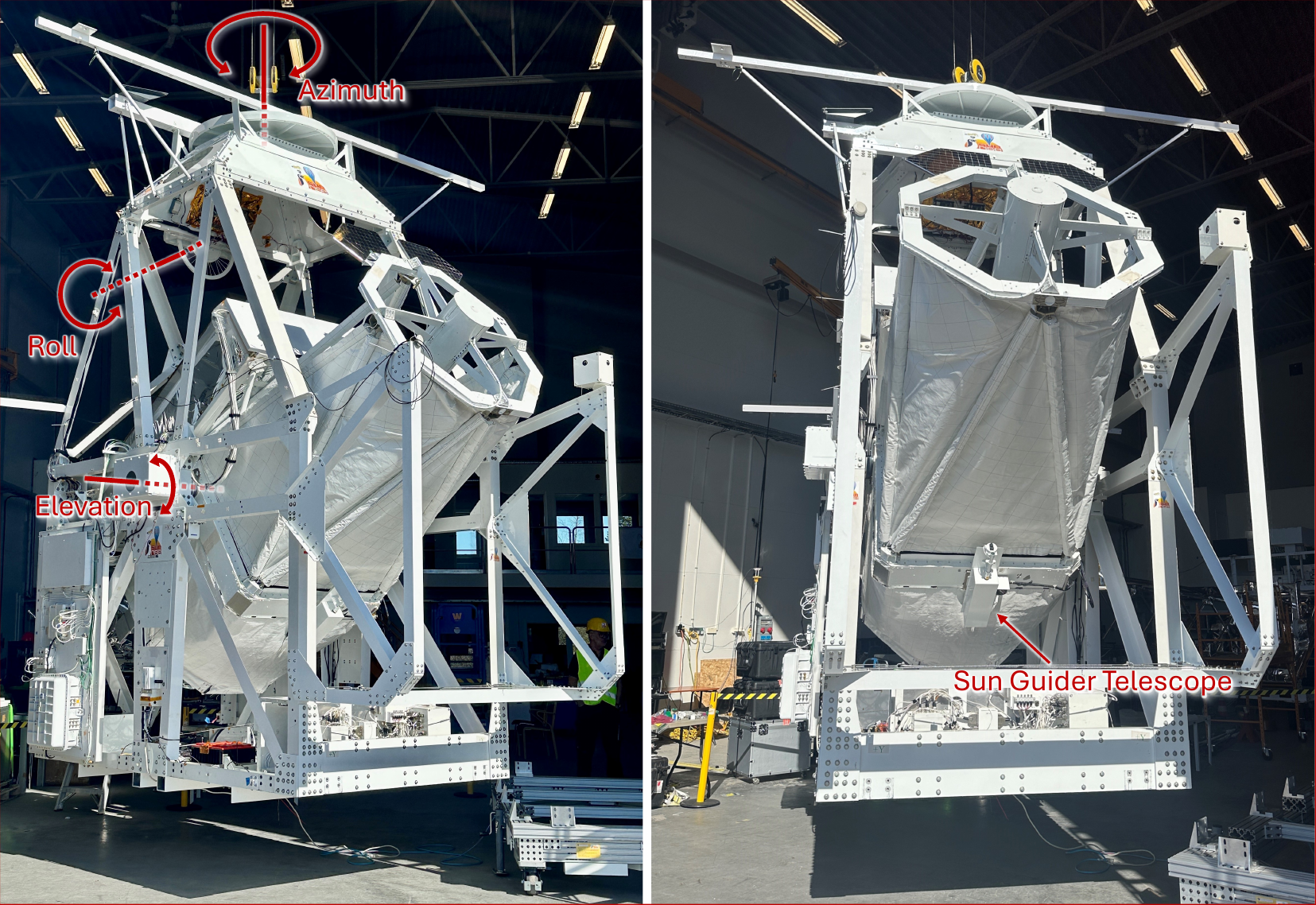}{\sunriseiii{} during the first-light hang test on 01 May 2024 at \ac{esrange}, illustrating the gondola \acf{pcs}: The elevation-azimuth actuation, and the roll wheel compensation. The \textsl{right image} shows the location of the \acf{sgt}.}

\subsubsection{Structure}\label{gondola_structure}
The gondola structure has three main assemblies.  The uppermost section, called the penthouse, supports the \acf{mtu}, azimuth reaction wheel, roll wheel and azimuth and roll control electronics.  It is a truncated pyramid structure constructed with a machined aluminum top plate, shear panels and edge support members. The middle section, called the H-Frame, supports the two elevation gimbals assemblies, the elevation drive motor, elevation hard stops, elevation brake, roll cage, and the elevation control electronics.  It is primarily constructed with aluminum tubes that are interconnected with machined aluminum fittings and gusset brackets. The elevation gimbals support the telescope and allow it to smoothly rotate in elevation.  The lower section is the electronics deck which houses the gondola control computers (avionics), \acf{pdu}, batteries, \acf{sip}, and ballast hopper.  This section is primarily constructed with aluminum tubes, aluminum angles, aluminum sheet metal and aluminum gussets. The three sections are connected to each other using aluminum angle beams.  All structural part connections are mated using high strength bolts.  The entire gondola frame is painted white to maintain it thermally stable.  Exterior to the primary structure are the photovoltaic arrays and roll cage.  The photovoltaic array support structure is a welded and bolted aluminum assembly that is bolted to the gondola structure.  The roll cage is a bolted aluminum assembly that protects the telescope should the gondola roll over upon landing, which happened at the end of the 2024 flight.

The gondola structure is designed to support the loads as described in \acf{bpo} structural requirements document \aplsrd{}
\cite[]{gondola-requirements}.  The gondola must show a positive margin of safety for both yield and ultimate strength of ductile metallic materials when oriented vertically at 8\,$g$ (gravitational force), oriented 45\dg{} at 4\,$g$, and oriented horizontally at 4\,$g$.  The loads are applied at the balloon flight train attachment point on the \ac{mtu}.  Design factors of safety are 1.25 for yield strength and 1.4 for ultimate strength. The method to perform the analysis uses inertia relief.
An additional load case was analyzed that was not documented in \aplsrd{}.  A high rotational load case that can occur at launch was analyzed as a result of a lesson learned from a failure experienced during a first launch attempt of \sunriseiii{} in 2022.  This rotational load caused a shear failure for the bolts securing the elevation drive to the brake mechanism.  The analysis was performed at twice the measured angular acceleration of 190\dg{}s$^{-2}$ and resulted in design updates to many of the brake parts. The launch in July 2024 likely exerted even larger rotational loads than in 2022, nevertheless the brake structure operated flawlessly.
The gondola structure was \acl{fem}ed (\acsu{fem}) using NASTRAN.  The structure \ac{fem} contains \lnum{97000} plate elements and \lnum{101000} nodes.  The bolts stresses were analyzed for axial and shear loads using loads from the NASTRAN analysis.  All structural parts meet the requirements in \aplsrd{}.

\subsubsection{\acf{pcs}}\label{pointing}
The attitude control system is colloquially referred to as the \acf{pcs}. The gondola utilizes elevation-azimuth actuation (often called altitude-azimuth or alt-az mount) to point the telescope (see \fig{FirstLight}, left) at the Sun. Telescope elevation is controlled using a direct-drive torque motor on the elevation axis. Azimuthal control is more complicated. It is accomplished by rotating the entire gondola about its vertical axis by means of a complex mechanism called the \ac{mtu}. 
The \ac{mtu} provides also the attachment point from the gondola to the flight train/balloon. Thus it must accommodate three rotating elements about the vertical, azimuth, axis: the gondola, the azimuth reaction wheel, and the suspension connection to the balloon flight train. Each of these elements is free to rotate in its own direction and speed independently of the other two. To control gondola azimuth, the gondola is torqued against the azimuth reaction wheel, and in turn, the azimuth reaction wheel sheds its momentum into the balloon flight train.

If the gondola were to remain perfectly leveled during flight, elevation-azimuth actuation control alone would be sufficient for providing the necessary stability for telescope pointing. However, as the gondola is suspended from a floating balloon, it is subject to several modes of pendulum motion. When this motion is front-to-back (or vice versa) with respect to the gondola (pitch), the elevation drive (with the inherent assistance of the telescope's inertia) is enough to compensate for the motion. When this motion causes side-to-side movements (roll), the azimuth drive must compensate, but it is only able to do so slowly due to the gondola's inertia. This slow correction leads to undesirable pointing errors that manifest in a side-to-side oscillation of the images. This effect is stronger when pointing the telescope at high elevation angles as the azimuth correction is less effective. To combat this, the \ac{pcs} utilizes another motor driven reaction wheel mounted along the gondola's roll axis. This is the roll reaction wheel that is seen as one component in \fig{siii_design}. This wheel is able to dampen frequency modes greater than 0.5\,Hz of side-to-side gondola motion allowing the gondola to continue pointing the telescope with high-stability even with pendulum induced movements and disturbances during flight.

To acquire the Sun in flight, the \ac{pcs} takes a methodical approach. First, the telescope's elevation is driven to match that of the Sun's. The telescope's elevation is verified utilizing a rotational encoder (also called a goniometer) on the elevation drive and knowing that the gondola always hangs close to vertical from the balloon. Once the telescope is at the proper elevation, the \ac{pcs} uses feedback from the \ac{imu} to rotate the gondola azimuthally in a circle at a rate of approximately 1° per second. The \ac{sgt} is equipped with wide-angle sun sensors that have a field of view of roughly $\pm$15\dg{}. When the gondola's slow rotation puts enough light onto these sun sensors, they provide the feedback for the \ac{pcs} to roughly center the telescope on the Sun, which allows for the higher precision pointing of the \ac{sgt} quad-cell to takeover (see \sref{sunguider}).

The \ac{pcs} was tested on the ground by suspending the gondola, including telescope and \ac{pfi}, from a crane (outside the  `Ballonhalle' at \ac{mps}, and the `Dome' at \ac{esrange}, see \fig{FirstLight}) and by pointing at the Sun. These tests demonstrated its stability and tested its limits in dealing with wind and cloud based disturbances. 

Excellent performance of the \ac{pcs} was achieved during the 2024 flight. The pointing worked accurately and stably throughout the whole flight, with only one major pointing loss on the last observing day due to an unexpected avionics computer reboot. A detailed analysis of the pointing performance will be provided after completion of the data reduction phase (see \sref{dataconcept}).

\subsection{Telescope}\label{telescope}

The \sunrise{} telescope (see \fig{telescope_labelled}) is a light-weight Gregory-type reflector with an aperture of 1000\,mm, 324\,mm central obscuration and 24.2\,m effective focal length originally manufactured by Kayser-Threde in Munich in close collaboration with \ac{mps}. Eight struts arranged in triangles form its Serrurier structure, which is connected to the central frame, made of riveted sheet steel. Front and rear rings as well as connecting struts are made of carbon-fiber-based composite materials for high stiffness and low thermal expansion. 
The parabolic main mirror (made from \href{https://www.schott.com/en-au/products/zerodur-p1000269}{Zerodur\textsuperscript{\textregistered}}  by SCHOTT AG in Mainz, Germany)
was realised as an extremely light-weight structure, fine-ground and polished by \href{https://www.safran-group.com/companies/safran-reosc}{Safran Reosc}
(formerly Sagem) in France.
At the primary focus, the \acf{hrw} acts as a field stop with a central hole defining the usable \ac{fov} of 3.4\arcmin{}, corresponding to about 150\,Mm on the solar surface. Approximately 1\,kW of solar radiation hits the field stop, of which 99\% are removed mostly by reflection (85\%) and partially by absorption, such that only 10\,W enter the science instrumentation. Dedicated radiators connected by ammonia heat pipes prevent overheating of the \ac{hrw}. 

The telescope of \sunriseiii{} is almost identical to the one flown in 2009 and 2013 with only minor modifications \cite[for a detailed description of the \sunrisei{} telescope see][]{barthol11}. The secondary mirror was already replaced for the 2013 flight, due to damage during the 2009 landing. New M3 and M4 mirrors (plane folding mirrors towards the \ac{pfi}) have been manufactured by the \ac{kbsi} prior to the 2022 flight.  Damaged parts of the carbon fiber structure were re-manufactured or replaced by spare parts. All mirrors (M1, M2, M3, M4) have been re-coated at the Calar Alto Astronomical Observatory (Spain) in November 2023 with 120\,nm bare aluminum without protection layer. 

The \ac{hrw} and its thermal subsystem were rebuilt. In contrast to the 2009 and 2013 flights, the surface of the primary focus field stop was diamond milled on the aluminum bulk by \href{https://www.imtek.de/}{\acsu{imtek}}
(\aclu{imtek},  University of Freiburg, Germany), avoiding the fragile \ac{ssm}, which was used before. A slightly higher absorptivity was accepted in favor of a reduced risk of failure. 
The thermal insulation of the telescope was adapted to the new gondola geometry. 

\colfig{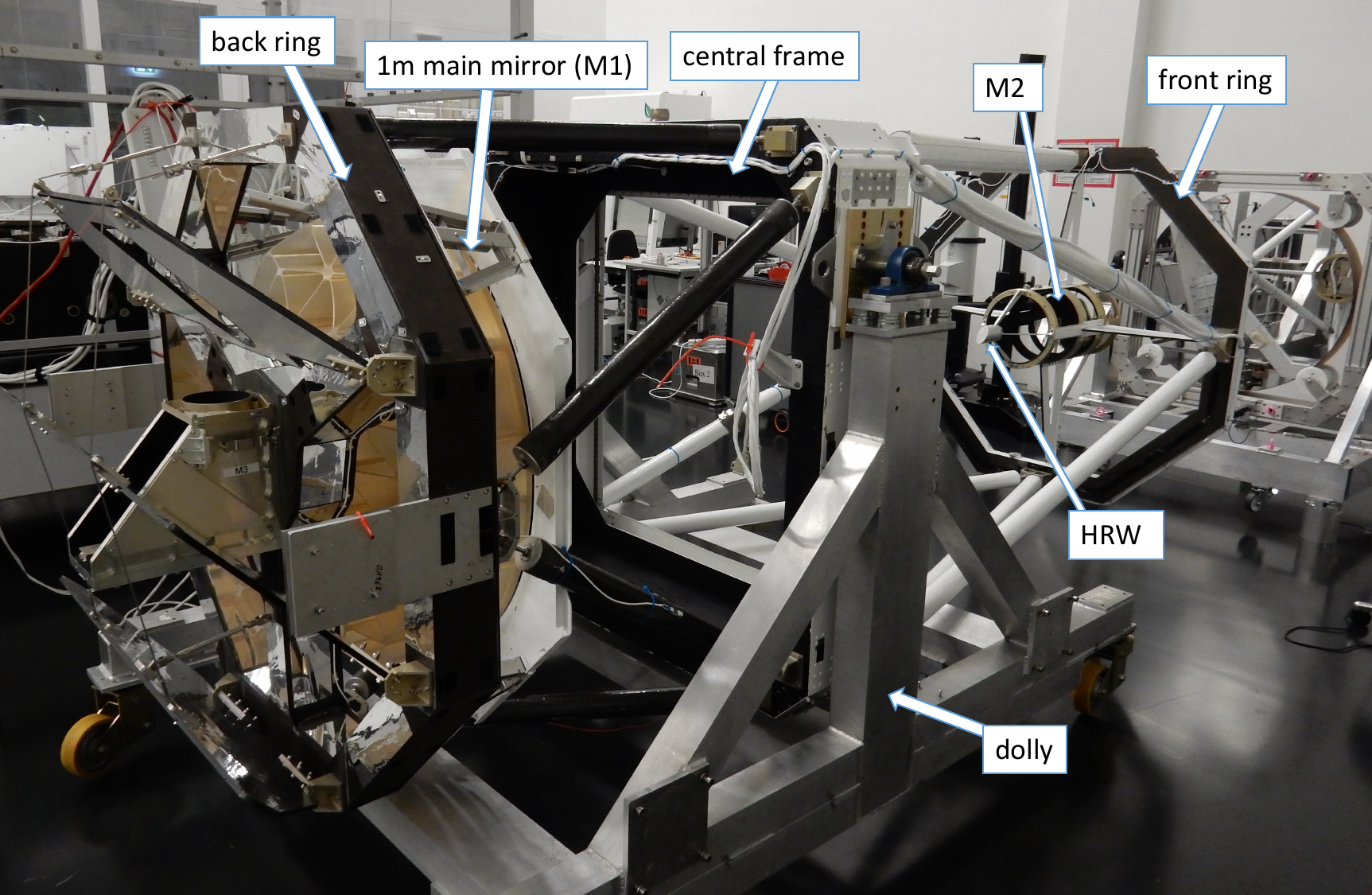}{\sunriseiii{} telescope during alignment in the ISO-8 cleanroom at \ac{mps}. The 1-m primary mirror is on the left and can be recognized by the brownish honeycomb structure, the central frame houses the aperture door (open in this image) and the spider to the right holds the \ac{hrw} at the primary focus of the telescope and the secondary mirror. The dolly is used for alignment and transport and is a non-flight item.}

The highest optical performance throughout the flight is achieved by the semi-active control system of the \sunrise{} telescope. A wavefront sensor located in the post-focus instrumentation constantly monitors the alignment status, generating control signals for mirror re-positioning. The M2 lateral and axial position can be fine-tuned to ensure the correct relative M1/M2 alignment for all telescope elevations and the expected thermal loads. While the lateral position is manually optimized (by minimizing coma), axial position (focus) is automatically corrected in closed loop (see \sref{cws}).

\subsubsection{\acf{sgt}}\label{sunguider}

The \acl{sgt} (see \fig{FirstLight}, right image) is the main source of feedback for the gondola \ac{pcs} when it is tracking the Sun. The \ac{sgt} has two sets of sensors for pointing: the intermediate sun sensors and the quad-cell photodiode. The intermediate sun sensors have a \ac{fov} of approximately $\pm$15\dg{} and are mounted in an azimuth/elevation configuration. They are used to coarsely align with the Sun, and are only used up until the point at which enough sunlight enters the \ac{sgt}'s aperture and hits the quad-cell photodiode. The optics of the \ac{sgt} result in a 1\,cm image of the Sun landing on the 14\,mm diameter quad-cell. Mounted on the center of the quad-cell is a circular occulting disk, which prevents the sensor from saturating and makes it more sensitive to pointing errors. Once sufficient sunlight hits the quad-cell, the \ac{pcs} works to keep the image of the Sun perfectly centered on the sensor. The quad-cell photodiode is mounted on an adjustable X-Y Stage, the movement of which causes an intentional misalignment between the \ac{sgt} and the main telescope. This enables to accurately aim the main telescope at any location on the solar disk, including the limb, while the \ac{pcs} constantly points the \ac{sgt} at Sun center.

\subsubsection{\acf{imu}}\label{imu}
A KVH 1725 \ac{imu} is mounted next to the \acsu{sgt}. It features three single-axis interferometric fiber optic gyroscopes as well as three single-axis accelerometers resulting in a six-degree of freedom inertial sensor. When tracking the Sun, the \ac{pcs} uses \ac{imu} readings to dampen the gondola's roll movement. When first acquiring the Sun, the \ac{imu} readings are used to control the 1\dg{} per second azimuth spin.

\subsection{\acf{pfi} Platform}\label{pfi}

\colfigtwo[0.49]{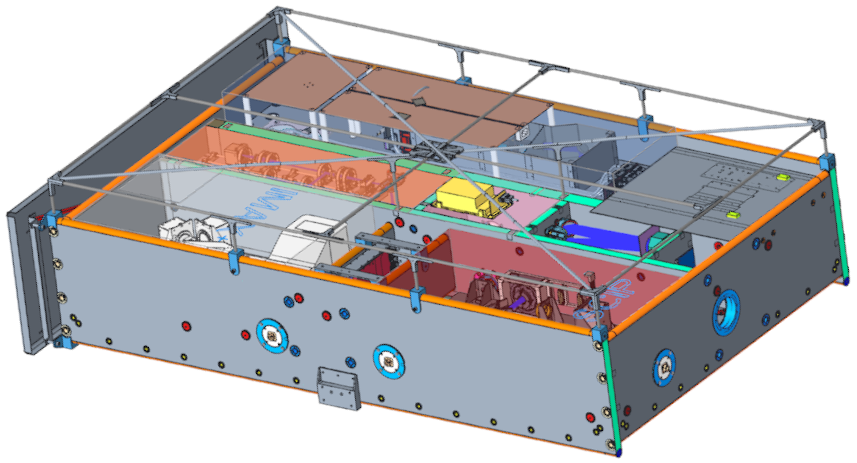}{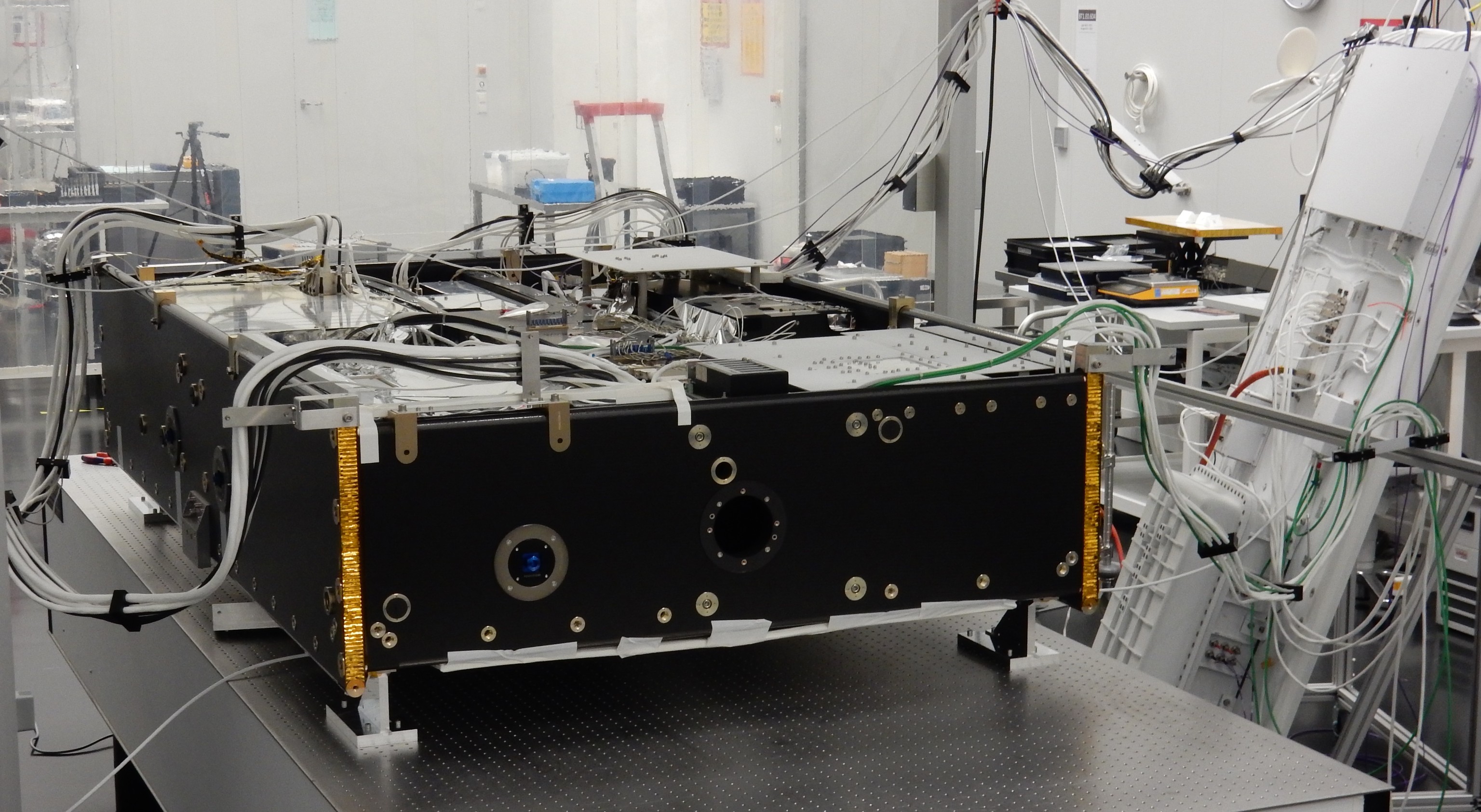}{\acs{cad} model of the \sunriseiii{} \ac{pfi} (\textsl{left}) and the flight-version of the \ac{pfi} under the clean tent in the \ac{mps} clean room (\textsl{right}). 
}

The \sunriseiii{} \ac{pfi} (see \fig{PFI1}) is mounted on top of the telescope, and houses the  four instrument modules with their proximity electronics. The compact and lightweight bookshelf-like structure of the \ac{pfi} is made from Carbon-fibre honeycomb elements and provides a stable platform for both, the scientific instruments, the image stabilizer and the light distribution optics. It also contains the \ac{islid} controllers responsible of commanding the \ac{islid} filter wheel in F2, and of the temperature control during the ascent and flight phases, as well as the \ac{mtc}. On top of its surface, the individual radiators of the instruments are located for maximum cooling efficiency of cameras and electronics. 

To achieve its science goals, \sunriseiii{} requires instruments that are capable of measuring with high precision the full polarization state of sunlight at a spatial resolution close to the diffraction limit of the 1-meter telescope and at a spectral resolution allowing to gain a maximum of information from the spectral lines. The basis for this lies in the \acf{islid} and the \acf{cws}, described in the following two subsections.

\subsubsection{\acf{islid}}\label{islid}

\ac{islid} is the central optical interface unit between the telescope and the scientific instruments.
It also contains the fast tip/tilt mirror, which acts as image motion compensator, and which is controlled in closed loop by the \ac{cws}. 
The second important task of \ac{islid} is to separate the wavelength bands in a photon efficient manner, such that each instrument gets the maximum available photon flux in its working wavelength range.

\colfig{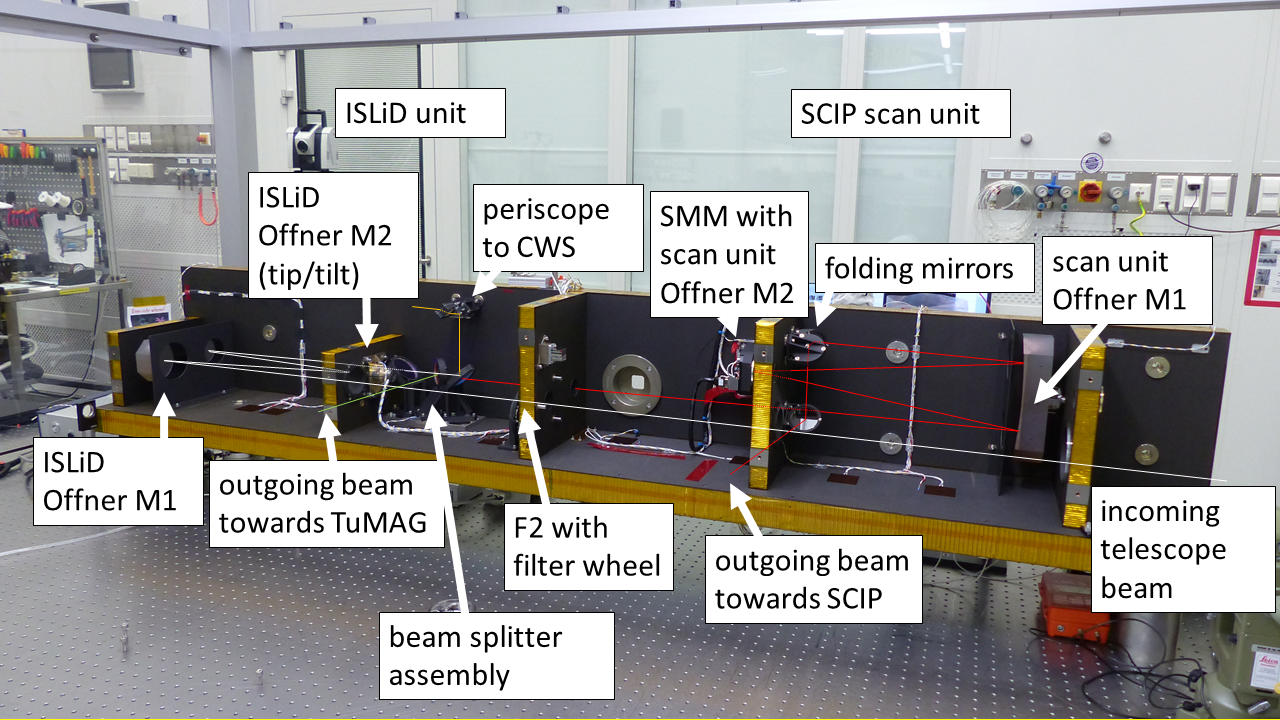}{The \ac{islid} system during assembly, before the left-hand side stiffening wall is mounted. The optical paths within the \ac{islid} system including the \ac{scip} scan unit are illustrated by \textsl{thin colored lines} (\textsl{white} means panchromatic, \textsl{green} symbolises the wavelength band sent to \ac{tumag}, \textsl{orange} for \ac{cws}, \textsl{red} goes to \ac{scip}. The \textsl{blue part} is not visible in this picture, since it is sent to the other side of the vertical bench, see \fig{ISLiD_center}). Major subunits are labeled.}

The design of the light distribution system of \sunriseiandii{} was driven by the need to observe below 250\,nm, and it was consequently  tailored to this purpose \cite[see][]{gandorfer11}. For \sunriseiii{}, we could concentrate on wavelengths above 300\,nm, which greatly reduces the complexity of the system. 

The new \ac{islid} is a 1:1 optical relay, which means that each instrument can be built in such a way as it would be directly fed by the F\#24.2 telescope beam. 
The optical arrangement is based on an Offner relay (see Figure~\ref{ISLiD_overview}), consisting of two spherical mirrors (Offner-M1 and Offner-M2), sharing a common center of curvature \citep{Offner73}. Their radii of curvature have a ratio of 2:1. Such a system is per design free from spherical aberration, astigmatism and coma. The common center of curvature is positioned in the geometrical center of the \ac{pfi}, minimizing any negative impact of (thermo-)elastic bending of the \ac{pfi} structure to optical misalignment of \ac{islid}.  The reflection geometry is chosen such that the three reflections (M1--M2--M1) compensate the polarimetric behaviour of the telescope mirrors M3 and M4 to a great extent. 
The Offner-M1 focal length is chosen such that it forms a real image of the telescope pupil on the Offner-M2 mirror, which can thus be used as a fast tip/tilt mirror for image motion compensation. It is mounted on a piezo actuator, which receives its control signal in closed loop by the \ac{cws} (see \sref{cws}). The Offner M1 and M2 mirrors of \ac{islid} were manufactured by \ac{kbsi}.

The light distribution and wavelength separation is achieved by 4 plane-parallel beam splitter plates, arranged in two pairs of anti-parallel orientation, such that the transmitted beam is free from any astigmatism and lateral chromatic aberration (see \fig{ISLiD_center}). They all  reflect the light by 90\dg{} in the following order:  
The first beam splitter plate reflects all wavelengths below 450\,nm towards the \ac{susi} instrument. 
The second plate compensates the astigmatism introduced by the oblique transmission through the first plate. The transmitted beam is therefore free of astigmatism, but shows chromatic lateral aberration. 
The third plate reflects 500--550\,nm towards \ac{tumag}, which is from a technical point of view monochromatic and thus insensitive to the chromatic error. 
The fourth plate compensates the astigmatism introduced by the oblique transmission through the \ac{tumag} beam splitter and reflects 590--650\,nm to the \ac{cws}. The \ac{cws} path is equipped with a narrowband filter, which is used as the astigmatism corrector for the uncompensated transmission through plate 3.
The transmitted beam through all plates is free from any chromatic lateral aberration and astigmatism and forms the interface focus to the \ac{scip} scan unit. 

\colfig{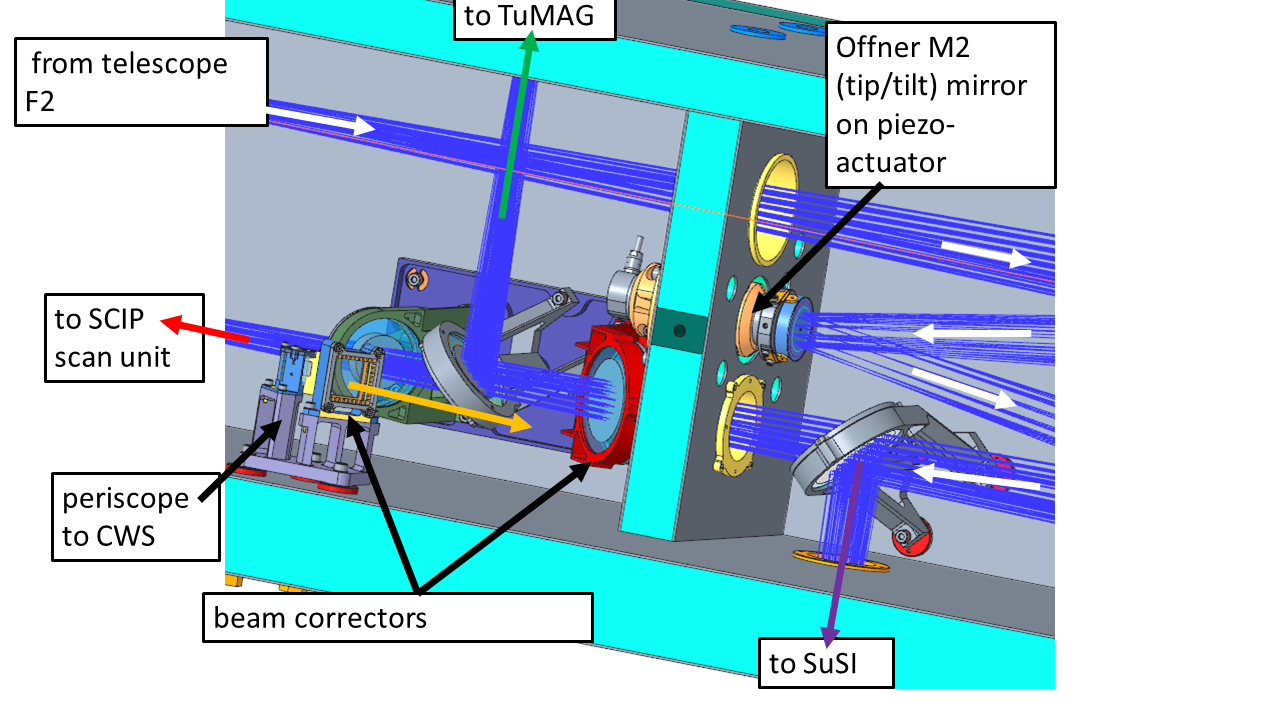}{Central part of the \ac{islid}  system, showing the Offner-M2 (tip/tilt) mirror on its piezo stage, and the beam splitter arrangement. \textsl{White arrows} mark the panchromatic beam from the telescope. The \textsl{colored arrows} indicate the wavelength bands sent towards the exit foci. \textsl{Black arrows} denote mechanical or optical devices. }

This scan unit for the \ac{scip} instrument contains the \acl{smm} \cite[\acsu{smm},][]{oba22} and is incorporated as a part of the \ac{islid} arrangement (see \fig{ISLiD_scan_unit}). 
It consists of a second Offner relay, which is identical to the one in the \ac{islid} system, but rotated by 90\dg{} around the \ac{los}. In between the two systems, a field stop is located, which minimizes false light contamination towards \ac{scip}. The mirrors of the \ac{scip} scan unit (first mirror fabricated by \ac{kbsi} and secondary mirror on \ac{smm} fabricated by Natsume Optical Corp.) are gold coated, making the system insensitive to back reflections of the \ac{tumag} prefilters and protecting \ac{tumag} against potential back-reflections of residual green light from the \ac{scip} polarization modulator.  Both Offner systems are equipped with a carefully designed set of vanes, in order to prevent any sneak paths. 

\colfig{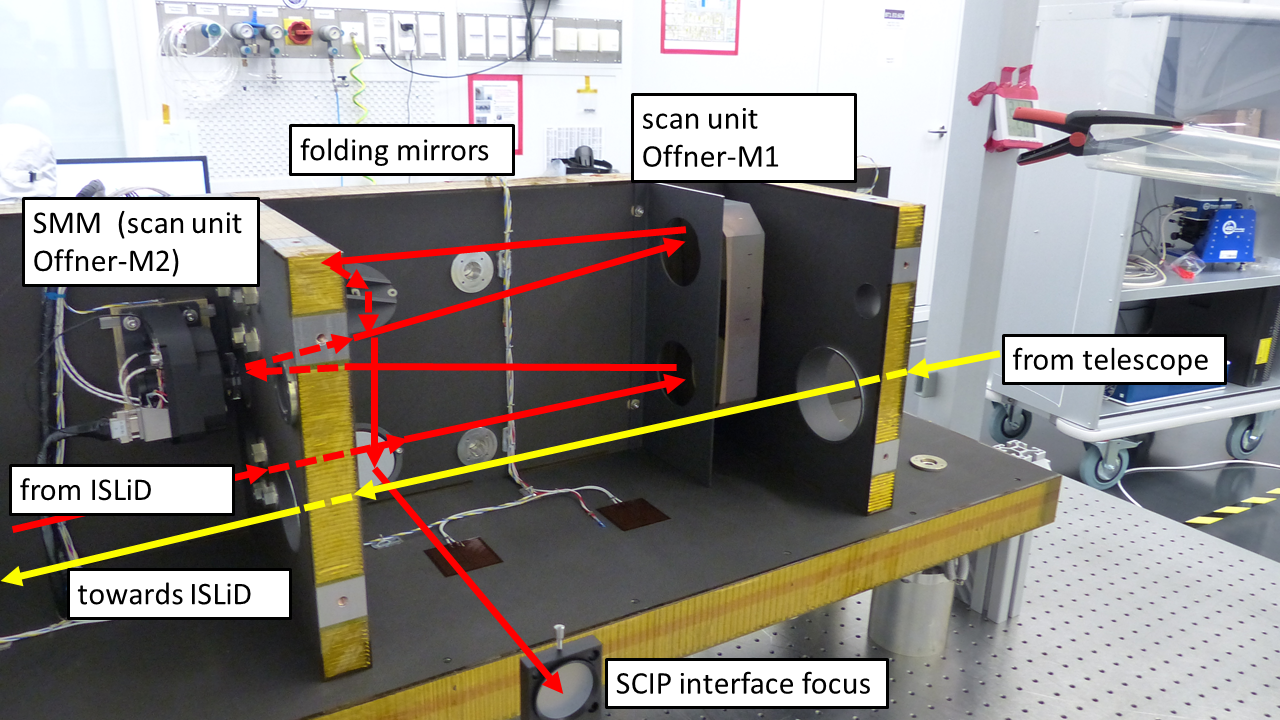}{\ac{scip} scan unit during assembly. The \textsl{red arrows} symbolize the light path from \ac{islid} towards the \ac{scip} interface focus. The Offner relay uses the \ac{smm} as its secondary mirror. A set of three folding mirrors (only two are visible here) acts as path length compensator and is used to bring the exit focus and exit pupil to the desired positions, respectively.}

At the F2 focus, the entrance of \ac{islid}, a filter wheel is located with 8 positions (see \sref{test:optical}). A field stop with a diameter of 180\arcsec{}, and a shutter for common dark exposures. The 6 remaining positions carry alignment and test targets, which can be used by all instruments in flight for calibration purposes. 
The filter wheel is based on piezo-technology and has been manufactured for \ac{mps} by SmarAct in Germany.  

A first glimpse of the excellent optical performance of telescope and \ac{islid} could already be obtained during the 2024 flight: all three science instruments transmitted near real-time full-resolution thumbnail images via satellite communication of superb quality. A detailed assessment of the image quality will only be possible after the data reduction phase (see \sref{dataconcept}).

\subsubsection{\acf{cws}}\label{cws}

The \ac{cws} provides high-frequency precision image stabilization and autofocus. 
In contrast to \sunriseiandii{}, where the \ac{cws} had a six-element Shack-Hartmann \ac{wfs} for fast tip-tilt correction, slow autofocus and coma measurement, \sunriseiii{} uses the \ac{wfs} only for the slow autofocus and coma measurement and adds a \ac{ct} for the fast tip-tilt-correction. This has the advantage of an improved tip-tilt measurement accuracy by using diffraction limited sampling of the full aperture resolution and a much faster readout of the \ac{ct} camera which results in an increased correction bandwidth.
The fast tip-tilt correction is done by the tip-tilt mirror (Offner-M2), the slow autofocus by positioning the secondary mirror M2 in the $z$-direction (see \fig{siii_design}, telescope coordinate system).
\fig{CWS} shows the optics of the \ac{cws} during final testing before integration into \ac{islid}, \tab{ounittable} lists its parameters.

\colfig{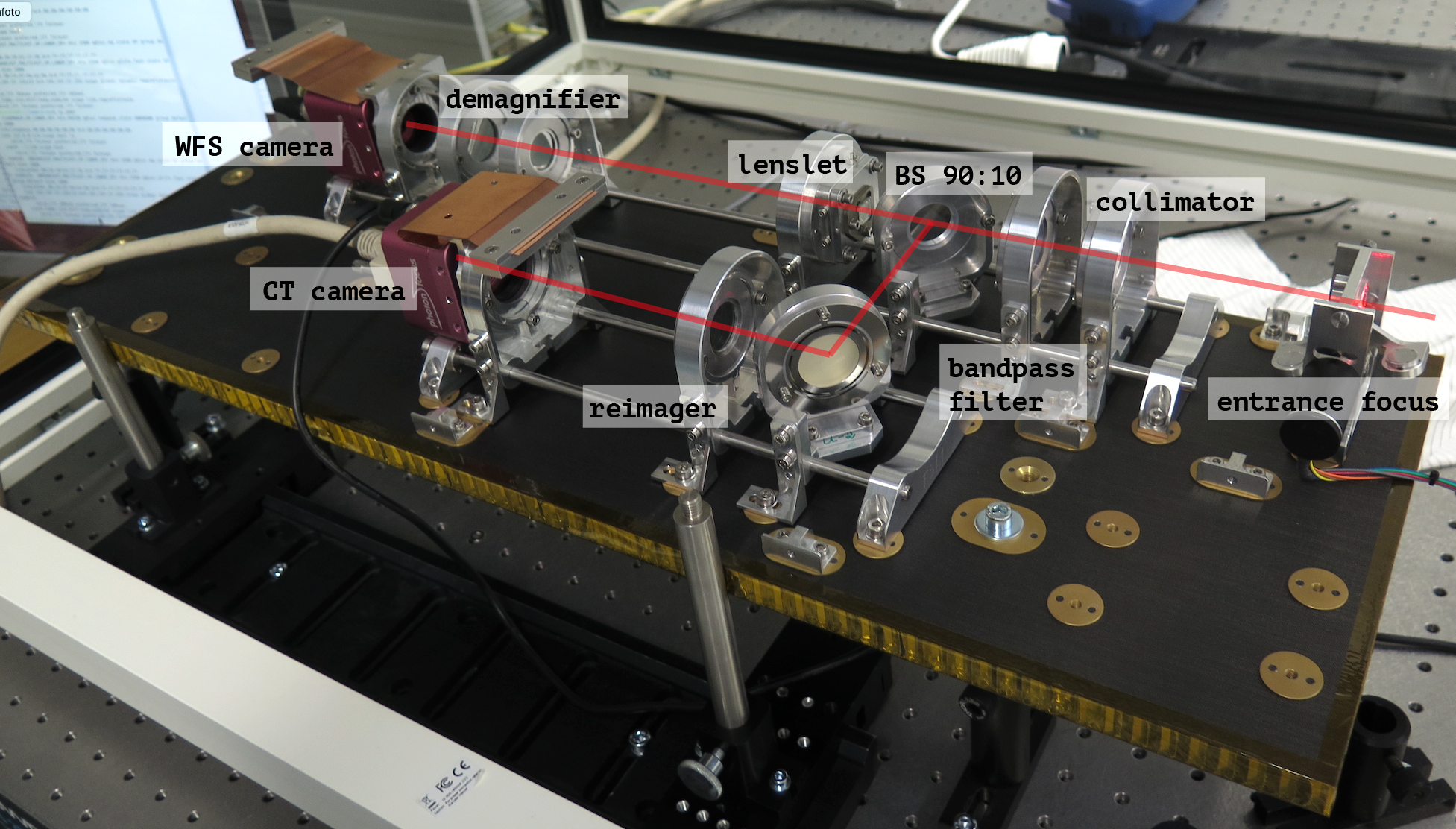}{Optics Unit of the \ac{cws}. The light enters the \ac{cws} at the entrance focus on the right.}

\begin{table}[ht]
\caption{Parameters of the \ac{wfs} and \ac{ct} inside the optics unit of the \ac{cws}.} 
\label{ounittable}
\begin{tabular}{lcc} %
\hline
   & \ac{wfs} & \ac{ct}  \\
\hline
  \# of subapertures (total) & 6 & 1  \\
  measured aberrations & tip-tilt, focus, coma & tip-tilt \\
  pupil size on sky [m] & 0.33 &  1  \\
  \ac{fov} on sky [arcsec] & 13 & 6.7  \\
  pixel scale [arcsec/pix] & 0.2 & 0.07   \\
  \# of pixels used for cross-correlation & 64$\times$64 & 96$\times$96   \\
  camera (manufacturer Photonfocus) & MV1-D1024E-80-G2 & MV1-D1024E-160-CL   \\
  typical framerate [Hz]& 500 & 7000 \\
\hline
\end{tabular}
\end{table}

Tests have shown a 0\,db correction bandwidth of 130\,Hz and 160\,Hz for tip and tilt correction, respectively. The tip-tilt measurement accuracy is of the order of 1 milli-arcsecond \ac{rms} per axis for a typical granulation contrast at the operating wavelength of 640\,nm.
The focus and coma measurement accuracy is better than $\lambda/100$ ($\approx\,$6\,nm \ac{rms}) per mode. The tip-tilt axes can be moved in steps of 1 milli-arcsecond over the full range of 60\arcsec{}.

The \ac{cws} electronic unit consists of a fast conduction-cooled 4-core compute board, 16-bit digital-to-analog conversion, two 100\,V voltage amplifier channels for the fast tip-tilt mirror and additional DC-DC converters and components such as temperature sensors or the motor controller for the \ac{cws} entrance focus unit. All electronics (including the two cameras of the \ac{cws} optics unit) have been modified and thermally vacuum tested to ensure operation at millibar pressure levels without requiring a pressurized box. The consumed power, i.e., the resulting heat load of typically 60\,W in total, is conducted away by heat straps attached to external radiation shields and radiated into space.

A more detailed description of the \ac{cws} can be found in \cite{cws25}. A preliminary assessment of the 2024 flight performance of the \ac{cws} confirmed excellent stability, enabling \sunriseiii{} to obtain a record-long time series of more than four hours without a single loss of the lock point, and to stably lock on a faculae region for limb observations.

\subsection{Scientific Instruments}\label{scienceinstruments}

Three scientific instruments analyze the light provided by the telescope and \ac{islid}, stabilized to milli-arcsecond accuracy by the \ac{cws}. The instruments are highly complementary and make use of the unique possibility of observing spectral lines from the \ac{nuv} to the \ac{nir} simultaneously, without suffering from the effects of seeing and the displacement introduced by the refraction in the Earth's atmosphere. All three instruments are measuring the full polarimetric information of the light (Stokes $I$, $Q$, $U$, $V$) and are optimized for photon efficiency to achieve the maximum possible \ac{s2n}, important for the accurate determination of the magnetic field vector. The instruments deliver images and spectra at the diffraction limit of the telescope at high spectral resolution. 

\tab{tab:specs} summarizes the most important specifications for the three instruments. The first three rows are self-explanatory (spectral range, optical scheme, and science targets). The fourth row describes the \ac{fov}: For the two slit-spectrographs, \ac{scip} and \ac{susi}, with one spatial and one spectral dimension, a 2D-spatial \ac{fov} is achieved by scanning, i.e., moving the slit over the solar image. This is done in a step-like fashion for \ac{scip}, and in a continuous way by \ac{susi} (see \sref{perf:slit}), with  the scanning range depending on the observing mode, but always lying within the 2D-\ac{fov} of \ac{tumag}.

The fifth row in \tab{tab:specs} specifies the maximum temporal resolution of the instruments. For the slit spectropolarimeters, it is derived from the time required to record the spectrum for one slit position. In intensity mode, this is determined by the frame rate of the cameras, and in full-Stokes polarimetric mode, the time it takes to record a full modulation cycle. For \ac{susi}, all frames were stored on disk and the data are available for scientific analysis at these high frame rates. However, to achieve a polarimetric sensitivity of \ten{-3} (see row number six), at least 5\,s accumulation time is needed, strongly dependent on the selected wavelength and observed solar feature. \ac{scip} followed an on-board accumulation strategy. Depending on the observing mode, frames are accumulated over \lnum{1024}, \lnum{5120}, or \lnum{10240}\,ms in full-Stokes mode, or 64\,ms in intensity-only mode. \ac{scip} reaches a polarimetric sensitivity of 3$\cdot$\ten{-4} after an integration time of less than 10\,s.
The temporal resolution of \ac{tumag} given in the sixth row of \tab{tab:specs}  refers to the time required to acquire a complete single spectral line scan in full-Stokes mode and intensity-only mode, respectively. This time is about doubled when scanning both \ac{tumag} spectral lines.

\begin{table*}
\caption{Main specifications of the scientific post-focus instruments.}
\label{tab:specs}
\begin{tabular}
{L{0.17\textwidth}L{0.225\textwidth}L{0.225\textwidth}L{0.225\textwidth}}
\hline
\footnotesize
   & \acs{susi} & \acs{scip}  &  \acs{tumag} \\ \hline
Spectral range & $309 - 417$\,nm & $765 - 855$\,nm & $517 - 525$\,nm \\ \hline
Optical scheme  & Modified Czerny-Turner spectrograph with tunable low-order grating; slit scan unit and slit-jaw imager & Littrow spectrograph with fixed high-order echelle grating; slit scan unit and slit-jaw imager & Collimated double-pass Fabry-P\'erot filtergraph \\ \hline
Scientific target, wavelength bands & 
Spectrograph:\newline many-line magnetic field diagnostics over entire spectral range\newline\vspace{-2ex}\newline
Slit-jaw imager:\newline
325 nm continuum (lower photosphere) & 
Photosphere:\newline
\fei{} 846.8 \& 851.4\,nm\newline\vspace{-2ex}
\newline
Upper photosphere: \newline
\ki{} 766.5 \& 769.9\,nm\newline\vspace{-2ex}\newline
Chromosphere:\newline 
\caii{} 849.8 \& 854.2\,nm\newline\vspace{-2ex} 
\newline 
Slit-jaw imager:\newline 
770.5\,nm continuum
& 
 Photosphere:\newline \fei{} 525.02 \& 525.06\,nm\newline\vspace{-2ex}
 \newline
 Temperature minimum:\newline
 \mgi{} 517.3\,nm
 \\ \hline
\acs{fov}      & 59\arcsec{}$\times$(17--27)~\AA{} & 58\arcsec{}$\times$(61--83)~\AA{} & 63\arcsec{}$\times$63\arcsec{} \\ 
Spatial sampling & 0.03\arcsec{}/px & 0.094\arcsec{}/px & 0.037\arcsec{}/px  \\ 
Resolution on Sun & 56--76\,km & 138--154\,km& 92\,km  \\ 
Spectral sampling & (8.5--13.5)\,m\AA{}/px & (39--42)\,m\AA{}/px &  66\,m\AA{} spec. bandwidth  \\ \hline
Time resolution: \newline full-Stokes \newline intensity-only & 
\mbox{}\newline
4\,\acs{fps} \newline
47\,\acs{fps}
&
\mbox{}\newline
1\,\acs{fps} \newline
{16\,\acs{fps}}
& 
\mbox{}\newline
29.07\,s for 8 WL-pos \newline
23.74\,s for 15 WL-pos  \\
\hline
Pol. sensitivity & $<1\cdot\ten{-3}$ & $<1\cdot\ten{-3}$ & $<1\cdot\ten{-3}$ \\ \hline
Max. data rate   & 600\,MByte/s & 100\,MByte/s & 100\,MByte/s \\ \hline
\end{tabular}
\end{table*}

\subsubsection{\acf{susi}}\label{susi}

The \ac{susi} spectropolarimeter has been designed, developed and built by \ac{mps}, with contributions from \ac{naoj}. It observes the \ac{nuv} wavelength range between 309\,nm  and 417\,nm. Compared to the visible or the \ac{nir}, this spectral domain exhibits a much larger density of spectral lines, formed in different layers of the solar photosphere and chromosphere (see \fig{formheights}). Due to the short wavelengths, \ac{susi} is the \sunriseiii{} instrument offering the highest spatial resolution of about 60\,km on the Sun. The combined effect of the high spatial resolution and the large number of lines leads to a significant improvement in the detectability of small-scale magnetic structures and their height dependence. In order to meet these specifications, \ac{susi} relies on a number of technical innovations.

The opto-mechanical layout of \ac{susi} is shown in \fig{SUSI_layout}. To measure the polarization of light, the polarization information is converted into a periodic intensity modulation and then demodulated with synchronized detectors. In \ac{susi}, the polarization modulation is accomplished by a rotating waveplate. In order to meet the challenging polarimetric accuracy requirements, the rotating mechanism of this \ac{pmu} is designed for a very high rotation uniformity \cite[deviations less than 0.3\%,][]{kubo20}. The details of the polarimetric calibration of \ac{susi} are described in \cite{iglesias2025}.

The three \ac{susi} cameras operate at a framerate of 47.040\,fps in synchronization with the \ac{pmu}. This implies that each modulation cycle (half \ac{pmu} rotation lasting 256\,ms) is sampled by 12 frames of the spectrograph cameras. During pre-flight tests, the synchronization error between cameras and \ac{pmu} was measured to be below 0.36\,ms, and the residual image motion due to \ac{pmu} misalignment (beam wobbling) was below the required 0.1 pixels \ac{rms}.

\colfig[0.8]{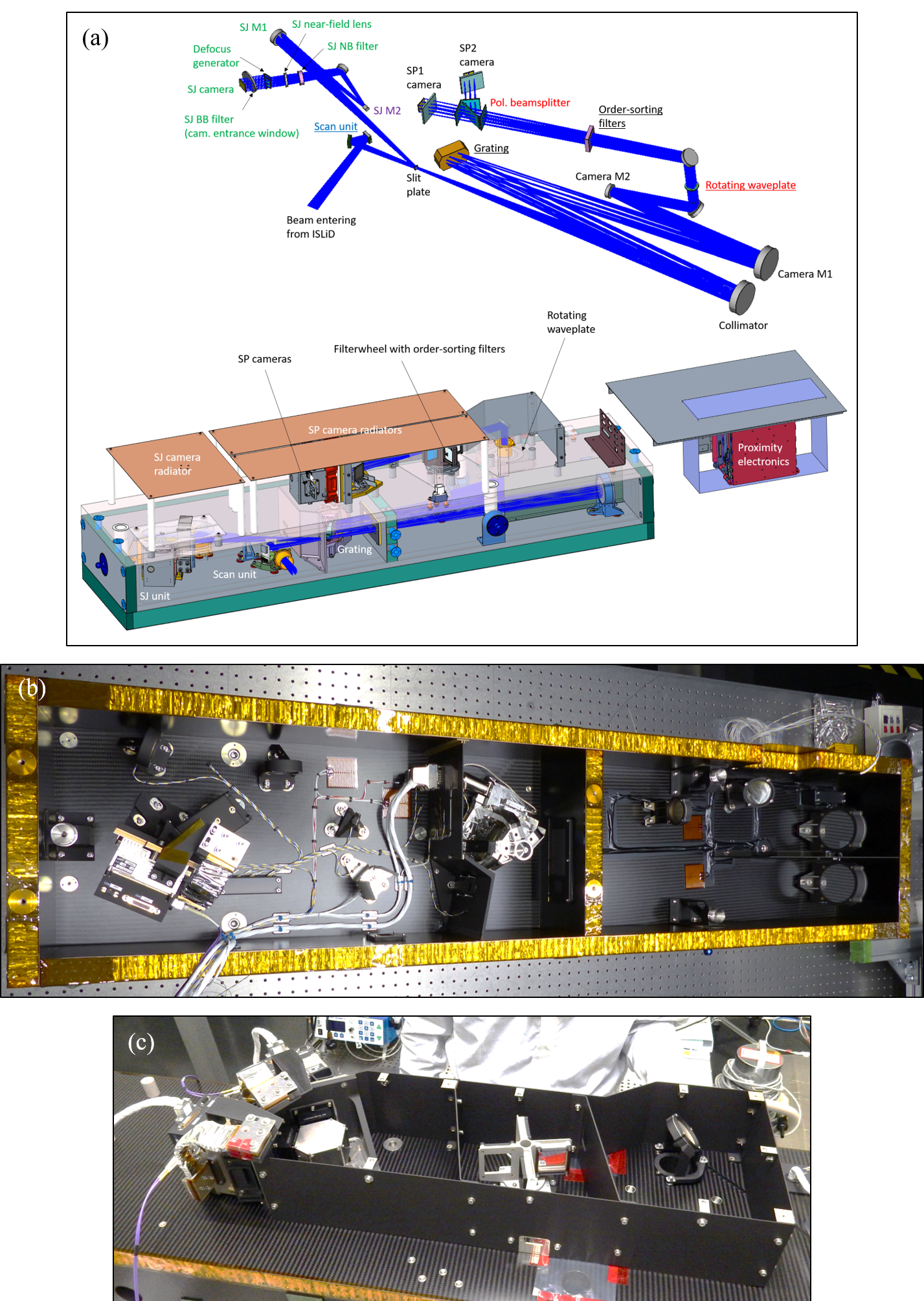}{Layout of \ac{susi}. Panel (a), \textsl{top}: basic optical layout; the components of the main functional units are labeled in different colors: scan unit (\textsl{sky blue}), spectrograph (\textsl{black}), polarization modulator (\textsl{red}) and slit-jaw unit (\textsl{green}); \textsl{underlined labels} denote a moving component involving a mechanism. Ray paths are marked in \textsl{navy blue}. Panel (a), \textsl{bottom}: Opto-mechanical layout with structural elements, radiators and proximity electronics. Panels (b, c): Lower and upper levels of \ac{susi}, during integration in the \ac{mps} cleanroom.}

The \ac{susi} spectrograph is equipped with two cameras in a dual-beam polarimetric configuration which allows suppressing measurement errors originating from residual jitter. The cameras, developed at \ac{mps}, meet the harsh environmental requirements of a stratospheric balloon flight and are based on \ac{uv} optimized image sensors with very low noise (2 e$^{-}$ \ac{rms}) to allow for efficient measurements in the deep, heavily photon-starved \ac{uv} spectral lines. The data are recorded at a high frame rate of 47 \ac{fps} to make them insensitive to residual large amplitude slow jitter of the system. All individual frames are stored on-board. This greatly  enhances our flexibility in the later data reduction, in particular in terms of trading spatial and temporal resolution against signal-to-noise, depending on the actual science question. 

High spatial resolution context images are obtained by re-imaging the 2D solar scene around the linear spectrograph entrance slit onto a separate camera. This setup is known as slit-jaw imaging. The slit-jaw imager, which is strictly synchronized to the two spectrograph cameras, detects residual jitter and optical wavefront errors. It allows for a post-facto numerical image restoration of the spectrograph scans following the technique developed by \cite{vannoort17} that has impressively stood the test in recent ground-based observations.

\subsubsection{\acf{tumag}}\label{tumag}

\ac{tumag} (see \fig{TuMag-both}) is a tunable, dual-beam imaging spectropolarimeter in visible wavelengths \cite[]{tumag25}. It has been designed, developed, and manufactured by the \ac{s3pc} to probe the \ac{los} velocity and vector magnetic field of the photosphere and the low chromosphere across a \ac{fov} of 63\arcsec{}$\times$63\arcsec{}. It builds upon technical solutions proven in the  \sunriseiandii{} flights of the \ac{imax}  instrument  \cite[]{martinezpillet11} and on the \ac{phi} instrument \cite[]{solanki20} aboard \acl{solo}, like the use of liquid crystal variable retarders as polarization modulators and a  LiNbO$_3$ solid Fabry-P\'erot etalon as spectrometer in collimated configuration, while \ac{phi} incorporates the etalon in telecentric configuration. \ac{tumag} implements a number of technical innovations, like a state-of-the-art double filter wheel \cite[]{2022SPIE12188E..3AS}, two custom-made cameras \cite[\aclp{spgcam}, \acsp{spgcam}\acused{spgcam},][]{orozcosuarez23}, a composite-material optical bench \cite[]{fernandezmedina23}, specialized thermal control means for critical subsystems \cite[]{2023TuM_ICES139...G}, and completely new electronics, software and ground support equipment.  Its optical design enables spatial sampling of the solar surface at frequencies above the Nyquist cut-off. Quasi-simultaneous, precise full-Stokes polarimetry is possible in multiple wavelengths across the line profiles of any two among the three Zeeman-sensitive \fei{} (525.02\,nm and 525.06\,nm) and \mgi{} line (517.3\,nm) spectral lines in less than 90\,s. 

\colfigwide{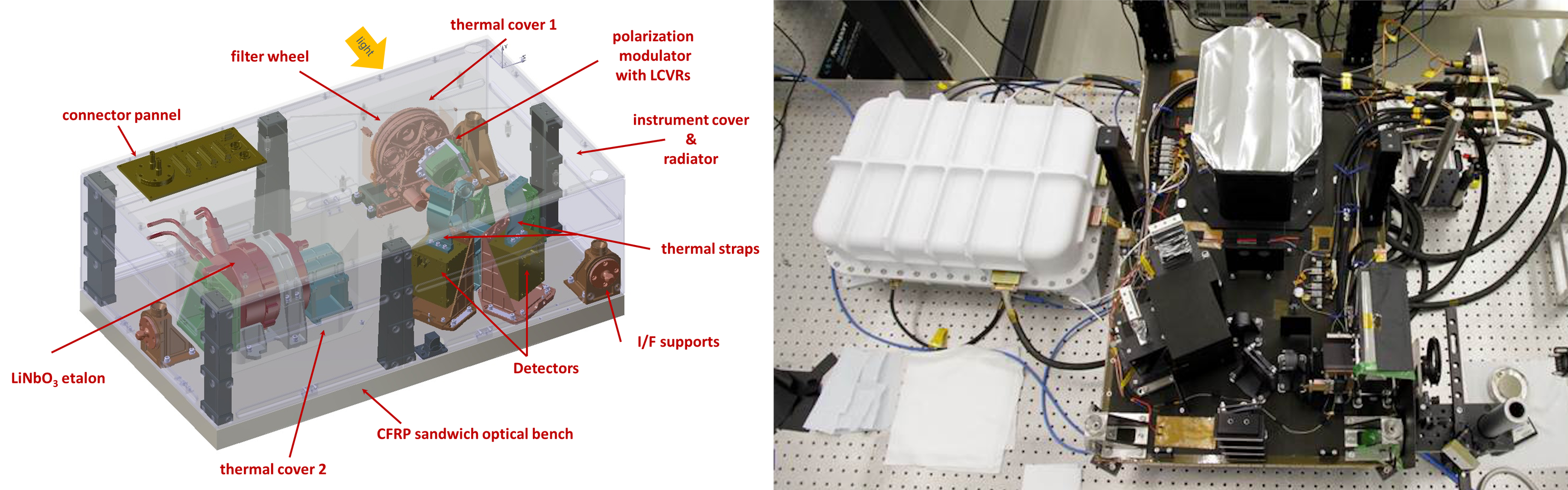}{Opto-mechanical design of the \ac{tumag} instrument (\textsl{left}), and flight instrument before closing the housing, including the electronic unit in its white housing (\textsl{right}).}

The optical unit is controlled by the electronic unit, which utilizes commercial off-the-shelf components within a pressurized box to ensure functionality in low-pressure environments. The electronic unit includes a power converter module, a \ac{hvps} to drive the etalon, an \ac{amhd} module to control the filter wheel, the \acp{lcvr}, the \ac{hvps}, and to thermally control the instrument and monitor its status, and a digital processing unit to manage and control the whole instrument. This processing unit is based on an NVIDIA\textsuperscript{\textregistered} Jetson TX2, with an embedded multimedia card with 8\,TByte \ac{ssd} external data storage, and it runs the different observation modes, compresses the data, and sends them to the \ac{ics}. The control software, firmware, and ground support equipment have been developed by \ac{s3pc} to maximize commonalities with \ac{scip}.
 
\ac{tumag}'s spatial resolution is determined by the aperture of the \sunrise{} telescope, and allows to resolve structures of $\approx\,$80\,km on the Sun, oversampled by about a factor three, so that the pixel size is about 0.037\arcsec{}, or 27\,km on the Sun. The narrow-band tunable filter is a LiNbO$_3$ solid-state Fabry-P\'erot etalon, an identical spare device to the one already flown on \sunriseiandii{}. It is used in double pass, which relaxes the demanding requirements on passband stability and saves mass and power. The spectral resolution is $\approx7\,$pm, corresponding to the \ac{fwhm} of a Gaussian function representing the spectral transmission profile of the instrument. To select the three different spectral lines, the filter wheel is equipped with narrowband interference prefilters, which can be switched at a speed better than 6\,s. The filter wheel also hosts a plate introducing a well-defined defocus for phase diversity measurements, allowing to reconstruct the \ac{psf} of the instrument \cite[]{bailen22,bailen22b}. The \acp{lcvr} provide optimum polarization modulation and guarantee a polarization efficiency exceeding 95\% in all four Stokes parameters. This assures achieving a polarimetric sensitivity better than \ten{-3}, in units of the continuum intensity, at a cadence of one observation in about 10\,s to 60\,s, depending on the number of wavelength points and the observation mode of the instrument \cite[see Tab. 8 in][]{tumag25}. The filter wheel also hosts two in-flight polarization calibration devices, a linear polarizer and a micro-polarizer calibration target\footnote{Patent pending. Patent application number: EP2282703.1 ”Calibration target and method for the snapshot calibration of imaging polarimeters”, Alberto Álvarez-Herrero and Pilar García  Parejo, 22 July 2022.}, allowing for the first time to perform in-flight polarization calibration measurements.

\subsubsection{\acf{scip}}\label{scip}

\ac{scip}, designed and built jointly by \ac{naoj} (lead) and \ac{s3pc} with contributions from \ac{jaxa} and \ac{mps}, performs spectropolarimetry in \ac{nir} photospheric and chromospheric lines with well-known properties. \ac{scip} emphasizes spectral resolution and polarimetric precision. The Stokes spectra of these lines are characterized by a high magnetic sensitivity, thanks to large Zeeman splitting at the long wavelengths at which \ac{scip} observes, and by the absence or at least greatly reduced strength of telluric blends. These characteristics greatly simplify the interpretation of \ac{scip} spectropolarimetry data compared to data recorded from ground-based observatories in this wavelength range. \ac{scip} is therefore highly complementary to the \ac{nuv} spectropolarimeter \ac{susi} (\sref{susi}), which offers increased resolution in the spatial and height domains at the expense of a lower photon flux and a more complex analysis.

\ac{scip} allows for simultaneous full Stokes measurements in two spectral regions. One of the spectral channels contains two of the \caii{} infrared triplet lines at 849.8\,nm and at 854.2\,nm that form at chromospheric heights (see \fig{SCIP_Ca8498}). This spectral channel is highly suitable for studying the stratification of magnetic fields in the solar chromosphere due to its sensitivity to atmospheric heights ranging from the photosphere to the middle chromosphere \cite[]{quinteronoda17}. The second spectral window is centered on the \ki{} D1 and D2 lines at 769.9 and 766.4\,nm, respectively. These spectral lines allow us to probe the upper photosphere and the temperature minimum heights, but are very poorly accessible from the ground due to absorption mainly by O$_2$ in the Earth's atmosphere at these wavelengths \cite[]{quinteronoda18}. 
When combined with highly Zeeman-sensitive \fei{} photospheric lines present in the \caii{}  spectral window, they allow to seamlessly retrieve the magnetic and thermodynamic parameters through all photospheric and most chromospheric heights.

The absence of atmospheric refraction and seeing at the float altitude of \sunrise{} ensures perfect co-pointing between the two \ac{scip} spectral regions, even though they are separated by almost 100\,nm. Together, and along with the information from \ac{susi}, the spectral lines in both spectral bands constrain the conditions in the solar atmosphere to unprecedented accuracy, well beyond what can be achieved by current ground-based solar telescopes.

\colfigwide{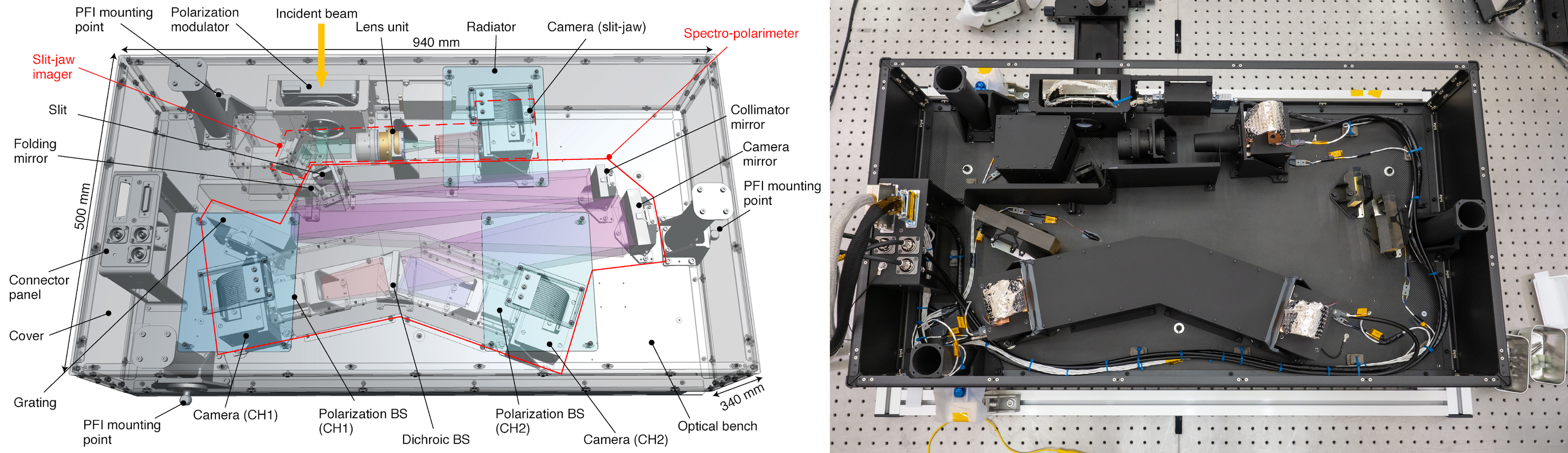}{Opto-mechanical layout of \ac{scip} (\textsl{left}) and pre-flight laboratory setup (\textsl{right}) showing opto-mechanical elements mounted on an optical bench made of a \ac{cfrp} sandwich panel with low thermal expansion.}

The opto-mechanical layout of \ac{scip} is shown in \fig{SCIP} \cite[]{uraguchi20}. The basic optical configuration of the \ac{scip} spectrograph is a reflective Littrow-configuration using a high-order echelle grating and aspheric mirrors \cite[]{tsuzuki20}. The slit width and the spatial pixel sampling are 0.094\arcsec{} to reach a spatial resolution of 0.19\arcsec{}, which is the diffraction limit at 770\,nm. The slit length and the scan mirror cover a \ac{fov} of 58\arcsec{}$\,\times\,$58\arcsec{}. Both  wavelength channels with different diffraction orders are separated by a dichroic beamsplitter and focused onto two cameras. Each camera employs a dual-beam polarimetric setup with a \ac{pbs} projecting both orthogonal polarization beams on the same sensor. The cameras employ a back-thinned, 2k$\,\times\,$2k  \ac{cmos} sensor \cite[\acp{spgcam},][]{orozcosuarez23}. 

As in \ac{susi}, the \ac{pmu} is based on a rotating waveplate whose retardation is optimized to have good polarization modulation in the spectral channels \cite[]{kubo20}. By synchronizing rotating phases of the waveplate and exposures of two \acp{spgcam}, modulated spectra are read-out every 32\,ms and are demodulated on-board to produce polarized spectra every 512\,ms as a minimum integration cycle at each slit position \cite[]{kubo23}. Because the waveplate consists of quartz and sapphire plates, anti-reflection coating is applied to all the surfaces and their parallelism is well controlled to suppress fringing and intensity variations.
The beam wobbling is attenuated by adjusting a tilt to the waveplate with respect to the rotating axis and is smaller than 0.01\arcsec{} \ac{rms}. The polarization measurement system of \ac{scip} provides 3$\times$\ten{-4} (1$\sigma$) polarization precision \cite[]{kawabata22} when integrated for 10\,s. 

The scan mirror mechanism \cite[\acsu{smm},][]{oba22} is placed in the \ac{scip} scan unit (see \sref{islid}). \ac{scip} has a slit-jaw imager whose wavelength is centered at around 770.5\,nm to obtain a continuum image. The read-out of the slit-jaw camera is also synchronized with the rotating waveplate and the two spectrograph cameras. In addition to the polarimetric measurements, \ac{scip} allows a rapid scanning observation only for intensities in which the scan mirror is stepped every 64\,ms to cover the \ac{fov} of 58\arcsec{}$\,\times\,$58\arcsec{} in 40\,s. In this mode, the waveplate continues to rotate. The resulting intensity modulation is corrected in the post-flight data processing pipeline.

\subsection{Electronic Units}\label{eunits}

\subsubsection{Electronic units of the science instruments}

The scientific instruments described above are controlled by their own electronic units. The units for \ac{scip}, \ac{tumag} and \ac{cws}  are located on the \acp{erack} on the $+x$ and $-x$ side of the gondola (see \fig{siii_design}), whereas the \ac{susi} is managed by the \ac{ics} directly and by a proximity electronic unit located in the \ac{pfi}. The electronic units of the instruments are connected to the corresponding optical units and detailed in the individual instrument sections and papers.

\subsubsection{\acf{ics}}\label{ics}

Unlike many ground-based solar observatories, the balloon-borne \sunriseiii{} observatory is designed for largely autonomous operation allowing for an automated execution of several-hour long scientific observing programs. For this purpose, the onboard hardware of \sunriseiii{} is hierarchically structured. The central node is the \ac{ics} that provides several Ethernet and serial interfaces for the communication with the scientific instruments, other onboard subsystems, and the ground operation center (see schematic representation in \fig{fig:siii_network}). The main elements of the \ac{ics} are a single-board computer, also named \acf{icu}, and two \acfp{dss} for onboard storage of science data and \ac{hk} data (see \fig{ICS-open}). 

\colfig{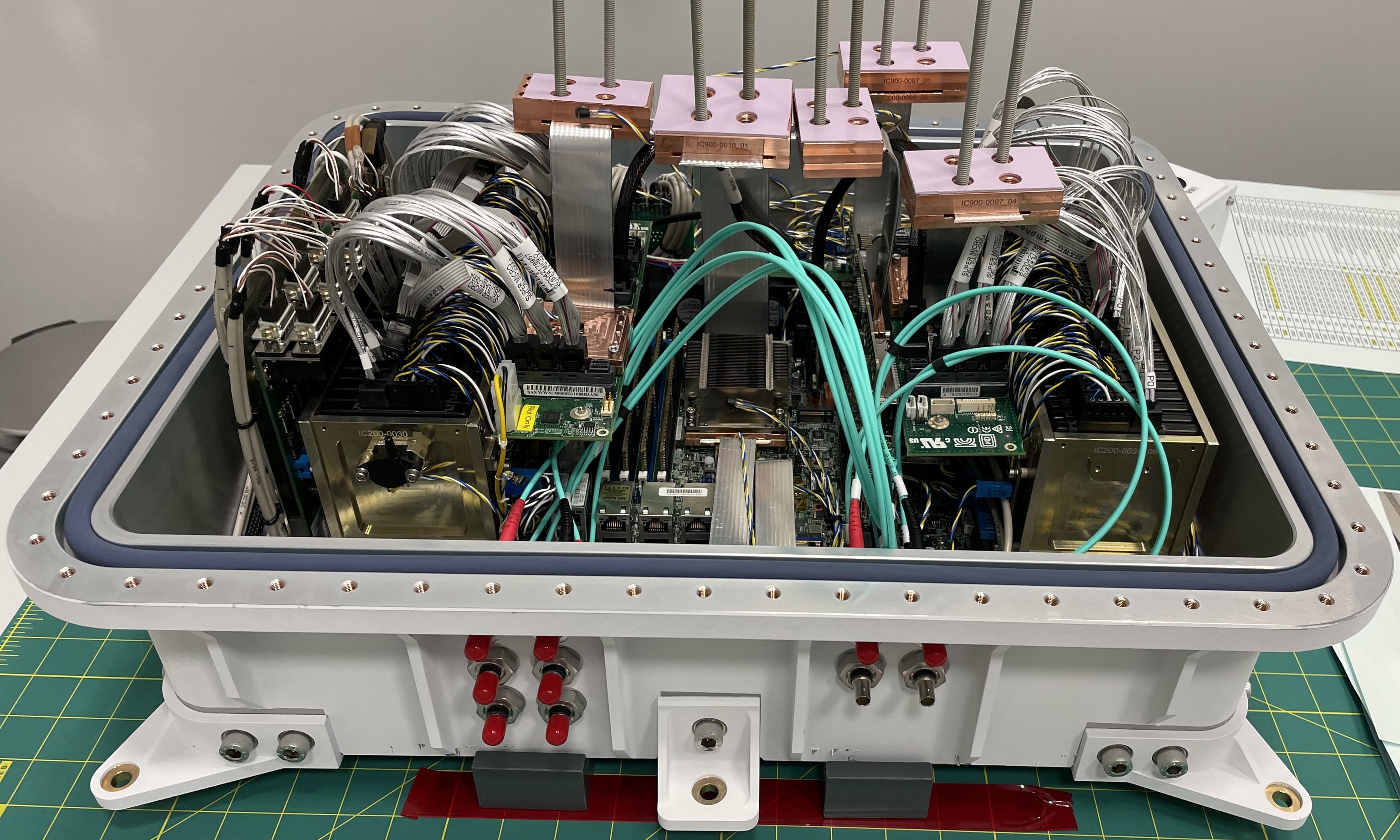}{\ac{ics} shortly before the final closing in October 2021. The mainboard of the \ac{icu}  is at the  bottom of the housing, left and right are the two \ac{dss} stacks containing the \ac{ssd} disks. Identified hot spots on the electronic boards are connected via heat pipes to copper blocks. During the closing procedure, the  guiding threading rods are replaced by screws that attach the copper blocks to the lid via gap pads.}

The \ac{ics} carries out the following tasks:
(i) Control the three science instruments \ac{scip}, \ac{tumag}, \ac{susi}, the image stabilization and autofocus system \ac{cws}, as well as technical subsystems,
(ii) store science and \ac{hk} data on the \ac{dss},
(iii) transfer \ac{hk} data and extracts of science data (thumbnails) to the ground station,
(iv) monitor critical \ac{hk} values and initiate pre-defined actions if lower or higher thresholds are hit,
(v) receive commands from the ground station and execute them, and
(vi) run the timeline, i.e., execute pre-defined observing programs in an coordinated way for all instruments.

The \ac{icu} is based on the mainboard Supermicro X10SDV-7TP8F, equipped with an Intel Xeon D-1587 processor with 16 cores, 128\,GByte of volatile memory in the form of four Samsung DRAM 32GB DDR4-2133 ECC modules, and two \acp{ssd} of 128\,GByte each in a RAID1 configuration for redundant storage of program code and configuration files. The \ac{ics} is built up mainly from commercial off-the-shelf components, whose operational reliability is guaranteed by housing the \ac{ics} in a pressurized vessel.

48 Samsung PM883 7.68\,TByte \acp{ssd} are available for storing \ac{hk} and science data of the instruments. The \acp{ssd} are distributed over two \acp{dss} that are identical in construction. Each \ac{dss} is equipped with an internal controller connected to the \ac{icu} via a serial interface. The controller provides \ac{hk} data (temperatures, pressure, humidity) and allows for switching the power lines of the \acp{ssd}. In order to avoid data losses in the case of an \ac{ssd} failure, the \acp{ssd} are organized in RAID5 chains of 4 \acp{ssd} each. The effective capacity of a chain is 20\,TByte, which leads to a total capacity of the two \acp{dss} of 240\,TByte. To minimize the power consumption, just one chain is powered at any moment in time. If the current chain is completely filled with data, the chain is switched off and the next chain is used.

An important design driver was the high total data rate generated by the three science instruments of about 700\,MByte/s, which is several orders of magnitude higher than the available telemetry bandwidth. Each science instrument is therefore equipped with an embedded PC acquiring and pre-processing the camera data as well as controlling its mechanisms. The camera data are transferred via a gigabit Ethernet to the \ac{icu} and stored on the \ac{dss}. A slightly different approach was chosen for the \ac{susi} instrument. Because the data rate generated by \ac{susi} is much higher than the network bandwidth of 1\,Gbit/s, the \ac{susi} software and the \ac{icu} software run on the same single-board computer of the \ac{ics}. It is thus possible to transfer the camera data directly from the \ac{susi} software to the \ac{icu} software via local-host communication without requiring a 10-gigabit Ethernet interface.

The \ac{ics} is also responsible for preparing the \ac{hk} data and quick-look thumbnails (see \sref{housekeeping}) for transfer to the ground station via the various communication channels (see \sref{comm}). Central node of the ground station is the system \ac{egse}, which displays system-relevant \ac{hk} data. A color coding helps the operator to easily identify critical out-of-range values during testing and flight. In addition, the system \ac{egse} distributes the instrument-specific \ac{hk} data to the individual instrument \acp{egse}.

\subsubsection{Electrical power consumption and infrastructure}\label{power}

The nominal power consumption of \sunriseiii{} during science operation amounts to $\approx\,$750\,W. The peak power consumption occurs during the ascent, where heaters are switched on to prevent the instruments and mechanisms to cool down below their non-operational limits, and can reach a maximum value of \lnum{1031}\,W. \sunriseiii{} can also be operated in an idle/safe mode, requiring 405\,W.

The \ac{pdu} is responsible for the power management. During the flight, it receives power from solar panels in combination with a battery buffer, and redistributes it to the various subsystems via ground controllable switched power lines.
Two solar panels are mounted in two frames on the $+x$ and $-x$ side of the gondola (see coordinate system in \fig{siii_design}). At float altitude they provide a peak power of 1824\,W and an average power of 1700\,W after stable sun pointing is achieved. They are connected to the battery bank via a Morningstar TS-MPPT-60 maximum peak power tracker solar charge controller. The battery bank uses a set of four Odyssey PC1700 pure lead batteries with two cells wired in series and two in parallel (2s2p configuration). Each battery has a capacity of 60\,Ah at room temperature, enough to operate the heaters and some selected electronic components (including the \ac{ics}) during the ascent phase when the gondola is free spinning and not constantly aimed at the Sun. 
The bus voltage ranges between 20.4\,V and 29.2\,V (nominal 27\,V). For ground operations and battery charge-up, the system stays connected to a ground power supply  until $\approx\,$30 minutes before launch.
The scientific instrumentation is powered by the \ac{pdu} via two separate 75\,A circuits to two \acp{ppd} ($+x$/$-x$) that provide individually switchable outputs for the instrument subsystems and mechanisms and a direct power output to the \ac{ics}. 

As an independent system, the \ac{sip}, provided by \ac{csbf}, is responsible for controlling the balloon (e.g., ballast dropping, apex Helium valve venting, termination) and to communicate to and from the ground. The \ac{sip} is powered by its own independent power system with its own set of solar panels, installed on the 4 sides of the gondola, to ensure continuous power also when the gondola is freely rotating and not constantly facing its front side at the Sun. One component of the \ac{sip} is the so called \acf{ss}, which provides an independent power switching and \ac{hk} interface to the \ac{pdu}.

The \ac{pdu} including the batteries and the solar panels was built by \ac{apl}. The \ac{sip} and \ac{ss} fall under the responsibility of \ac{csbf}, whereas the \acp{ppd} were designed and built by \ac{mps}.

\subsubsection{Network infrastructure and communication systems}\label{comm}

\begin{figure}
    \centering
    \includegraphics[width=\columnwidth]{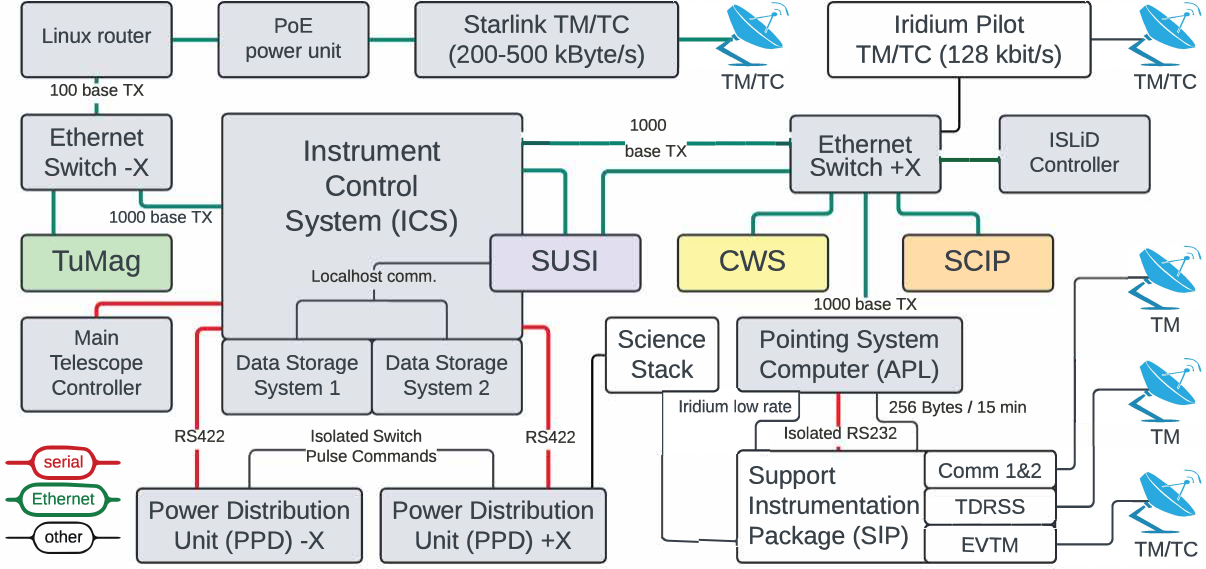}
    \caption{\sunriseiii{} network setup illustrating the data connections within the \sunriseiii{} observatory and to the \acsu{tmtc} links to the ground stations. \textsl{Colored boxes} indicate the scientific instruments, \textsl{gray boxes} the central units provided by \ac{mps} and \ac{apl}, and \textsl{white boxes} show the \ac{csbf} equipment.}
    \label{fig:siii_network}
\end{figure}

The internal communication  within the \sunriseiii{} observatory is based on an Ethernet network (see \fig{fig:siii_network}). This network ensures the communication between the \ac{ics} on one side and the \acf{pcs}, the in-flight communication systems and the scientific instruments on the other side. A bandwidth of 1\,Gbit/s (1000 base TX)  is needed for the data transfer from  \ac{scip}, \ac{tumag}, and the \ac{cws} to the \ac{ics}. The high frame rates of the \ac{susi} cameras combined with the necessity to store all frames for on-ground spectral restoration require a different setup, with a local-host communication between \ac{susi} and \ac{ics}. The \ac{susi} cameras are directly connected to the \ac{ics} with fiber-optics cables, and the \ac{susi} software for data acquisition and storage runs on the \ac{ics}. In addition, \ac{susi} is connected via a 1\,Gbit/s Ethernet line for commanding and communication with the \ac{susi} hardware.

For the inflight communication (ground to gondola, gondola to ground),  \sunriseiii{} uses \ac{los} and satellite \acf{tmtc} systems (i.e., systems for up- and downlink), provided by \ac{csbf} and \ac{apl}, and for the 2022 flight also by \ac{ssc} (see summary in \tab{tmcoverview}). They are used  for commanding, house-keeping and thumbnail image transmissions.

\begin{center}
\begin{table}[!ht]
\addtolength{\tabcolsep}{-4pt}    
\begin{tabular}{L{0.18\linewidth}C{0.10\linewidth}C{0.10\linewidth}C{0.10\linewidth}C{0.10\linewidth}C{0.10\linewidth}C{0.10\linewidth}C{0.10\linewidth}}
\hline
 & Comm& Iridum & \multicolumn{2}{c}{\acs{tdrss}}& \acs{elink} & \acs{starlink} & \acs{evtm} \\
 & 1\&2 & Pilot & omni & high-gain & & & \\
\hline
2022/2024 & \cmark/\cmark & \cmark/\cmark& \cmark/\cmark& \cmark/\xmark& \cmark/\xmark& \xmark/\cmark& \cmark/\cmark\\
Bandwidth (kbit/s) 
& 1.2 & 128 & 6 & 92 & \lnum{2000} & \lnum{8000} & \lnum{12500} \\
Direction & \acs{tmtc} & \acs{tmtc} & \acs{tm} & \acs{tm} &\acs{tmtc} & \acs{tmtc} & \acs{tm} \\
Availability & near 100\% & near 100\% & $\ge$85\% &  $\ge$85\%  & \ac{los} & 99.9\% & \ac{los} \\
\hline
\end{tabular}
\addtolength{\tabcolsep}{4pt}    
\caption{Communication channel overview used for the \sunriseiii{} flights in 2022 and 2024 with directional information (\ac{tmtc} = up- and downlink, \acs{tm} = only downlink), achieved bandwidths and the availability during float phase (in percent) or during the \ac{los} phase only. }
\label{tmcoverview}
\end{table}
\end{center}

The \ac{csbf} systems (white boxes in \fig{fig:siii_network}) include the \ac{sip} on the flight deck below the telescope, the antenna boom for \ac{gps} navigation and ground- and satellite communication antennas, and provide the following communication channels available during the full flight phase: 
(i)  the \ac{tmtc} satellite communication system \href{https://www.iridium.com/products/iridium-pilot/}{\aclu{iridium}}
with up to 128\,kbit/s, (ii) two low-rate science port  \ac{tmtc} channels primarily for commanding redundancy (Comm1 \& Comm2, 1200\,bit/s; \acs{tc} availability depends on satellite schedule),  and (iii) the \acl{tm} (\acsu{tm}, downlink-only) \acf{tdrss} omni-directional channel with 6\,kbit/s. For the 2022 flight, also the high-gain \ac{tdrss}  \ac{tm} channel (92\,kbit/s) was available.
For the early phase of the mission, when the balloon is still visible from the ground station, the downlink-only \acf{evtm} with up to 12.5\,Mbits/s allows for an efficient data transfer during the commissioning phase. The duration of this so-called \ac{los} phase depends strongly on the wind conditions on the launch day and was for the 2024 flight in the range of 12 hours. 

As a last-minute upgrade to the 2024 flight, \ac{apl} initiated the installation of  \acsu{starlink}
from the U.S. company SpaceX. It proved to be the most reliable and fastest \ac{tmtc} communication during the \ac{gusto} flight in Antarctica \cite[2023/2024 season,][]{gusto22}, offering higher bandwidth and coverage than the previously flown systems. \ac{starlink} on \sunriseiii{} is installed and  managed by \ac{apl} and \ac{mps}. The \acs{starlink} antenna is mounted on the \ac{csbf} antenna boom, and is powered via a proprietary \ac{starlink} cable connected to a \ac{poe} injector. 
A single-board Linux computer with two \acp{nic} acts as a router, providing the gateway from the gondola network to the wide-area network provided by Starlink.

During the 2022 flight, the \ac{elink} system was rented from \ac{ssc}. It promised a high-bandwidth, bi-directional communication during the \ac{los} phase, to be extended to up to 20 hours after launch by the use of a mobile transmitter station placed in Andenes (Norway). This system was not used for the 2024 reflight, since \ac{elink} is no longer maintained by \ac{ssc}, and the costs for the system were too high for the unreliable performance during the 2022 campaign. Also, the \ac{tdrss} high-data rate channel with up to 92\,kbit/s, used during the 2022 flight, was not available for the 2024 flight. The pointable high-gain antenna required by this system had to make space for the antenna of the \ac{starlink} system.

Connection to the observatory was maintained for 99.9\% of the flight time, achieving data rates via \ac{starlink} consistently exceeding 1\,MByte/s.
The stable and high-speed connection allowed operating \sunriseiii{} almost like a ground-based telescope, with continuous housekeeping data and near real-time thumbnail image transmission (see next section). Commands sent from the \ac{goc} were executed with almost no delay, with the response visible on the housekeeping monitors after a few seconds only. The other communication channels also worked well, and were kept alive as a backup option in case \ac{starlink} would drop out, which did not happen.

\subsubsection{Housekeeping and thumbnail concept}\label{housekeeping}

The \ac{icu} collects more than 1200 different \ac{hk} values from the gondola and \ac{pcs}, the telescope, \ac{pfi}, \ac{islid}, the \ac{cws}, and the three science instruments. The values were recorded every 2 seconds and sent to the ground station via the various communication channels (see \sref{comm}).
The \ac{hk} values were displayed on a dedicated screen in the \ac{goc} and continuously monitored. A color-coding indicated for every \ac{hk} value if it was within the expected range (green), close to the critical value (yellow), or outside the allowed range (red).

Of special interest were values about the power consumption (currents and voltages) of the instruments and the temperature of critical components. Under special surveillance were the temperatures of the \ac{hrw} and inside the \ac{ics}. The \ac{hrw} temperatures were lower than expected (40--45$^\circ$C), showing that the new concept of using a diamond-milled surface worked very well (see \sref{telescope}). Similarly, the \ac{ics} temperatures stayed within the allowed boundaries. Power monitoring turned out to be crucial for the success of the mission. 
Two  major power anomalies occurred when the charge controller stopped delivering power from the solar panels to the batteries, and the observatory had to operate solely on battery power (see \sref{sunriseiii2024}).
The early discovery of this failure allowed the operators to develop a rescue strategy before the batteries had fully drained (4--5 hours). Reducing the load on the charge controller by turning the observatory away from the Sun and by switching off instruments brought the charge controller back to the nominal state, and observations could be resumed with only a minimal disruption.

The fast up- and downlink channel via \ac{starlink} enabled the download of around 118\,GByte of data during the 6.5-day-long flight, including high-quality thumbnail images from the three science instruments nearly continuously. The accurate pointing to the targets defined during the science planning could be guaranteed by fine-adjusting the position using the near real-time thumbnail images, increasing the scientific value of the data significantly and enabling good overlap with the regions observed by the co-observing facilities. The quality of the thumbnail images could be tuned to almost native resolution, providing excellent information about the in-flight optical performance of telescope and instruments.

\subsection{Piggybacks}{}

\sunriseiii{} offers the opportunity of  a stratospheric flight  for two piggyback payloads: the \acf{iris2} and the \acf{ramon}. \ac{iris2}  delivered spectacular images and videos of the observatory during all flight phases, and \ac{ramon} is a space radiation monitor dedicated to measure high-energy particle showers. The piggyback payloads were designed to have minimal interference with the observatory and to work completely autonomously.

\colfig{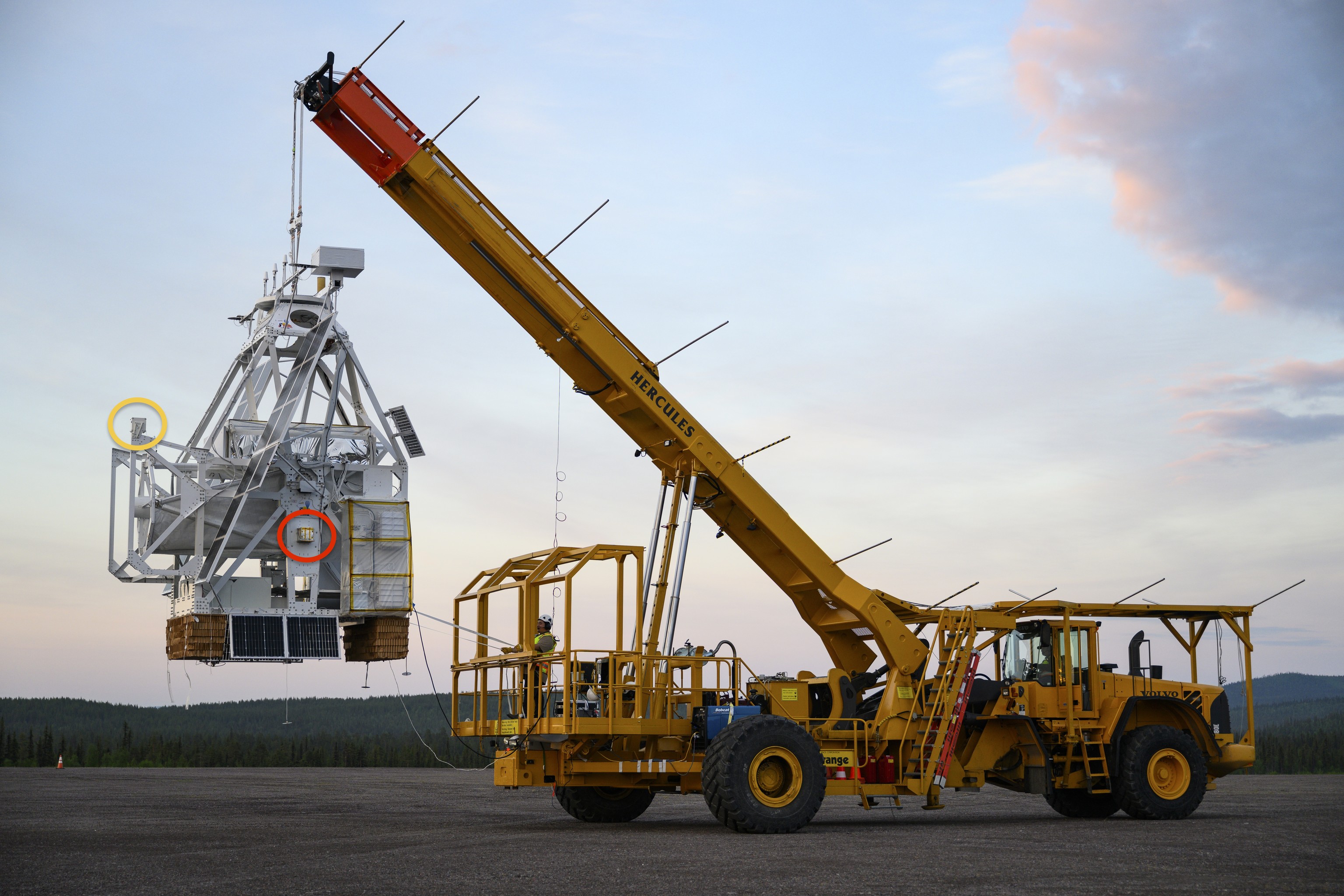}{\sunriseiii{} shortly before the launch on 10 July 2024, hanging on the launch vehicle Hercules. The locations of the piggy back payloads \ac{iris2} and \ac{ramon} are marked with the \textsl{yellow} and the \textsl{red circles}, respectively. Image courtesy: \ac{ssc} / Mattias Forsberg. }

\subsubsection{\ac{iris2}}\label{iris}

\ac{iris2} is a video recording and picture capturing instrument. It was designed and built by a Spanish team of enthusiastic amateur astronomers, engineers and technicians, with heritage from \acs{iris1}, which flew on \sunriseii{} in 2013. The instrument's primary goal is to provide imagery of the flight for communication and public outreach purposes, and additionally for monitoring and improving the performance of mechanical interfaces and the \ac{pcs} of the observatory from launch to landing and recovery.

\ac{iris2} is attached to the top of the gondola's left roll-cage (see yellow circle in \fig{piggylocation}), and boasts four cameras pointing in four directions: to the front (horizon - sun-side), the rear (telescope mirror), lateral (horizon and part of the gondola), and upwards (sky, balloon, and part of the gondola). The cameras (GoPro Hero 4 Black) record 4K-resolution 30-\ac{fps} video for 2 hours during launch, 1 hour during landing, 15 minutes at 120-\ac{fps} during balloon release, and then capture pictures with a 120\,s interval at other moments of the flight for time-lapse creation. The instrument can cover a flight duration of 6 to 12 days depending on the configured time-lapse interval.

\begin{figure}%
    \centering
    \subfloat[\raggedright Launch from \ac{esrange}.]{{\includegraphics[width=0.49\textwidth]{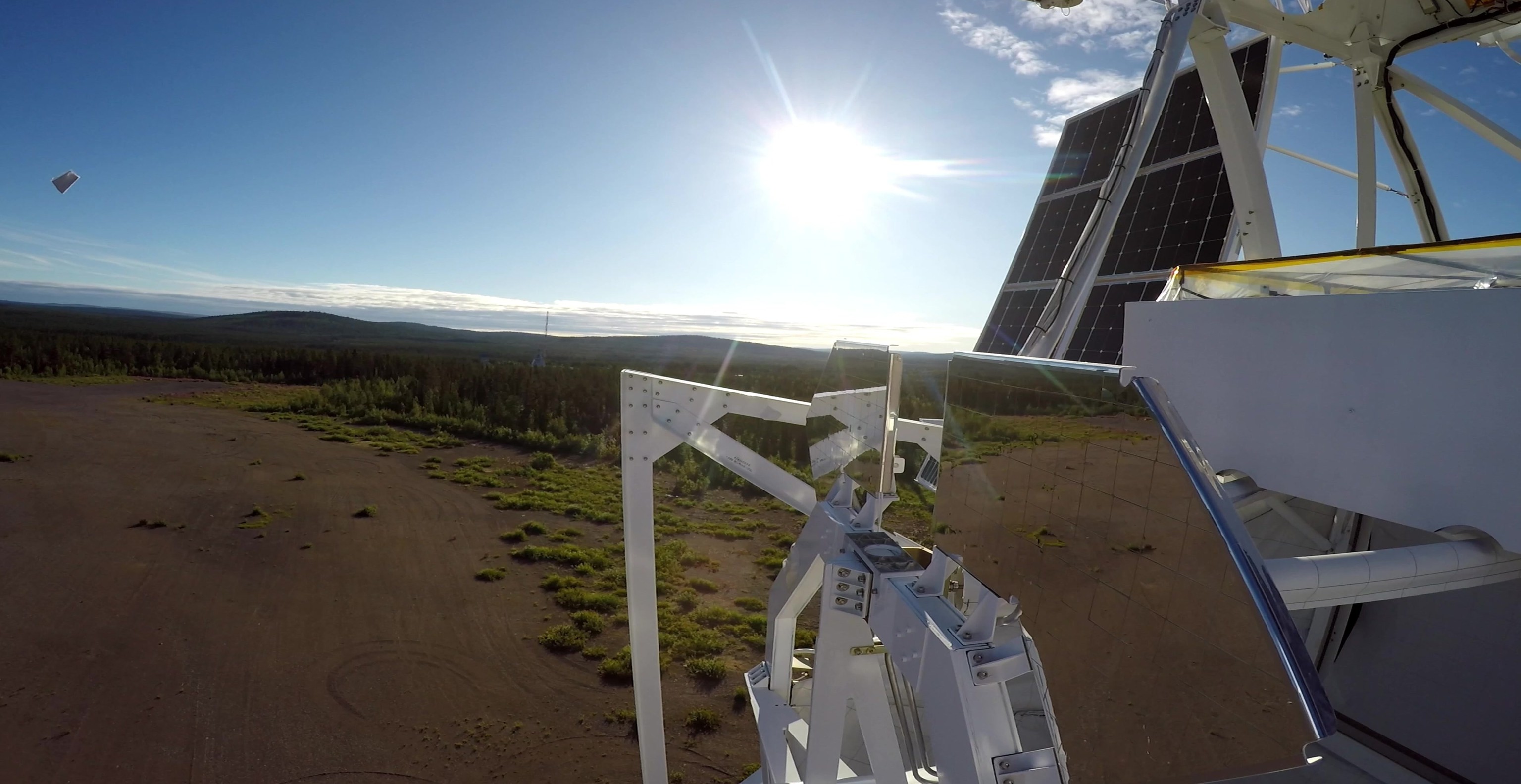} }}\hfill
    \subfloat[\raggedright Upward view with inflated balloon, antenna boom.]{{\includegraphics[width=0.49\textwidth]{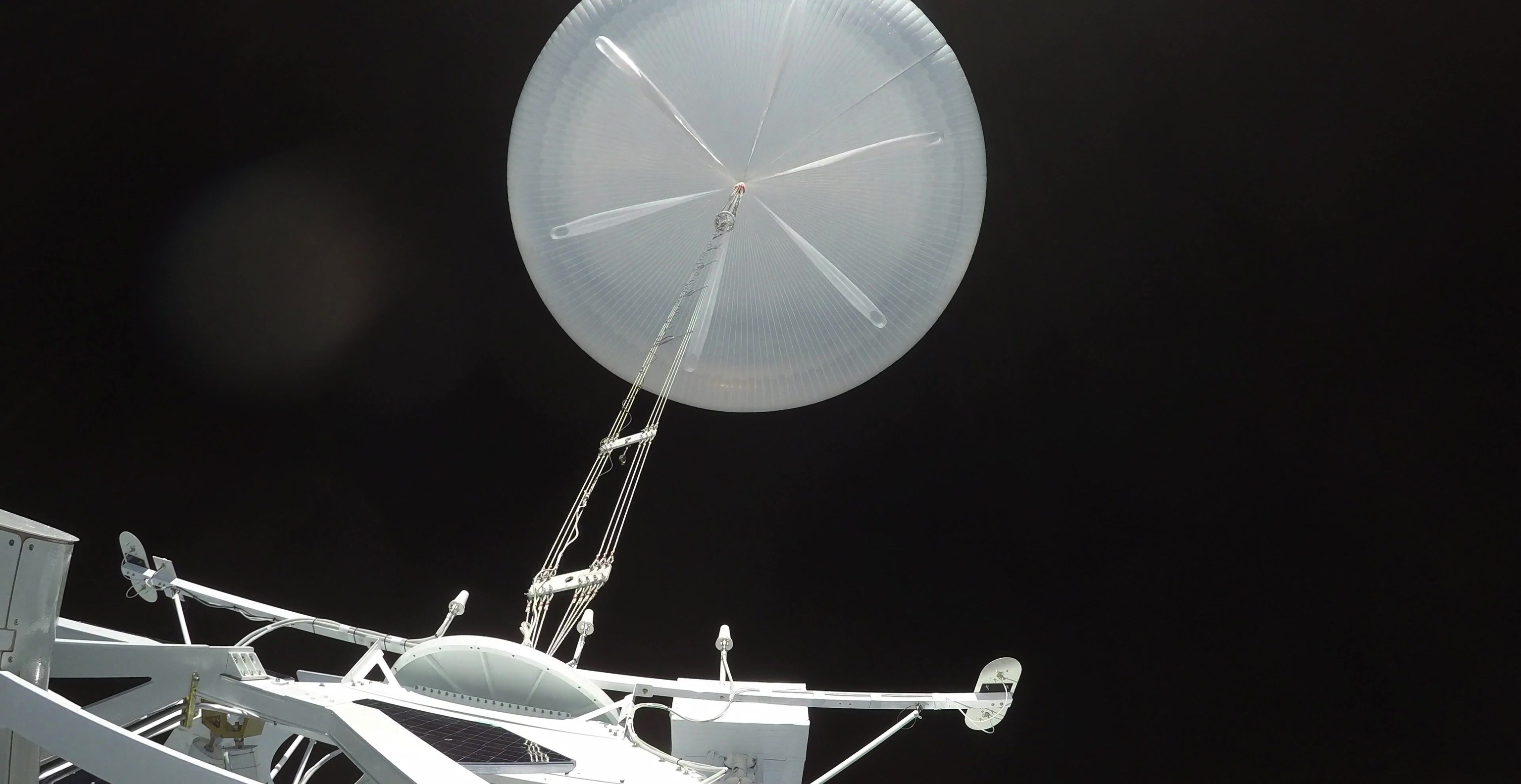} }}\\
    \subfloat[\raggedright Telescope front part during science operations.]{{\includegraphics[width=0.49\textwidth]{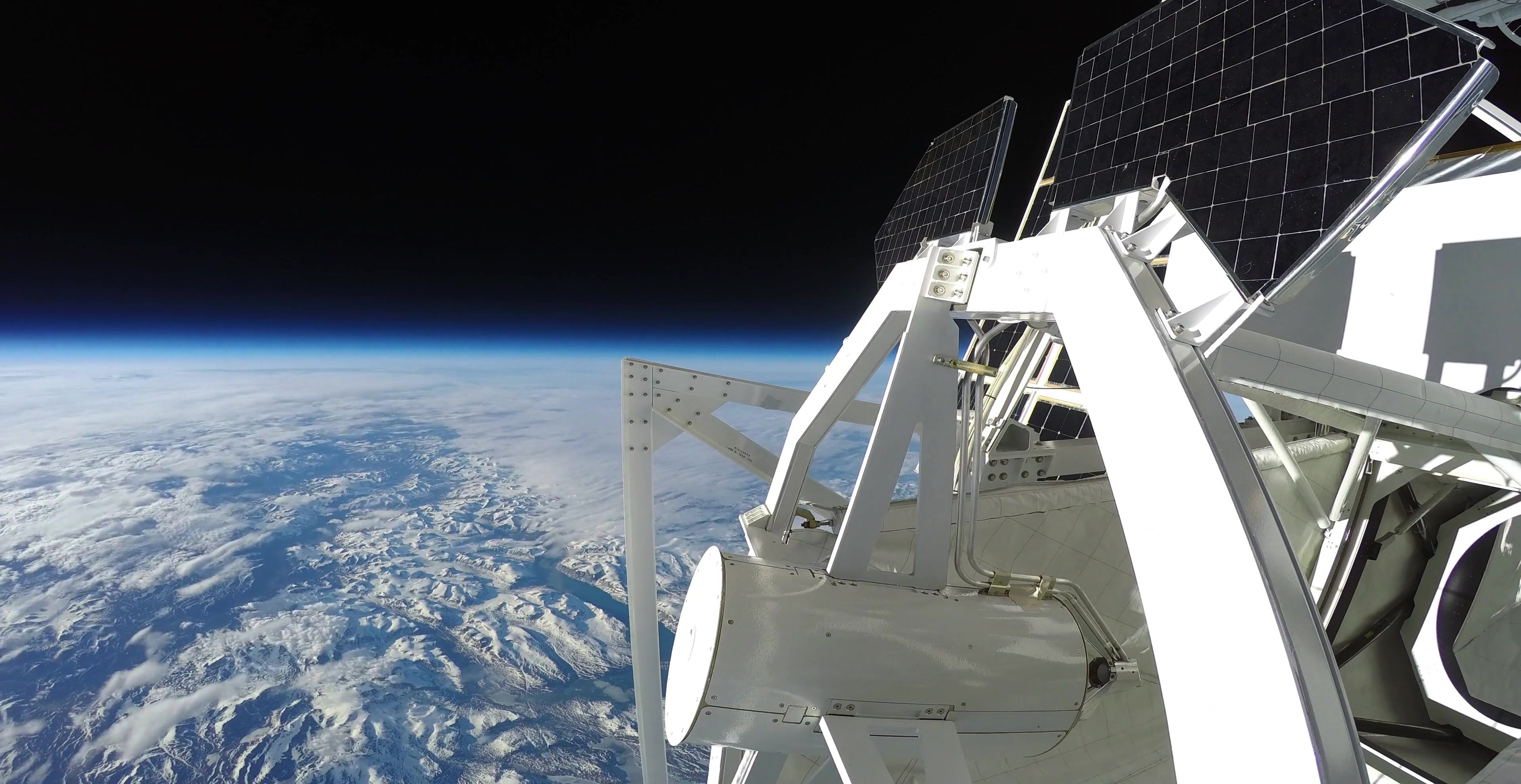} }}\hfill
    \subfloat[\raggedright Front camera view.]{{\includegraphics[width=0.49\textwidth]{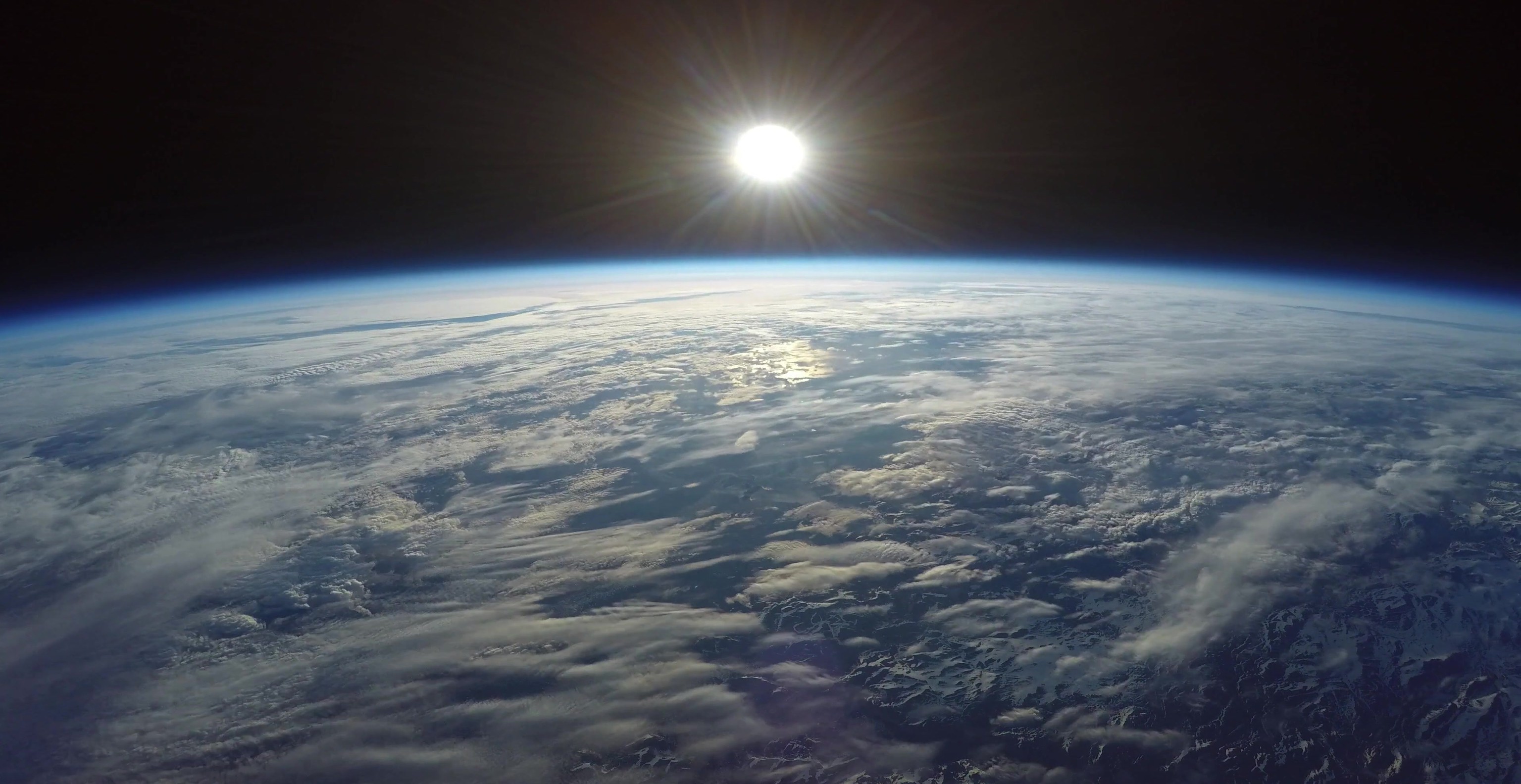} }}
    \caption{\ac{iris2} images recorded during the 2024 flight.}%
    \label{fig:IRIS}%
\end{figure}

The instrument's onboard computer is based on an MSP430 \ac{fram}-based microcontroller, with spaceflight heritage, coupled with external memories for telemetry, \ac{rtc}, \ac{imu}, and pressure (altimeter) sensors on a custom-designed \ac{pcb}. The flight software is coded in C on an \ac{fsm} structure with custom-made libraries for external bus communications. The power system, based on Nickel metal-hydride batteries with a capacity of 160\,Wh, does not rely on external power from the gondola. The instrument is only interfaced to the \ac{ics} through a digital pin to signal flight events from mission control, with the gondola being isolated by an optocoupler. The \ac{imu} and altimeters are used as redundant triggers for the different phases of the mission, as the instrument does not provide feedback to mission control. All the telemetry from the instrument must be dumped from its internal memories once it is recovered. The structure, built on extruded aluminum profiles and panels, is internally protected by low thermal conductivity foam panels, and has internal camera holders that allow for fine-tuning of the pointing. \ac{iris2} has a size of $200\times200\times150$\,mm$^3$, a mass of about 5\,kg, and consumes around 0.02\,W in standby and a peak of around 18.1\,W at launch recording. The instrument was tested with a \ac{tv} test and underwent several full-duration-flight runs simulating expected flight conditions.

The performance of \ac{iris2} during the first \sunriseiii{} flight in 2022 was nominal, taking over 8 hours of video and over \lnum{16000} still frames. The videos and pictures of the launch, flight, landing, and recovery have allowed the \sunrise{} team to better understand the events that happened during the different phases of the flight, confirming the usefulness of a relatively small and inexpensive instrument for such flights.

\ac{iris2} was flown again on the \sunriseiii{} flight in 2024 and produced a stunning view from the observatory during all phases of the mission, from launch until the landing and the recovery in Canada (see \fig{fig:IRIS}).  In total, \ac{iris2} recorded 8.5 hours of video and captured over \lnum{16000} still frames, \lnum{14576} of which were taken during the flight and the remainder after landing.  The only anomaly during the \ac{iris2} mission was with the rear-facing camera, whose power was automatically cut three minutes before landing as a self-protection measure while passing through a thick layer of clouds and heavy rain. The other three cameras successfully captured the landing.

\subsubsection{\acf{ramon}}\label{ramon}

When a high level energy particle coming from space impacts the higher layers of the Earth's atmosphere, atmospheric showers are produced. They consist of a cascade of secondary particles produced by the interaction of the incoming particle with the air molecules. As the showers go down through the atmosphere, more secondary particles are produced, but with less energy, becoming a very useful natural mechanism to avoid dangerous space radiation from hitting the Earth's surface and thus to preserve life as we know it.
Atmospheric showers have been frequently measured at aeronautic heights, as it is relatively easy to include radiation detectors in airplanes. However, there are less measurements at stratospheric heights, and this is becoming a very important layer for use cases like scientific balloons, communication platforms and high altitude pseudo satellites.

\ac{ramon} was developed by the Spanish \href{http://hidronav.com/page2.html}{SME Hidronav Technologies}
under the leadership of \ac{inta}, for a flight on the \sunriseiii{} mission (see red circle in \fig{piggylocation}). It delivered already valuable data during the short 2022 flight, and was again part of the payload for the 2024 mission. Unfortunately, the 2024 flight did not deliver any data, since the long waiting time for the launch window resulted in the complete drainage of the batteries, which was discovered only a few hours before the launch with no possibility anymore to react.

\ac{ramon} is based on  the \href{http://www.cosmicwatch.lns.mit.edu/}{Cosmic Watch} radiation monitor
developed by \ac{mit}, with modifications by Hidronav Technologies to improve its performance. 
Each monitor consists of two 5$\times$5$\times$1\,cm$^3$ scintillators, each with its associated \ac{sipm}, its Arduino microcontroller, and the necessary electronics to treat the electric pulse generated when an ionizing particle hits the scintillator. The scintillator converts the energy from the ionizing particle into visible light and the \ac{sipm} converts it into an electric pulse which can be conveniently managed.
Every time this process occurs, an event is recorded on the corresponding storage card (one for every scintillator). Particles exciting both scintillators trigger a so-called coincidence event, useful for detecting primary particles which are more energetic than the secondary ones.

\subsection{Thermal Concept}\label{thermal}

The \sunriseiii{} thermal design concept is similar to that of previous missions, with system-level solutions akin to those employed in the second flight \cite[]{perezgrande:11}. However, advancements have been made in thermal analysis and modelling. A distinctive thermal environment characterization \cite[]{gonzelezllana:18}, based on real-data observations, considers local and temporal variability in key parameters (albedo, \ac{olr}, and solar radiation). The analysis assumes a quasi-steady state for a stratospheric balloon at 40\,km altitude, allowing to focus on steady-state cases during which the combination of environmental parameters is maximized and minimized to determine maximum and minimum temperatures \cite[]{gonzalezbarcena:19}.

A thorough analysis of the ascent phase, critical for thermal considerations in platforms like \sunriseiii{}, has been conducted \cite[]{gonzalezbarcena:22}. Unlike satellites, the ascent phase means a challenge due to low temperatures in the tropopause and jet streams at around 12\,km altitude. The system's minimum temperature may occur during this phase, thus, a quantification of convection heat transfer is needed \cite[]{fernandezsoler:23}. A characterization of the thermal environment throughout the atmosphere, specific to the launch location and epoch, was also conducted for polar long duration balloon missions launched from \ac{esrange} in summer \cite[]{gonzalezbarcena:20}.

\sunriseiii{}'s thermal design can be summarized as follows:
The electronic units are mounted on two white-painted \acp{erack} on both sides of the gondola, offering a direct view to the cold sky and Earth. The rack does not provide a controlled interface to the attached units, therefore their thermal control is their own responsibility.
To avoid the freezing of the instruments during the ascent phase, the \acp{erack} are shielded from external winds with transparent windshields made of a 1.5\,mil thick \ac{lldpe} film, allowing solar and infrared radiation to pass through. During the initial phase of ascent, the gondola rotates, exposing the \acp{erack} to direct sunlight, which contributes to heating the units.

The telescope thermal design is based on the concept of the previous flights with some improvements. The surface of the \ac{hrw} is now diamond milled on the aluminum bulk (see \sref{telescope}), reflecting around 85\% of the incoming solar radiation. To mitigate the blurring effect caused by too high temperatures, two ammonia-based heat pipes are used to transfer the absorbed heat flux by the \ac{hrw} to two \ac{ssm} radiators situated in the upper section of the telescope's front ring. A \acl{mli} (\acs{mli}) made of Betacloth 500 GW and \ac{vda}-coated Mylar is used to prevent additional heat loads on M2 and the \ac{hrw} coming from the Earth or the lower deck equipment. It is attached to the front ring struts of the telescope, differing from its attachment to the roll cage in previous missions.

The science instruments are placed in the \ac{pfi} module. Its sun-facing front part is protected with a sun shield. High solar reflectivity ($\varrho=0.8$) for the whole structure is achieved by a styrofoam cover with aluminized Mylar on the outer face. The thermal control of the instrumentation is based on the use of radiators coated with white paint on the top to dissipate the internal power. 
Every instrument is insulated from the structure with a single layer of \ac{vda}-coated Mylar foil, providing them with more thermal stability. Finally, similarly to the \acp{erack}, an \ac{lldpe} film, has been used to cover the top part of the \ac{pfi} to act as a wind shield to avoid freezing of the instruments during the ascent phase.  

\begin{figure}
   \centering
   \includegraphics[width=1.0\linewidth]{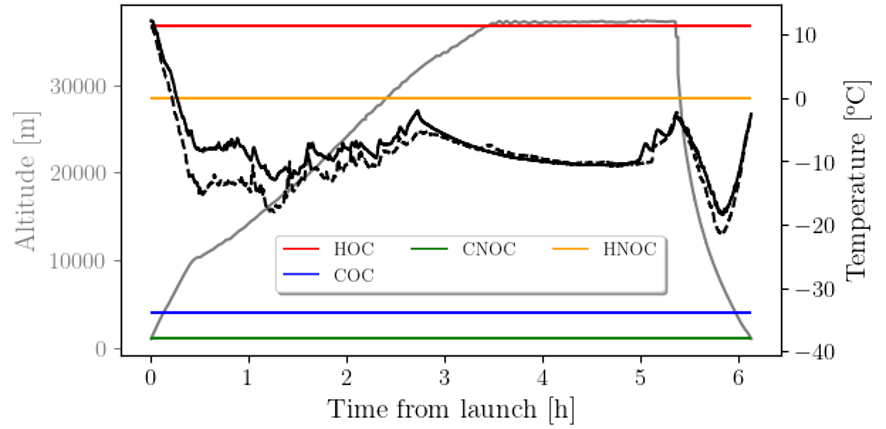}
   \caption{\ac{erack} $-x$ temperature measurements during the 2022 flight of \sunriseiii{} (\textsl{solid} and \textsl{dashed black lines}) compared to operational (\textsl{red} and \textsl{blue}) and non-operational (\textsl{green} and \textsl{yellow}) case analyses.}
   \label{fig:thermal}
\end{figure}

During the flight in July 2022, the temperatures reached at the floating altitude were in-line with steady-state predictions. Due to the malfunction during the launch, the thermal analyses were re-executed considering that the instruments were in a non-operational mode with the telescope elevation angle at 60\dg{}. The sensor's measurements are shown in \fig{fig:thermal}. The thermal analysis includes results from extreme hot and cold operational cases 
and hot and cold non-operational cases 
in which instruments were off and the telescope curtain was closed. During the initial phase of the ascent, the gondola rotation exposed the \acp{erack} to direct solar radiation, preventing excessive cooling. At an altitude of around 26\,km, the gondola started pointing to the Sun, and the solar panels blocked direct sunlight, leading to a noticeable temperature reduction in the \acp{erack}, and visible in \fig{fig:thermal} in the smooth, exponential temperature decrease between 2.8 and 5 hours after launch. The temperature gap between the hot and cold non-operational cases was found to be influenced by the Sun's relative position over the horizon and the changing environmental conditions (albedo and outgoing longwave radiation) during the flight. 
The thermal analysis for the 2024 flight will only be available after the data reduction phase.

\section{Observatory Testing and Flight Preparation}\label{testing}

All components and instruments of the observatory underwent thorough testing on component/subsystem level and also in the fully assembled state. The tests on instrument level are described in the individual instrument papers, the description of the tests in this section concentrates on the ones performed at system level and after the integration of the instruments into the \ac{pfi}.

\subsection{Test of Mechanisms}\label{test:mechanisms}

A crucial part in a remotely operated observatory is the reliable operation of all its mechanisms, requiring extensive testing under realistic conditions. The mechanisms can be divided into four groups: 

\begin{enumerate}
	\item \label{mech:gondola} Gondola mechanisms: \acf{mtu}, elevation drive, roll wheel, brake (launch-lock), and X-Y Stage mechanism. These mechanisms guarantee the correct pointing of the observatory and secure the telescope during launch, ascent, descent and landing. The mechanisms and their tests are described in the gondola paper \cite[]{gondola25}. 
	\item \label{mech:telescope} Telescope mechanisms: The M2 mechanism allows for an adjustment of M2 in three degrees of freedom (translational motions in $x$, $y$, $z$, in the telescope coordinate system, see \fig{siii_design}). M3 and M4 are equipped with mechanisms for axial movement, and the curtain mechanism opens and closes the protective curtain for M1 in the central frame of the telescope.
	\item \label{mech:islid} \Ac{islid} mechanism: The filter wheel in F2 (see \sref{islid}) allows to insert 6 different targets and a closed (dark) position into the light beam. They are used for the calibration of the instruments, and to determine their in-flight performance and alignment.
	\item \label{mech:instrument} Instrument mechanisms: Every instrument has its own mechanisms, e.g. for rotating filter wheels, moving the grating and the slit, or rotating the waveplate in the \ac{pmu}. These mechanisms and their testing are described in the individual instrument papers \cite[]{cws25,susi25,tumag25,scip25}.
\end{enumerate}

The telescope mechanisms to control the mirrors M2, M3 and M4, and for opening and closing the curtain were already flown in \sunriseiandii{}. Thorough thermal vacuum testing was performed prior to those flights. For the re-use of the mechanisms in \sunriseiii{}  a thorough maintenance and cleaning was performed, and their functional and electrical reliability was verified by extensive ground testing. Functional tests were repeated after the 2022 flight.

The \ac{islid} filter wheel is a new development for \sunriseiii{}. It allows placing the following targets into the light beam for calibration measurements: open beam (nominal observations), closed (common dark), 2 random dot targets (co-focus check), \ac{usaf} target (resolution and image orientation), polka dot target (near-field straylight), rectangular grid (distortion), and a knife edge (far-field straylight). These targets 
allowed to assess the instruments' in-flight performance. Both, the filter wheel mechanism and its controller, were tested under flight conditions in the \ac{mps} thermal vacuum chambers. The detailed test report is available in the \docdb{SR3-MPS-RP-LD700-004}.

\subsection{Optical Tests}\label{test:optical}

All optical instruments and \ac{islid} were individually optically tested and characterized prior to integration into the \ac{pfi}. 
\ac{islid} was tested by interferometry between F2 and the exit foci, respectively, in double pass autocollimation against a high precision reference sphere. 
The interface positions were verified by a laser tracker. In addition, images of the F2 targets were taken in the exit foci by a camera.  
After the integration of the scientific instruments and the \ac{cws} into the \ac{pfi}, co-alignment and optical performance between F2 and the science foci were tested using a stimulation system, which mimicks the telescope beam. High power \acp{led} were used as light sources, optimized for the different spectral channels.

For the optical tests of the instruments the filter wheel unit of \ac{islid} in F2 was used. An accurate determination of the performances of all instruments could be achieved, and the proper co-alignment between the different science foci could be verified \cite[see instrument papers:][]{cws25,susi25,scip25,tumag25}.

\subsection{Polarimetric Calibrations}\label{polcal}

All three science instruments were designed to measure the polarization state of the incoming solar radiation to a level of \ten{-3} of the continuum intensity or better. To achieve this precision, polarimetric calibration campaigns were executed at various stages of the project. The most important ones were the calibration measurements in the fully assembled state, with the instruments in the \ac{pfi} mounted on top of the telescope, and the calibration light sources feeding the instruments from the prime focus (F1, at the position of the \ac{hrw}) and the secondary focus (F2, inside the \ac{pfi}). Calibration measurements were performed before and after the hang tests, with the observatory suspended from a crane (see \sref{hangtests}), performed in the \ac{mps} balloon hall (2021 and 2022), and in the integration hall `The Dome' in Kiruna (2022 and 2024).

The polarimetric calibration measurements for \ac{susi} were all performed on the ground. A total of 35 independent measurements were carried out at different wavelengths by placing 
the \ac{psg}, a motorized device generating 40 well-defined polarization states from \ac{led}  light sources for different wavelength ranges,
at three locations within the beam path. These locations were the \sunrise{} foci F1 and F2, and the entrance port of \ac{susi}. 
A detailed description of the entrance-port calibrations is given in \cite{iglesias2025}, that uses eight \ac{susi} standalone measurements to prove the instrument polarimetric response is stable enough to achieve the required sensitivity of \ten{-3} of the continuum intensity. As part of the ongoing data reduction process, the \ac{susi} team is analyzing the polarimetric calibrations performed at \sunrise{} foci F1 and F2 to derive the final end-to-end calibration strategy. We note that the measurements performed at the primary telescope focus F1 account for the polarimetric effects of the full beampath, except for the main mirror M1, thus no independent telescope calibration is foreseen. Instead, a correction of residual polarimetric crosstalks is performed using solar signals of known polarization properties acquired during the flight. These include zero-polarization targets such as spectral continuum, average quiet Sun and the magnetically insensitive \fei{} line at 406.538 nm \cite[see][for extra details]{iglesias2025}.

The \ac{scip} calibration measurements were first performed at the instrument level at \ac{naoj} before the integration into the \ac{pfi}, in which a wire-grid linear polarizer and achromatic waveplate in the \ac{scip} wavelength range were placed in front of the \ac{scip} beam entrance to produce the known polarization states from a halogen lamp. Polarization response matrices and their spatial and wavelength dependence were obtained, including the effects of synchronization with the \ac{pmu} (rotating waveplate), cameras with rolling shutter, and onboard demodulation \cite[see][for details]{kawabata22}. After \ac{scip} was mounted into the \ac{pfi} and the telescope, the polarization response matrices were updated by the polarization calibration measurements from F1, accounting for the polarization effect of the telescope except for M1. The F1 calibration configuration was the same as for \ac{susi} except for the achromatic wave plate and the \ac{led}, which were specific for the \ac{scip} wavelength range.
For both, \ac{scip} and \ac{susi}, no in-flight calibration measurements are possible.

\ac{tumag} carried out the polarimetric calibrations for the three wavelengths  and both cameras of the instrument with a modulation scheme by setting the retardances of the two \acp{lcvr} to [225\dg{}, 225\dg{}, 315\dg{}, 315\dg{}] (\ac{lcvr}1) and [234.74\dg{}, 125.26\dg{}, 54.74\dg{}, 305.26\dg{}] (\ac{lcvr}2), chosen to maximize the polarimetric efficiencies. The modulation was performed by applying voltages to the two \acp{lcvr}, corresponding to the calibration carried out with a variable angle spectroscopic ellipsometer from J. A. Woollam Co. The polarimetric efficiencies were further optimized by a fine-tuning procedure \cite[]{Alvarez-Herrero:18} to achieve polarimetric efficiencies over the whole \ac{fov} of $\epsilon\ge$~[0.95, 0.45, 0.45, 0.45], exceeding the requirements.

\ac{tumag} allows for an in-flight, field-dependent polarimetric calibration. The \ac{tumag} filter wheel contains a linear polarizer and an array of 16$\times$16 patches consisting of 3$\times$3 micro-polarizers each, with polarization axes at 0\dg{}, 45\dg{}, 90\dg{}, and 135\dg{}. During in-flight calibration campaigns, the insertion of these targets allows to: (i) monitor the polarization modulation behavior of the instrument to identify deviations in polarimetric performance, and (ii) fine-tune the polarimetric modulation matrices over the entire \ac{fov} of \ac{tumag}.

\subsection{Vacuum Tests}\label{test:vacuum}

\subsubsection{\Ac{ics} thermal vacuum testing}\label{vacuum:ics}

The high requirements on computing power and disk storage capacity made it mandatory to assemble the \ac{ics}, \sunriseiii{}'s main computer (see \sref{ics}), from standard electronic components contained in a pressurized housing. The air-tightness of the housing and the reliable cooling of the components inside the closed compartment had to be tested thoroughly under simulated flight conditions, including the thermally demanding ascent phase. The \ac{ics} was placed in a vacuum chamber at \ac{mps}, and over a time period of 5 days thermal cycles with the expected temperature ranges were executed, while the \ac{ics} was operated under full load and in idle state. It could be confirmed that the pressurized box was air-tight and that all temperatures measured with the temperature sensors inside the \ac{ics} as well as the mainboard temperature values (e.g. the CPU temperatures) stayed within the nominal values. The test procedure ('as-run') is described in the \docdb{SR3-MPS-PR-IC000-001}.

\subsubsection{\ac{pfi} vacuum test}\label{vacuum:pfi}

The \ac{pfi} was tested in the big vacuum chamber of \ac{mps} (`Big Mac') in February 2022. 
The test setup, described in the \docdb{SR3-MPS-PR-AV200-001}, was as follows: The \ac{pfi} with all scientific instruments installed in flight configuration was placed inside the vacuum chamber. The electronic units and \ac{egse} systems to control the instruments and the mechanisms in \ac{islid} were placed outside the chamber, and connected to the \ac{pfi} via vacuum feedthroughs. 
Illumination was done with both, artificial light sources and real sunlight, using the coelostat system on the roof of the \ac{mps} laboratory building, through a fused silica viewport located on the grey-room side of the vacuum chamber.
During the 1-week long test the subsystems \ac{islid}, \ac{cws}, \ac{susi}, \ac{scip}, and \ac{tumag} were first operated individually, and on the last day together. The optical performance in vacuum  between the F2 focus (optical entrance of the \ac{pfi}) and the various science foci could be verified. The sunlight allowed to confirm the spectral performance of all instruments, and closed-loop locking of the \ac{cws} on the F2 target could be achieved.  
Since all sub-units had seen thermal vacuum testing before, the full \ac{pfi} test was done at ambient temperature only. 
The \ac{pfi} vacuum test confirmed that all subsystems performed within the specifications under the conditions expected during the flight.

\subsection{Hang Tests}\label{hangtests}

The so-called hang tests offer the possibility to test the observatory in full flight configuration. A crane takes over the role of the ladder/balloon and lifts the gondola a few tens of centimeters above the ground.  This simulates well the flight configuration and allows to fully test all three degrees of freedom for the gondola and telescope pointing (azimuth, elevation, roll compensation). The tests were performed with solar illumination allowing full characterization of the \ac{pcs}'s Sun acquiring and tracking capabilities while simultaneously feeding all science instruments with real sunlight. 

Hang tests were performed in October 2021 in the balloon hall at \ac{mps}, and during the preparations for the 2022 and the 2024 flights of \sunriseiii{} in and outside of the integration hall `The Dome' at \ac{esrange}. 
The stable sun-pointing allowed for the opening of the curtain, and the illumination of the telescope and the \ac{pfi} in a similar way as during the flight. Several hour-long stable sun-pointing could be achieved at both locations, proving the excellent performance of the \ac{pcs}. Despite the poor seeing conditions for these locations, an \ac{rms} pointing accuracy in the range of a few arcseconds could be achieved (see \sref{perf:pointing}).

All three scientific instruments were operated in the planned orchestrated mode during these tests, and important inter-instrument alignment properties could be verified and confirmed for varying telescope elevation. The sunlight enabled the instruments to record spectra of the solar absorption lines, and to assess their spectral resolution. The achieved pointing stability and the performance parameters of the instruments are summarized in the gondola paper \cite[]{gondola25} and the individual instrument papers \cite[]{cws25,susi25,tumag25,scip25}. 

\subsection{Communication Tests}\label{test:comm}

Communication tests between the observatory and the ground station had been performed during the launch preparations in 2022 and 2024 at \ac{esrange}. For these tests, \sunriseiii{} was put into flight configuration with all communication hardware attached (including the \ac{los} antennas for \ac{elink} (2022 only) and \ac{evtm} and the antennas for the satellite communication systems \acsu{iridium} and \ac{starlink}), and driven to a location on the balloon pad where satellite communication and ground radio station communication could be established.

The tests revealed some difficulties for the \ac{los} communication system, especially for the \ac{elink} service provided by \ac{ssc}. The performance during the short 2022 flight was acceptable, albeit not optimal. Since the \ac{elink} system is no longer maintained by \ac{ssc}, and the high cost to benefit ratio, \ac{elink} was not be used for the reflight in 2024.
Similar tests were repeated prior to the \ac{frr} for the 2024 flight, including tests of the \ac{starlink} system used for the first time on \sunriseiii{}.

\subsection{Timeline Tests}\label{test:timeline}

The operation of \sunriseiii{} during the flight is based on the timeline concept described in \sref{timeline}: all instruments are directly controlled by the \ac{ics} with minimal required interference from the ground station. A thorough testing of the timelines and the observing blocks was therefore a mandatory step to ensure reliable and hassle-free operation.

The individual timelines from the instruments and the \acf{pcs} were compiled manually. A software simulator was used to verify the consistency of the timelines and to guarantee the accurate timing of the command sequences. Additionally, all timeline scripts where peer-reviewed to ensure consistency. After this consistency check, all the observing blocks were uploaded to the \ac{ics} and then executed on the flight hardware on ground. Some very long timeline sequences were  replaced by shorter test versions for these ground tests. This procedure allowed to verify that the fully autonomous operation of the instruments by the \ac{ics} worked as expected. 

The tests also included the simulation of unforeseen failures, like for example, a temporary error of an instrument. The individual instrument teams trained their capabilities of operating the instrument manually to recover it from the failure mode, and to re-join the timeline, which continued to run for the other instruments. Also, pointing control losses were simulated and efficient mitigation procedures, like safe entry points into the timeline, were implemented.

\subsection{Path to Flight Readiness} \label{test:flight}

\ac{csbf} requires to follow a strict procedure to achieve flight readiness\footnote{See website of \ac{nasa}'s \ac{bpo}: \href{https://sites.wff.nasa.gov/code820/}{https://sites.wff.nasa.gov/code820/}} for zero-pressure \ac{ldb} flights like \sunriseiii{}. The observatory needed to pass several reviews during the construction and testing phase, with the \ac{mrr} being the last one before the payload is being shipped to Kiruna. In the \ac{mrr}, the \sunrise{} team had to present, amongst others, a science status review, a deployment and integration schedule, requirements for integration and operation, and to prove that the current status of the payload is ready for shipment to \ac{esrange}.

After arrival at \ac{esrange} early April 2024, the first milestone was achieved when the telescope and the \ac{pfi} were mated, followed by functional tests and calibration measurements, for example, the final polarimetric calibration of the science instruments. On 01 May 2024,  `first light' was achieved: telescope and \ac{pfi} were mounted in the gondola, and a hang test allowed for the first time to point the telescope towards the Sun at \ac{esrange}. In the following three weeks extensive tests of all the subsystems were performed by the \sunrise{} team (timeline testing, operation training), and the \ac{csbf} hardware (e.g., \ac{sip}, antennas) for in-flight communication between observatory and ground station was installed. One week before the launch window opened on 25 May 2024, \sunriseiii{} completed the \ac{ssct}. This is a final hang test to check on mission compliance with the stated requirements, and to confirm that integration is complete and \sunriseiii{} is ready for launch. It was followed by the \acf{frr} where permission to proceed with the launch preparations was granted. Flight readiness for the 2024 flight was achieved on 21 May 2024.

\subsection{Flight Operations}\label{goc}

The 2024 flight of \sunriseiii{} was controlled by three flight operations centers: The flight control (i.e., launch, \ac{los} communication, ballast control balloon venting, flight termination) was handled by \ac{csbf} from the \aclu{eoc} (\acsu{eoc}, Kiruna, Sweden) and the \ac{occ} in Palestine (Texas, USA), whereas the science operations were executed from the \acf{goc}. The \ac{goc} turned out to be a very efficient way to control \sunriseiii{} during the 2024 flight: it  enabled optimum personnel support under the tight logistics situation at \ac{esrange}, caused by the dense 2024 balloon launch schedule, and it allowed to use the top-notch infrastructure of \ac{mps}. The communication between all three operations centers and the observatory in flight worked flawlessly. The operations center setup and the data flow concept is illustrated in \fig{SunriseIII_GroundSegment_DataFlow}.

\colfig{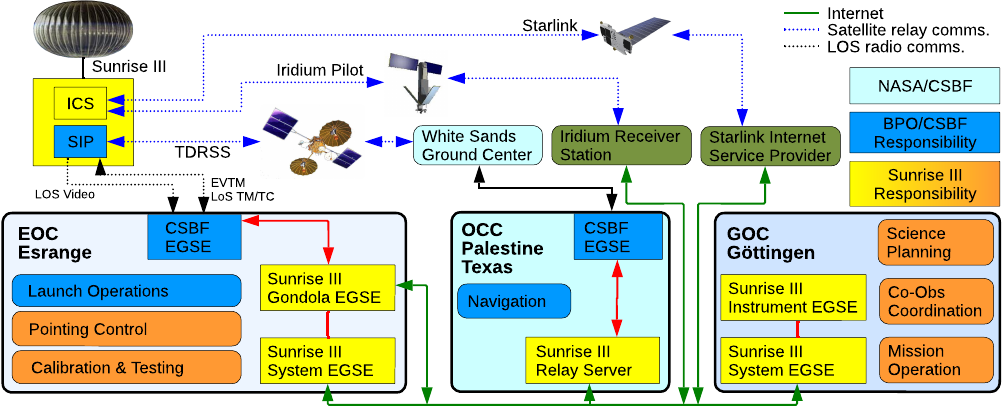}{\sunriseiii{} ground segment, data flow and communication concept for the 2024 flight.}

The system \ac{egse} and all the instrument \acp{egse} were located at the \ac{goc} in a dedicated room at \ac{mps}. The control of the observatory was handed over from the \ac{eoc} to the \ac{goc} already after first light on 01 May 2024, i.e., when the fully assembled observatory achieved sun-pointing during a hang test at \ac{esrange}. All subsequent tests, and the flight, were controlled from the \ac{goc}. 
It was connected via a \ac{vpn} to the \ac{eoc} where a second set of \ac{egse} computers allowed for a mirrored operation of the system \ac{egse} and all science instruments. The \ac{eoc} provided the data analysis and storage computer for the scientific instruments used during their calibration and testing prior to the flight, when the produced data volume was too large for a direct transfer to the \ac{goc}.
During the flight, the available data rates between balloon and ground station (see \sref{comm}) allowed for very efficient remote operations, including the exact positioning of the telescope to the science targets using real-time thumbnails. In addition, the data from the satellite communication systems (\ac{starlink}, \ac{iridium}, \ac{tdrss}) were directly available from both operations centers. 

A small flight-operations team at \ac{esrange} ensured the efficient communication between the operations centers in both, the technical aspects and the human interaction between science team and \ac{csbf}.

\subsection{Fault Protection Measures}\label{safety}

\sunriseiii{} was equipped with several fault protection measures, ensuring that the observatory is brought into a safe state in the case of a failure. Three important mechanical and thermal measures are briefly described here.

Aperture door: In the event of a loss of pointing, the solar image in F1 would not hit the \ac{hrw} anymore, but could deposit the full radiative power collected by the main mirror M1 ($\approx$1\,kW) onto an element not designed to withstand this enormous power. Therefore, an emergency system was implemented, which closes the aperture door automatically when the \ac{pcs} reports a loss of pointing. A curtain in the central frame in front of M1 is closed within a few seconds, fast enough to ensure that pointing loss does not cause any damage.

Survival heaters: When the temperature drops below 0$^\circ$C, survival heaters in various critical components are automatically turned on. These components include, e.g., the \ac{ics}, or the electronic units of the instruments. 
This case can, for example, be triggered by a failure in the unit itself, when the power consumption of the unit drops and the radiative cooling of the electronic box dominates. 
It can also occur when the predictions from the thermal modeling were not correct (see \sref{thermal}). This is likely to be the case especially during the ascent, given the unknown  weather conditions and the uncontrolled spinning of the observatory. Survival heaters were added to critical components, where the thermal modeling predicted large uncertainties.

Elevation brake and hard stops: The elevation brake is a critical component during launch and landing of the observatory, holding the telescope in the safe 0\dg{}-elevation position. The strong torsional acceleration at the release from the Hercules launch vehicle caused this brake to fail during the 2022 launch. A significantly strengthened launch-lock mechanism ensured a safe launch under even harsher conditions in 2024. As an additional fault protection measure, two mechanical hard stops at -5\dg{} and +50\dg{} elevation ensure that the telescope stays within the clear area, to avoid damage in the case of a wrong elevation command.

However, the vast majority of fault protection measures are implemented in software.
All subsystems have a plethora of fault protection code in place, reacting to out-of-bounds \ac{hk}-values in an automated way. The last, very important element in the fault protection chain  was the 24\,h/day constant monitoring of the observatory operations and health by the ground operators.

\section{Scientific Performance and Data Concept}\label{performance}

The intensive testing of \sunriseiii{} in the laboratories, the hang tests and during the 2022 campaign provided insight into the scientific performance of the observatory. 
Here we summarize some selected parameters of the global performance of the observatory, whereas instrument specific scientific performance is described in the  individual instrument sections and papers \cite[]{cws25,susi25,tumag25,scip25,gondola25}. A detailed post-flight evaluation of the observatory performance will only be available after the data reduction phase is completed, and is intended to be published with the first scientific papers of the project.

\subsection{Pointing Stability and Accuracy}\label{perf:pointing}

Stable sun-pointing was achieved prior to the flights during the hang tests at \ac{mps} and \ac{esrange}. The performance estimates from those tests suffer from the fact that the atmospheric disturbances, the so-called ``seeing'', do not provide a stable and contrast-rich image of the Sun, but only a highly variable, blurry one. Nevertheless, the \ac{pcs} could be tested and proved to work even under those unfavorable conditions. Surprisingly it was even possible, during the hang test at the \ac{mps} in Göttingen, to achieve closed-loop locking on a sunspot with the \ac{cws}.

At both locations, an \ac{rms} jitter in the range of 1\arcsec{} could be achieved by using the \ac{sgt}. 
During the 2022 flight, the azimuth pointing stability could be measured, and it exceeded the expected performance and achieved the same \ac{rms} jitter (1\arcsec{}) albeit with only the \ac{imu} of the \ac{pcs} being available. The elevation pointing jitter and the expected improvement of the pointing stability using the \ac{sgt} could not be tested because of the failure of the launch-lock mechanism.

Pointing accuracy, i.e., how well the telescope can be pointed at a desired target on the Sun, depends on a limb-seeking calibration routine. During hang tests on the ground, atmospheric seeing reduces the clarity of the Sun's limbs resulting in reduced absolute pointing accuracy. During those ground tests, by aiming the telescope at various prominent positions on the Sun, e.g. active regions or the solar limb, we estimated the pointing accuracy to be in the range of 10--15\arcsec{}.

During the flight in 2024, the gondola \ac{pcs} achieved stable sun-pointing continuously, with only one major interruption due to an unexpected avionics computer reboot. The \ac{rms} pointing stability was determined to be on average better than 3\arcsec{}, exceeding the specifications by a factor of 5. Locking on solar features with the \ac{cws} in closed loop was achieved almost continuously during the science observing programs, with a record-long time series of more than four hours at milli-arcsecond \ac{rms} accuracy. 

The absolute pointing accuracy of the gondola \ac{pcs}, directly after performing the limb-seeking calibration, was better than the values found during the hang tests. With time, due to elevation dependence of gravitational force and thermal stress on the telescope, an offset was introduced,  mainly along the elevation direction. A new limb-seeking calibration was performed when this offset in absolute pointing was larger than half the 60\arcsec{}-\ac{fov} of the science instruments, or when absolute pointing accuracy was crucial for the science program.

\subsection{Slit Alignment}\label{perf:slit}

The near-space like observing conditions in the stratosphere are not only advantageous for the image quality, stability and the access to new wavelength regimes, but also for removing the effects of differential refraction in the Earth's atmosphere, enabling the co-alignment of instruments operating at different wavelength bands. Both, \ac{scip} and \ac{susi} are equipped with their own slit-scanning mechanisms, allowing for a near-perfect alignment of the slits at any position within the \ac{tumag} \ac{fov}. Simultaneous spectra over long time periods in the \ac{nir} and \ac{nuv} with sub-arcsecond alignment accuracy (0.1\arcsec{}) over hours become possible. The slit-alignment procedure could be tested during the hang tests using the grid target in the F2 filter wheel, and was repeated during the commissioning phase of the flight. Similarly, the scan speeds for both slits were calibrated, allowing for simultaneous scans of the full \acp{fov} of both slit instruments.

Many observing programs required strictly co-aligned scanning. Therefore, a common scan speed for both instruments was agreed upon within the \ac{sswg} during the phase of defining the observation blocks. The exact alignment of the slits and the calibration of the scan speeds was part of the in-flight calibration procedure. Using near-real-time slit-jaw images of \ac{susi} and \ac{scip} of the grid target in F2, which were continuously sent to the \ac{goc} through the various communication channels (see \sref{comm}), the position of the slit could be accurately determined. The \ac{sppt} software (see \sref{sppt}) computed offsets and speed based on cross-correlating the two slit-jaw images.

The angle alignment of the \ac{susi} and the \ac{scip} slits was performed during the installation of the instruments into the \ac{pfi}. A very accurate angle alignment with a difference of only 0.1\dg{} could be achieved and maintained during the flight. A preliminary assessment of the data shows that the common scanning was successful, but the detailed performance analysis and alignment accuracy can only be performed during the data reduction phase.

\subsection{\sunriseiii{} Open-Access Data Concept}\label{dataconcept}

All data obtained with the \sunriseiii{} observatory will be treated as a public resource. The tentative schedule for the data recovery, processing, and publishing and the expected duration of the individual steps are sketched here. 
\begin{enumerate}[(i)]
\item After landing (16 July 2024): Recovery of the \ac{dss} from the landing site in Canada and transport to \ac{mps} (duration: 1 month).
\item Raw-data archiving and backup to the storage servers hosted by the \ac{mps} computing center (1 month).
\item \label{reduction} Data reduction: The raw data will be processed with the \sunriseiii{} data pipeline software to produce calibrated spectropolarimetric data (Stokes $I$, $Q$, $U$, and $V$) (3--6 months).
\item Embargo period: During this period the partner institutes and invited collaborators (e.g. co-observers) have prioritized access to the data. This period is dedicated to achieve the highest possible data quality prior to the public data release. It also allows the participating institutes to execute the scientific analysis according to their submitted ideas which formed the baseline of the observation timeline (6 months).
\item Public release of the science-ready \sunriseiii{} data in the \ac{fits} file format \cite[]{wells81}, accessible through a website hosted at the \ac{mps}.
\end{enumerate}
According to this schedule, the date for the public release of the data of the successful \sunriseiii{} 2024 flight is expected for the end of 2025.
However, delays in this schedule are possible. Especially the duration of step \ref{reduction} is influenced by the in-flight performance of the instruments, the telescope, and the \ac{pcs}. Although the data reduction software has been prepared thoroughly, the complexity of the observations does not allow to cover all possible effects in advance, and might even require post-calibration measurements using the flown instruments in the \ac{mps} laboratory.
A delay in the data reduction process will automatically shift the public release date by the same amount.
The timeline of the data reduction process and the data policy are described in the \docdb{SR3-MPS-CO-GEN-002}.

\section{Summary and Outlook}\label{outlook}

The \sunriseiii{} stratospheric observatory has been designed to deepen our understanding of the magnetism and the dynamic processes of the solar atmosphere. Equipped with instrumentation of a complexity comparable to world-class ground-based solar observatories, the 1-meter aperture telescope allows to resolve solar structures as small as 60\,km on the Sun.
Three scientific instruments allow for diffraction limited, spectropolarimetric observations at high spectral resolution: (i) \ac{susi}  performs many-line observations in the vastly unexplored \ac{nuv}, (ii) \ac{tumag} provides 2-dimensional maps over a \ac{fov} of $46\,\times\,46$\,Mm$^2$ in the visible, and (iii) \ac{scip}  concentrates on the \ac{nir} wavelength range.

The common goal of all three instruments is to provide highly complementary data to determine the dynamics and the magnetism in the solar atmosphere, seamlessly from the photosphere to the chromosphere in multi-hour, uninterrupted data sets with constant high resolution and quality. The diffraction limited observations reach a polarimetric sensitivity of better than \ten{-3} of the continuum intensity for all science instruments. This requires the stabilization of the telescope and the instrumentation to an accuracy of better than 0.005\arcsec{} \ac{rms}, achieved with the combination of the 3-axis stabilized gondola \acl{pcs} and an internal image stabilization system based on correlation tracking (\ac{cws}), which also controls the autofocus mechanism of the telescope and allows for an in-flight coma correction.

The science flight of \sunriseiii{} took place from 10 to 16 July 2024. Stratospheric winds carried the observatory on a long-distance zero-pressure balloon from \acl{esrange} in Sweden to the landing site west of Great Bear Lake (Canada) at a float altitude between 33--37\,km in 6.5 days. The preliminary analysis of in-flight data indicates excellent performance of the observatory, exceeding the high expectations of the \sunriseiii{} team. A detailed performance assessment will only be possible after the completion of the data reduction phase. 
We hope that the rich data gathered by \sunriseiii{} will help to solve many mysteries of the solar atmosphere, and to discover new ones.

\begin{acknowledgments}

We thank the Max Planck Society for the financial support. A special thanks goes to the Max-Planck-Förderstiftung, especially to Frau Pfündl, Frau Philipp and Prof. Pöllath, for their enthusiasm for \sunriseiii{} and the generous funding of this research. We thank the employees of \ac{csbf} and \ac{esrange} / \ac{ssc} for their support and hospitality during the launch campaigns. 
We warmly thank the kind and professional help of the personnel at Calar Alto Astronomical Observatory (Spain) who re-aluminized the \sunrise{} mirrors. 
We are grateful to C.~Müller and M.~Wölfle of \ac{imtek} (Freiburg, Germany) for excellent technical support in diamond milling of the \ac{hrw}, and Mike Smith from Aerostar International (TX, USA) for the excellent support during the 2024 campaign for providing the wind shield film for the \acp{erack} and the \ac{pfi}.

This version of the article has been accepted for publication, after peer review but is not the Version of Record and does not reflect post-acceptance improvements, or any corrections. The Version of Record is available online at: \href{https://doi.org/10.1007/s11207-025-02485-1}{https://doi.org/10.1007/s11207-025-02485-1}.
\end{acknowledgments}

\begin{fundinginformation}
\sunriseiii{} is supported by funding from the Max-Planck-Förderstiftung (Max Planck Foundation), NASA under Grant \#80NSSC18K0934 and \#80NSSC24M0024 (``Heliophysics Low Cost Access to Space' program),  and the ISAS/JAXA Small Mission-of-Opportunity program and JSPS KAKENHI JP18H05234/JP23H01220. 
This research has received financial support from the European Union’s Horizon 2020 research and innovation program under grant agreement No. 824135 (SOLARNET) and No. 101097844 (WINSUN) from the European Research Council (ERC). It has also been funded by the Deutsches Zentrum für Luft- und Raumfahrt e.V. (DLR, grant no. 50 OO 1608).
The Spanish contributions have been funded by the Spanish MCIN/AEI under projects RTI2018-096886-B-C5, and PID2021-125325OB-C5, and from ``Center of Excellence Severo Ochoa'' awards to IAA-CSIC (SEV-2017-0709, CEX2021-001131-S), all co-funded by European REDEF funds, ``A way of making Europe''. 

D.~Orozco~Suárez acknowledges financial support from a {\em Ram\'on y Cajal} fellowship.
The contributions of J.~H\"olken, D.~Vukadinović, E.~Harnes and K.~Sant have been supported by the International Max Planck Research School for Solar System Science at the University of Göttingen (IMPRS), Germany.      
F.\,A.~Iglesias is a member of the ``Carrera del Investigador Cient\'ifico" of CONICET and supported by \ac{mps} through the Max Planck Partner Group between \ac{mps} and the University of Mendoza, Argentina.
The research activities and the flight operation of the SCIP team members, R.~T. ~Ishikawa, M.~Kubo, Y.~Kawabata, and T.~Oba, have been supported from the JSPS KAKENHI grants No. 23KJ0299, No. 24K07105, No. 23K13152, and No. 21K13972, respectively.
\end{fundinginformation}

\bibliographystyle{spr-mp-sola}
\bibliography{main} %

\newpage
\appendix
\section{Acronyms}

\begin{acronym}
\setlength{\itemsep}{-4ex}

\acro{aia}[AIA]{Atmospheric Imaging Assembly}
\acro{alma}[ALMA]{Atacama Large Millimeter/sub-millimeter Array}
\acro{amhd}[AMHD]{{Analog, Motor, and Heaters Drivers}}
\acro{ao}[AO]{adaptive optics}
\acro{bbso}[BBSO]{Big Bear Solar Observatory} 
\acro{bpo}[BPO]{Balloon Program Office}
\acro{cad}[CAD]{Computer Aided Design}
\acro{cfrp}[CFRP]{carbon fiber-reinforced polymers}
\acro{chase}[CHASE]{Chinese \halpha{} Solar Explorer}
\acro{cmos}[CMOS]{complementary metal oxide semiconductor}
\acro{cnoc}[CNOC]{cold non-operational case}
\acro{coc}[COC]{cold operational case}
\acro{coi}[CoI]{Co-Investigator}
\acro{csbf}[CSBF]{Columbia Scientific Ballooning Facility}
\acro{ct}[CT]{correlation tracker}
\acro{cws}[CWS]{Correlating Wavefront Sensor}
\acro{dkist}[DKIST]{Daniel K. Inouye Solar Telescope}
\acro{docdb}[S3-DocDB]{\sunriseiii{} Document Database}
\acro{dss}[DSS]{Data Storage System}
\acro{dst}[DST]{Dunn Solar Telescope}
\acro{egse}[EGSE]{electrical ground support equipment}
\acro{elink}[E-Link]{Esrange Airborne Data Link}
\acro{eoc}[EOC]{Esrange Operations Center}
\acro{erack}[E-rack]{electronic rack}
\acro{esrange}[Esrange]{Esrange Space Center}
\acro{evtm}[EVTM]{Ethernet Via Telemetry}
\acro{fem}[FEM]{finite element model}
\acro{fits}[FITS]{Flexible Image Transport System}
\acro{fov}[FoV]{field-of-view}
\acro{fps}[fps]{frames per second}
\acro{fram}[FRAM]{ferroelectric random access memory}
\acro{frr}[FRR]{Flight Readiness Review}
\acro{fsm}[FSM]{finite state machine}
\acro{fwhm}[FWHM]{full width at half maximum}
\acro{goc}[GOC]{Göttingen Operations Center}
\acro{gps}[GPS]{Global Positioning System}
\acro{gregor}[GREGOR]{GREGOR}
\acro{gst}[GST]{Goode Solar Telescope}
\acro{gusto}[GUSTO]{Galactic/Extragalactic ULDB Spectroscopic Terahertz Observatory}
\acro{hmi}[HMI]{Helioseismic Magnetic Imager}
\acro{hnoc}[HNOC]{hot non-operational case}
\acro{hoc}[HOC]{hot operational case}
\acro{hrw}[HRW]{Heat Rejection Wedge}
\acro{hvps}[HVPS]{high voltage power supply}
\acro{iaa}[IAA]{Instituto de Astrofísica de Andalucía}
\acro{iac}[IAC]{Instituto de Astrofísica de Canarias}
\acro{ics}[ICS]{Instrument Control System}
\acro{icu}[ICU]{Instrument Control Unit}
\acro{ihop}[IHOP]{IRIS/\hinode{} Operations Plan}
\acro{imax}[\textsc{IMaX}]{Imaging Magnetograph eXperiment}
\acro{imtek}[IMTEK]{Department of Microsystems Engineering}
\acro{imu}[IMU]{Inertial Measurement Unit}
\acro{inta}[INTA]{Instituto Nacional de Técnica Aeroespacial}
\acro{iridium}[Iridium Pilot]{Iridium Pilot\textsuperscript{\textregistered}}
\acro{iris1}[IRIS-1]{Image Recording Instrument for \sunriseii{}}
\acro{iris2}[IRIS-2]{Image Recording Instrument for \sunriseiii{}}
\acro{iris}[IRIS]{Interface Region Imaging Spectrograph}
\acro{ir}[IR]{infrared}
\acro{islid}[\textsc{ISLiD}]{Image Stabilization and Light Distribution unit}
\acro{jaxa}[JAXA]{Japan Aerospace Exploration Agency}
\acro{apl}[APL]{Johns Hopkins University Applied Physics Laboratory}
\acro{kbsi}[KBSI]{Korea Basic Science Institute}
\acro{kis}[KIS]{Institut für Sonnenphysik}
\acro{lcvr}[LCVR]{liquid crystal variable retarder}
\acro{ldb}[LDB]{long-distance balloon}
\acro{ld}[LD]{launch date}
\acro{led}[LED]{light-emitting diode}
\acro{lldpe}[LLDPE]{linear low-density polyethylene}
\acro{los}[LoS]{line-of-sight}
\acro{lte}[LTE]{local thermodynamic equilibrium}
\acro{mast}[MAST]{Multi-Application Solar Telescope}
\acro{mhd}[MHD]{magneto-hydrodynamic}
\acro{MHD}[MHD]{Magneto-Hydrodynamic}
\acro{mit}[MIT]{Massachusetts Institute of Technology}
\acro{mli}[MLI]{multilayer insulation}
\acro{mpae}[MPAe]{Max-Planck-Institut für Aeronomie}
\acro{mpg}[MPG]{Max-Planck-Gesellschaft}
\acro{mps}[MPS]{Max Planck Institute for Solar System Research}
\acro{mrr}[MRR]{Mission Readiness Review}
\acro{mtc}[MTC]{Main Telescope Controller}
\acro{mtu}[MTU]{Momentum Transfer Unit}
\acro{muramce}[MURaM-ChE]{MURaM Chromospheric Extension}
\acro{muram}[MURaM]{MPS/University of Chicago Radiative MHD}
\acro{naoj}[NAOJ]{National Astronomical Observatory of Japan}
\acro{nasa}[NASA]{National Aeronautics and Space Administration}
\acro{nic}[NIC]{network interface controller}
\acro{nir}[near-IR]{near-infrared}
\acro{nlte}[non-LTE]{non-local thermodynamic equilibrium}
\acro{nuv}[near-UV]{near-ultraviolet}
\acro{nvst}[NVST]{New Vacuum Solar Telescope}
\acro{occ}[OCC]{Operations Control Center}
\acro{olr}[OLR]{outgoing longwave radiation}
\acro{pbs}[PBS]{polarising beam-splitter}
\acro{pcb}[PCB]{printed circuit board}
\acro{pcu}[PCU]{Power Converter Unit}
\acro{pdu}[PDU]{Power Distribution Unit}
\acro{pfi}[PFI]{Post Focus Instrumentation}
\acro{phi}[PHI]{Polarimetric and Helioseismic Imager}
\acro{pi}[PI]{Principal Investigator}
\acro{pmu}[PMU]{Polarization Modulation Unit}
\acro{poe}[PoE]{Power over Ethernet}
\acro{ppd}[PPD]{PFI Power Distribution Unit}
\acro{psg}[PSG]{Polarization State Generator}
\acro{psf}[PSF]{point spread function}
\acro{pcs}[PCS]{Pointing Control System}
\acro{hk}[HK]{housekeeping}
\acro{ramon}[RAMON]{RAdiation MONitor}
\acro{rms}[rms]{root-mean-square}
\acro{rtc}[RTC]{real-time clock}
\acro{s2n}[S/N]{signal-to-noise}
\acro{s3pc}[S$^3$PC]{Spanish Space Solar Physics Consortium}
\acro{scip}[\textsc{SCIP}]{\sunrise{} Chromospheric Infrared spectro-Polarimeter}
\acro{sdo}[SDO]{Solar Dynamics Observatory}
\acro{sgt}[SGT]{Sun Guider Telescope}
\acro{sipm}[SiPM]{Silicon PhotoMultiplier}
\acro{sip}[SIP]{Support Instrumentation Package}
\acro{smart}[SMART]{Solar Magnetic Activity Research Telescope}
\acro{smm}[SMM]{scan mirror mechanism}
\acro{solo}[Solar Orbiter]{Solar Orbiter}
\acro{sophi}[SO/PHI]{Solar Orbiter Polarimetric and Helioseismic Imager}
\acro{sotnfi}[NFI]{Narrowband Filter Imager}
\acro{sotsp}[SP]{Spectro-Polarimeter}
\acro{sot}[SOT]{Solar Optical Telescope}
\acro{spg}[SPG]{Solar Physics Group}
\acro{spgcam}[SPGCam]{Solar Physics Group Camera}
\acro{ssct}[SSCT]{Science/Support Compatibility Test}    
\acro{ssc}[SSC]{Swedish Space Corporation}
\acro{ssd}[SSD]{solid state disk}
\acro{ssh}[SSH]{Secure Shell}
\acro{ssm}[SSM]{second surface mirror}
\acro{sst}[SST]{Swedish Solar Telescope}
\acro{sswg}[SSWG]{\sunrise{} Science Working Group}
\acro{ss}[SS]{Science Stack}
\acro{starlink}[Starlink]{Starlink (a division of SpaceX)}
\acro{sufi}[\textsc{SuFI}]{\sunrise{} Filter Imager}
\acro{sppt}[\textsc{SPPT}]{\sunriseiii{} Science Planning and Pointing Tool}
\acro{sunrise}[\textsc{Sunrise}]{{\textbf{S}olar \textbf{U}V to \textbf{N}ear-infra\textbf{R}ed \textbf{I}maging and \textbf{S}pectropolarimetric \textbf{E}xploration}}
\acro{susi}[\textsc{SUSI}]{\sunrise{} \acs{uv} Spectropolarimeter and Imager}
\acro{suit}[SUIT]{Solar Ultraviolet Imaging Telescope}
\acro{suvi}[SUVI]{Solar Ultraviolet Imager}
\acro{sutri}[SUTRI]{Solar Upper Transition Region Imager}
\acro{tac}[TAC]{Telescope Allocation Committee}
\acro{tc}[TC]{Telecommand}
\acro{tdrss}[TDRSS]{Tracking \& Data Relay Satellite System}
\acro{tmtc}[TM/TC]{Telemetry and Telecommand}
\acro{tm}[TM]{Telemetry}
\acro{tumag}[\textsc{TuMag}]{Tunable Magnetograph}
\acro{tv}[TV]{thermal-vacuum}
\acro{unival}[UV]{Universitat de València}
\acro{upm}[UPM]{Universidad Politécnica de Madrid}
\acro{usaf}[USAF]{U.S. Air Force}
\acro{uv}[UV]{ultraviolet}
\acro{vda}[VDA]{vapor deposited aluminum}
\acro{vpn}[VPN]{virtual private network}
\acro{vtt}[VTT]{German Vacuum Tower Telescope} 
\acro{wfs}[WFS]{wavefront sensor}

\end{acronym}

\end{document}